\documentclass{article}
\usepackage{arxiv}
\usepackage{lipsum}
\usepackage{amsfonts}
\usepackage{graphicx}
\usepackage{epstopdf}
\usepackage{algorithmic}
\ifpdf
  \DeclareGraphicsExtensions{.eps,.pdf,.png,.jpg}
\else
  \DeclareGraphicsExtensions{.eps}
\fi

\usepackage{enumitem}
\setlist[enumerate]{leftmargin=.5in}
\setlist[itemize]{leftmargin=.5in}

\newtheorem{theorem}{Theorem}
\newtheorem{corollary}{Corollary}[theorem]
\newtheorem{lemma}[theorem]{Lemma}
\newtheorem{definition}{Definition}[section]
\newtheorem{remark}{Remark}
\usepackage{authblk}
\usepackage{setspace}
\usepackage{verbatim}
\usepackage{longtable}
\usepackage{afterpage}
\usepackage{latexsym}
\usepackage{amsmath}
\usepackage{amsfonts}
\usepackage{amssymb}
\usepackage{amsbsy}
\newcommand{\nn}{\nonumber}
\newcommand{\Ind}{1\!\mathrm{l}}
\usepackage{mathrsfs}
\usepackage{algorithm}
\usepackage{wrapfig}
\usepackage{float}
\usepackage{hyperref}
\usepackage{soul}
\usepackage[normalem]{ulem}

\usepackage{multirow}

\usepackage{color}
\usepackage{graphicx}
\usepackage{subfigure}
\usepackage{colortbl}
\usepackage{comment}

\newcommand{\INDSTATE}[1][1]{\STATE\hspace{#1\algorithmicindent}}

\begin{document}

\title{\Large A Nonparametric Bayesian Framework for Uncertainty Quantification in Stochastic Simulation}
\author[1]{Wei Xie}
\author[2]{Cheng Li}
\author[3]{Yuefeng Wu}
\author[4]{Pu Zhang}

\affil[1]{Department of Mechanical and Industrial Engineering, Northeastern University
  (w.xie@northeastern.edu)}
\affil[2]{Department of Statistics and Data Science, National University of Singapore
  (stalic@nus.edu.sg)}
\affil[3]{Department of Mathematics and Computer Science, University of Missouri St. Louis
  (wuyue@umsl.edu)}
\affil[3]{Department of Industrial and Systems Engineering, Rensselaer Polytechnic Institute
  (zhangpu0703@gmail.com)} 

\maketitle

\begin{abstract}
  When we use simulation to assess the performance of stochastic systems, the input models used to drive simulation experiments are often estimated from finite real-world data. There exist both input model and simulation estimation uncertainties in the system performance estimates. Without strong prior information on the input models and the system mean response surface, in this paper, we propose a Bayesian nonparametric framework to quantify the impact from both sources of uncertainty. Specifically, since the real-world data often represent the variability caused by various latent sources of uncertainty, Dirichlet Processes Mixtures (DPM) based nonparametric input models are introduced to model
a mixture of heterogeneous
distributions, which can
  faithfully capture the important features of real-world data, such as multi-modality and skewness. Bayesian posteriors of flexible input models characterize the input model estimation uncertainty, which automatically accounts for both model selection and parameter value uncertainty. Then, input model estimation uncertainty is propagated to outputs by using direct simulation. Thus, under very general conditions, our framework delivers an empirical credible interval accounting for both input and simulation uncertainties. A variance decomposition is further developed to quantify the relative contributions from both sources of uncertainty. Our approach is supported by rigorous theoretical and empirical study. 
\end{abstract}

\keywords{Nonparametric Bayesian approach, design of experiments, stochastic simulation, uncertainty quantification, input uncertainty, Dirichlet processes mixtures}


\section{Introduction}

Stochastic simulation is widely used in many applications to assess the performance of complex stochastic systems, e.g., manufacturing, supply chain and health care systems. 
For example, when we simulate a healthcare service system, the random patients interarrival and service times are characterized by input distributions.
However, the \textit{input models}, defined as the driving stochastic processes in simulation experiments, are often estimated from finite real-world data. Therefore, there exist two sources of system performance estimation uncertainty, including: (1) the \textit{input uncertainty} which is the output variation due to input model estimation error, and (2) the \textit{simulation uncertainty} which is induced by the simulation estimation error due to finite simulation budget. Ignoring either source of uncertainty can lead to unfounded confidence in the simulation assessment of system performance.

Various approaches have been proposed in the literature to quantify the input and simulation uncertainties; see \cite{Barton_2012,Song_Nelson_2014wsc,Lam_2016wsc,CananAlpWei_2020} for a comprehensive review. Based on methodologies developed to quantify the input model estimation uncertainty, they can be divided into frequentist and Bayesian approaches. The frequentist approaches typically study the sampling distributions of point estimators of underlying input models. Since it could be hard to get the exact sampling distributions in many situations, the asymptotic approximation, including the normal approximation and the bootstrap, is often used to quantify the input model estimation uncertainty, which is valid when the amount of real-world data is large.
\textit{However, even in the current big data world, we often face the situations where the amount of real-world data is limited, especially for high-tech products with short life cycles.} For example, biopharmaceutical manufacturing requires 9 to 12 months from raw materials sourcing to the finished drug products, and it requires another 2 to 3 months for quality testing. However, the drug substances typically expire after 18 to 36 months; see \cite{Otto_Santagostino_2014}. 
Compared to frequentist methods, Bayesian approaches derive the posterior distributions quantifying the input model estimation uncertainty and they do not need a large-sample asymptotic approximation for their validation. It is also straightforward for Bayesian approaches to incorporate the prior information about the underlying input models; see \cite{Xie_Bayesian_2014} for the discussion of frequentist v.s. Bayesian approaches for input uncertainty.

{Thus, in this paper, we focus on developing a Bayesian nonparametric framework to quantify the estimation uncertainty of system mean performance, especially when we do not have strong prior information on the underlying input models and system mean response surface.} We consider univariate input models, which model independent and identically distributed (i.i.d.) data by mutually independent input distributions. Since we often assume that multivariate input models are characterized by marginal distributions and the dependence in the simulation literature \cite{biller2006multivariate}, the proposed univariate input models can be used to construct input models with dependence through copula-based approaches; see for example \cite{biller2009copula, XieLiZhang2017}.

Many existing methods assume specific parametric families for input models with unknown parameter values estimated from finite real-world data; see the review in \cite{Barton_2012}. The input model estimation uncertainty can be quantified by the posteriors of model parameters.
However, the standard parametric distributions cannot always capture the rich features in the real-world data, such as skewness and multi-modality.
If the selected parametric families do not have sufficient flexibility and cannot represent the underlying input models well, there always exists the distribution family selection error which does not vanish as the amount of real-world data becomes large. This inconsistent estimation could lead to incorrect inference even for the moderate size of real-world data \cite{Hjort_Holmes_2011}.

One possible remedy for the inconsistency of parametric approaches is to introduce the {\it family uncertainty}, which accounts for the input model selection error among a \textit{pre-specified} pool of candidate parametric families. The study in \cite{Chick_2001} proposed Bayesian Model Averaging (BMA) to quantify input model estimation uncertainty from both families and parameter values, where the family uncertainty is characterized by the posterior probabilities of different candidate parametric models. 
However, \textit{BMA is based on the assumption that all data come from one of candidate distributions; see Section~14.1 in \cite{Bishop_2006}.} In other words, BMA relies on the assumption that all data are generated from a single underlying true parametric family, and this family must be included as a candidate \it priori. 
Since it is difficult for any standard parametric family to capture the rich features in the real-world data, it could be challenging to select the appropriate candidate models for BMA.
Furthermore, if the selected families are not mutually exclusive, such as exponential and Gamma distributions, it can potentially lead to model identification problems.



Our study is motivated by the facts: (1) the real-world data represent the variability caused by various latent sources of uncertainty, which can lead to rich features, e.g., heterogeneity, multi-modality, skewness, and tails; and (2) we often have very limited real-world data in many applications.
Flexible Dirichlet Processes Mixtures (DPM) based Bayesian nonparametric approach is introduced for simulation input modeling and uncertainty quantification, which can capture important features in the real-world data. 
The DPM with Gaussian kernel was introduced in the statistics community; see for example \cite{west_1990,Escobar_West_1995}, etc.
It is extended to other kernel functions; see for example \cite{hanson2006,Kottas2006, Wu_Ghosal_2008}. 
In general, DPM has demonstrated robust performance in terms of density estimation (\cite{Escobar_West_1995,Ghosh_2003,Gorur2010}, etc.).
The Markov chain Monte Carlo (MCMC) method enables efficient sampling of mixture distributions from the posterior; see for example \cite{Escobar_West_1995,Neal2000,hanson2006,Kottas2006,Wang2011}.


From the modeling perspective, DPM has clear advantages over
standard parametric families because the variability across different mixing components naturally represents various latent sources of uncertainty, which makes it straightforward to capture the important properties in the real-world data.
Different from parametric approaches, the number of mixing components and parameters can automatically adjust to the complex features of real data, such as multi-modality, skewness, and tails.
Thus, our empirical study demonstrates that DPM has better and more robust finite sample performance.
From the theoretical perspective, DPM is able to consistently estimate a wide class of distributions under relatively general conditions (\cite{Ghosal1999,Wu_Ghosal_2008}, etc.). 
Compared to BMA, our approach 
avoids the difficulty of selecting the ``appropriate" candidate distributions. From the computational perspective, one can develop efficient posterior samplers for DPM with popular exponential family kernel density (see our Section~\ref{subsec:gibbssampler}, \cite{Escobar_West_1995,Neal2000}, etc.).

Among frequentist approaches, empirical distribution is the most commonly used
nonparametric approach in the simulation literature, and the bootstrap is typically used to quantify the input model estimation uncertainty; see for example \cite{Barton_Schruben_1993,Barton_2007}. Empirical distribution is simple and easy to implement. However, DPM has some important advantages compared to empirical distribution. 
First, even though the underlying true distribution is continuous, empirical distribution is always discrete. Smoothing methods, such as kernel estimators \cite{parzen1962estimation,baszczynska2016kernel} and splines \cite{cheng2002regression}, can be used to smooth the empirical cumulative distribution functions (ecdf) at the cost of losing unbiasedness. Second, with limited real-world data, the empirical distribution could overlook some important properties in the underlying input models. For example, in the presence of extreme values, DPM with infinite mixture components can provide a better fit to the tails.
Third, the validity of using the bootstrap to quantify the input uncertainty relies on asymptotic approximation, and therefore it requires large samples of real-world data. As we mentioned above, the decision makers often face the situations where the amount of real-world data is limited. As a Bayesian approach, DPM can overcome these limitations. Our empirical study demonstrates that DPM has better finite sample performance compared to frequentist competitors, especially when the sample size of real-world data is limited. 
The DPM-based input models have the \textit{potential} to detect the latent sources of uncertainty, 
and the selection of kernel function may impact the performance.



Therefore, in this paper, we develop a flexible Bayesian nonparametric framework to quantify the system mean response estimation uncertainty. We first introduce DPM-based Bayesian nonparametric input modeling and uncertainty quantification, which can capture the important properties in the real-world data. \textit{The samples drawn from posteriors of flexible input models can automatically quantify both model selection and parameters value uncertainty.}
Then, the input model estimation uncertainty is propagated to the output through direct simulation which runs simulations at each posterior sample of input models to estimate the system mean response. Our framework leads to a sampling procedure that delivers a percentile \textit{empirical credible interval} (CrI) quantifying the overall uncertainty of system performance estimation. 

In sum, the main contributions of our paper are as follows.
\begin{enumerate}
	
	\item 
	We propose a DPM-based Bayesian nonparametric framework accounting for both input and simulation uncertainties. It delivers a percentile empirical CrI quantifying the overall estimation uncertainty of system mean response. Furthermore, a variance decomposition is developed to quantify the relative contributions from input and simulation uncertainties.

	\item We provide the theoretical support for our nonparametric 
	framework. 
	The theory includes the consistency of univariate nonparametric continuous input models with supports, including (i) $\Re^+ \equiv [0,\infty)$, the nonnegative half real line; (ii) $\Re \equiv (-\infty,\infty)$, the entire real line; or 
(iii) a bounded interval $[a_1,a_2]\in \Re$. Beyond the existing consistency results of DPM with Gaussian and Gamma kernels for continuous distributions with support $\Re$ and $\Re^+$, we further prove the new posterior consistency of DPM with Beta kernel for continuous distributions supported on bounded intervals. 
In addition, we show the consistency of the proposed empirical CrI accounting for both input and simulation uncertainties.
	{As the amount of real-world data increases, without strong prior information on the distribution family, the posterior distributions of input models can converge to the underlying distributions.}
	Given the {finite} real-world input data, as the simulation budget increases, this interval converges to the CrI quantifying the impact of input uncertainty with the true mean response surface known. Further, as the amount of real-world data and the
	simulation budget go to infinity, the CrI converges to the true system performance.
	
	\item Since real-world data often represent the variability caused by various latent sources of uncertainty in many situations, 
	the DPM-based input models provide sufficient flexibility to capture the important features in the real-world data. 
	The empirical study demonstrates that the proposed framework has better performance than existing approaches in terms of both input density estimation and system mean response assessment. In addition, the finite sample performance of proposed framework is robust to the violation of sufficient conditions required by the asymptotic consistency proof for DPM-based input models.
\end{enumerate}

The remainder of the paper is organized as follows.
In Section~\ref{sec:BayesianFramework}, a Bayesian nonparametric framework is introduced to quantify the overall uncertainty of the system performance estimates.
We then report results of finite sample behaviors on both input models and system mean response estimation in Section~\ref{sec:EmpiricalStudy}, and we conclude this paper in Section~\ref{sec:Conclusion}. All proofs, derivations and other supplementary studies are included in the online Supplement.

\section{A Bayesian Nonparametric Framework} 
\label{sec:BayesianFramework}

When we use simulation to assess the stochastic system performance, the output from the $j$-th replication with input models, denoted by $F$, can be written as
\begin{equation}
Y_j(F) = \mu(F)+e_j(F)
\label{eq.responseY} \nonumber
\end{equation}
where $\mu(F)$ denotes the system mean response and $e_j(F)$ represents the simulation error {with mean zero and variance $\sigma_e^2(F)$.}
The input models in general could include multiple distributions, e.g., $F\equiv \{F_1,F_2,\ldots,F_L\}$. For ease of presentation, we assume that $F$ consists of a single univariate model; otherwise our DPM-based input modeling and Bayesian uncertainty quantification can be applied to each distribution in $F$. Denote the underlying unknown true input model by $F^c$.
{{We are interested in the system mean response at the true input model, denoted by $\mu^c\equiv\mu(F^c)$.}
\textit{
In this paper, we introduce a Bayesian nonparametric framework to quantify the overall estimation uncertainty of $\mu^c$.}}

Since the simulation output depends on the choice of input distribution $F$, the input model failing to capture important features of $F^c$ can lead to poor estimates of system performance. It is desirable to construct the input model that can faithfully capture the important properties (e.g., heterogeneity, multi-modality, and skewness). Thus, in Section~\ref{subsec:inputModel}, we present the flexible nonparametric DPM-based input models, in which we use a mixture of heterogeneous distributions to capture these rich properties induced by various latent sources of uncertainty. 

The underlying true input distribution $F^c$ is estimated by finite real-world data of size $m$, denoted by $\mathbf{X}_m\equiv\{X_1,X_2,\ldots,X_m\}$ with $X_i\stackrel{i.i.d.}\sim F^c$. The posterior distribution of the \textit{flexible} input model, denoted by $p(F|\mathbf{X}_m)$, can be used to quantify the model estimation uncertainty.
Since the DPM model does not have closed form distributions for analytical posterior analysis, we provide Gibbs samplers in Section~\ref{subsec:gibbssampler} to efficiently draw posterior samples of input models, $\mathcal{F}_B\equiv \{\widetilde{F}^{(1)},\widetilde{F}^{(2)},\ldots,\widetilde{F}^{(B)}\}$, from $p(F|\mathbf{X}_m)$ quantifying the input model estimation uncertainty. We {show} the asymptotic consistency of $p(F|\mathbf{X}_m)$ in Section~\ref{subsec:inputSamplingConvergence} and study its finite sample performance in Section~\ref{sec:EmpiricalStudy}. In this paper, the notation $\widetilde{\cdot}$ denotes posterior samples or random variables characterizing our belief on input model or parameters.



Then, the \textit{direct simulation} is used to propagate the input uncertainty to the output. At each sample $\widetilde{F}^{(b)}\sim p(F|\mathbf{X}_m)$ with $b=1,2,\ldots,B$, we generate $n_b$ replications and obtain the outputs $\mathbf{Y}_b\equiv \{Y_1(\widetilde{F}^{(b)}),Y_2(\widetilde{F}^{(b)}),
\ldots,Y_{n_b}(\widetilde{F}^{(b)})\}$. We estimate the mean response with sample mean $\bar{Y}_b\equiv\bar{Y}(\widetilde{F}^{(b)})=\sum_{j=1}^{n_b} Y_j(\widetilde{F}^{(b)})/n_b$
{and quantify the simulation uncertainty with the sampling distribution of $\bar{Y}_b|\widetilde{F}^{(b)}$.}
The overall uncertainty of system mean response estimation is characterized by the conditional distribution of the compound random variable $U\equiv \bar{Y}(\widetilde{F})$, denoted by $F_U(\cdot|\mathbf{X}_m)$, given the information obtained from the real-world data $\mathbf{X}_m$.
In Section~\ref{subsec:gibbssampler}, we propose a sampling procedure to construct a $(1-\alpha^*)100\%$ \textit{percentile empirical CrI} quantifying the overall estimation uncertainty of $\mu^c$, denoted by
$
\mbox{CrI}
=\left[\bar{Y}_{\left(\lceil(\alpha^*/2)B\rceil\right)},
\bar{Y}_{\left(\lceil(1-\alpha^*/2)B\rceil\right)}\right],$
{based on the order statistics $\bar{Y}_{(1)}\leq\bar{Y}_{(2)}\leq\ldots\leq\bar{Y}_{(B)}$.
}

\begin{sloppypar}

This empirical CrI accounts for both input and simulation uncertainties.
We study the asymptotic properties of this interval in Section~\ref{subsubsec:asymptoticProperties}. Define the random variable $W\equiv \mu(\widetilde{F})$ with $\widetilde{F}\sim p(F|\mathbf{X}_m)$, which is the true system mean response evaluated at the posterior sample of input model. Denote the conditional Cumulative Distribution Function (c.d.f.) of $W$ by $F_W(\cdot |\mathbf{X}_m)$. Let $q_W(\gamma|\mathbf{X}_m)\equiv \inf\{q:F_W (q|\mathbf{X}_m)\geq \gamma\}$ be the conditional $\gamma$-quantile of $W$. We prove that given the input data $\mathbf{X}_m$, as the simulation budget increases, the interval $\left[\bar{Y}_{\left(\lceil(\alpha^*/2)B\rceil\right)},
	\bar{Y}_{\left(\lceil(1-\alpha^*/2)B\rceil\right)}\right]$ converges to the true underlying CrI quantifying the impact of input uncertainty, $[q_W(\alpha^*/2|\mathbf{X}_m), q_W(1-\alpha^*/2|\mathbf{X}_m)]$.
We also show that as the size of real-world data and the simulation budget go to infinity, it converges to the true mean system response $\mu^c$.
In addition, if the interval $\left[\bar{Y}_{\left(\lceil(\alpha^*/2)B\rceil\right)},
	\bar{Y}_{\left(\lceil(1-\alpha^*/2)B\rceil\right)}\right]$ is too wide, the decision maker needs to know if the additional simulation could improve the estimation accuracy of $\mu^c$. For this practical consideration, we derive a variance decomposition to estimate the relative contributions from input and simulation uncertainties, and further study the asymptotic property of the corresponding variance components in Section~\ref{subsubsec:varianceDecomposition}.

\end{sloppypar}

\subsection{Input Modeling by Dirichlet Process Mixtures}
\label{subsec:inputModel}


According to \cite{Lo1984,Barrios2013}, given a kernel density function $h(\cdot)$, the input density {defined on the sample space $\mathfrak X$} from DPM can be represented as
$
f(x)=\int h(x|\pmb{\psi})dG,
$
where $\pmb{\psi}$ denotes the parameters of kernel density function, and an infinite mixture distribution $G$ on the parameter space of $\pmb{\psi}$ follows a Dirichlet process (DP), $G\sim DP(\alpha,G_0)$, with $G_0$ denoting the base distribution and $\alpha >0$ denoting the dispersion parameter. {We say that $DP(\alpha, G_0)$ assigns probability on $\mathscr G$, the space of all mixing distribution $G$.}
According to the definition of DP in \cite{Ferguson1973}, the random distribution $G$ over any finite measurable partitions, $A_1,\ldots,A_r$, of the space of $\pmb{\psi}$ follows a Dirichlet distribution, $\big(G(A_1),\ldots,G(A_r)\big)\sim \mbox{Dirichlet}\big(\alpha G_0(A_1),\ldots,\alpha G_0(A_r)\big)$.
Thus, the data $X_i$ drawn from DPM can be represented as
		\begin{equation} 
\label{DPM}
	X_i|\pmb{\psi}_i \sim  h(\cdot|\pmb{\psi}_i), \quad
	\pmb{\psi}_i {\mid G} \sim G, \quad
	G \sim DP(\alpha,G_0).
	\end{equation}
	According to \cite{Neal2000}, by integrating over $G$, we have the conditional prior distribution for $\pmb{\psi}_i$,
	\begin{equation}\label{CondPrior}
	\pmb{\psi}_i|\pmb{\psi}_1,\ldots,\pmb{\psi}_{i-1} \sim \frac{1}{i-1+\alpha}\sum_{i'=1}^{i-1}\delta(\pmb{\psi}_{i'})+\frac{\alpha}{i-1+\alpha}G_0
	\end{equation}
	where $\delta(\pmb{\psi})$ is the distribution concentrated at $\pmb{\psi}$.

DPM is specified by three key components: the dispersion parameter $\alpha$,  the kernel density $h(\cdot)$, and the base distribution $G_0$. The dispersion parameter $\alpha$ is related to the number of clusters generated in the DPM posterior. Based on the right side of (\ref{CondPrior}), given the cluster parameters of previous $X_1,X_2, \ldots,X_{i-1}$, the probability that $X_i$ is associated to a new cluster is $\frac{\alpha}{i-1+\alpha}$.
Thus, the DPM with a larger value of $\alpha$ tends to generate samples of input density $f(\cdot)$ with more distinct active components; see more explanation in  Section~\ref{subsec:gibbssampler}. The appropriate value of $\alpha$ can be inferred from the real-world data.

The choice of the kernel density $h(\cdot)$ is based on the support of $F^c$, and meanwhile it should account for the feasibility of implementation in posterior computation. We present DPM models with three kernel densities, including Gamma, Gaussian, and Beta, which account for the real-world data that are supported on $\mathfrak X$, where $\mathfrak X$ could be the half real line $\Re^+$, the whole real line $\Re$, or a finite interval $[a_1,a_2]$ with $-\infty<a_1<a_2<\infty$. The scaled version of DPM with Beta kernel is applicable to model continuous distributions with a {known} finite support interval $[a_1,a_2]$ through the transformation $X_i=(X_i'-a_1)/(a_2-a_1)$, where $X_i'$ denotes the raw data. 
Since Gamma, Gaussian, and Beta distributions belong to the exponential family and allow conjugate priors for the parameters $\pmb{\psi}$ of each cluster, we derive efficient samplers to generate posterior samples of input model. 
{Notice that even though these three kernels allow us to model many input models commonly used in the simulation applications, we can also select other kernels, such as exponential family density functions (e.g., Weibull) \cite{canu2006kernel}.
If the computational efficiency is not concerned, we can use kernel densities that do not belong to exponential family, such as student t} kernel density function.


To simplify the posterior inference and sampling, we consider the conjugate prior $G_0$ for $\pmb{\psi}$, the parameters of each cluster or kernel density function. For DPM with Gamma kernel, we let $\pmb{\psi}=\left(V,u\right)^\top$ with $V$ and $u$ denoting the shape and mean parameters. Motivated by the study on Gamma mixture distributions in \cite{Wiper2001}, we consider a conditional conjugate prior for $V$ and $u$,
\begin{equation}
V \sim \mbox{Exponential}(\theta) \mbox{  and  }
u \sim  \mbox{Inv-Gamma}(r,s). \label{eq.priorGamma}
\end{equation}
Equation~\eqref{eq.priorGamma} specifies $G_0(V,u)$ with the hyper-parameters $\pmb{\theta}_G=(\theta,r,s)$.

For DPM with Gaussian kernel, we let $\pmb{\psi}=(u,\sigma^2)^\top$ with $u$ and $\sigma^2$ denoting the mean and variance parameters. Following \cite{Gelman_2004}, we choose the conjugate prior,
\begin{equation}\label{eq.GaussPrior}
u|\sigma^2 \sim \mathcal{N}(u_0,\sigma^2/m_0) \mbox{  and  }
\sigma^2/\sigma_0^2 \sim \mbox{Inv-Gamma}\left(\frac{v_0}{2},\frac{1}{2}\right).
\end{equation}
Equation~(\ref{eq.GaussPrior}) specifies $G_0(u,\sigma^2)$ with hyper-parameters $\pmb{\theta}_G=(u_0,m_0,v_0,\sigma_0)^\top$.

For DPM with Beta kernel, we let $\pmb{\psi}=\left(\omega,\beta\right)^\top$ with $\omega$ and $\beta$ denoting the two shape parameters. Since the Beta distribution belongs to the exponential family, we choose the conjugate prior,
\begin{equation}\label{eq.Betaprior}
\omega,\beta|\lambda_0,\lambda_1,\lambda_2 \propto \exp\left\{-\lambda_1\omega-\lambda_2\beta-\lambda_0
\log\left[\frac{\Gamma(\omega)\Gamma(\beta)}
{\Gamma(\omega+\beta)}\right]\right\}.
\end{equation}
Equation~(\ref{eq.Betaprior}) specifies $G_0(\omega,\beta)$ with the hyper-parameters $\pmb{\theta}_G=(\lambda_0,\lambda_1,\lambda_2)^\top$.



\subsection{{Gibbs Sampler for DPM and Uncertainty Quantification Procedure for $\mu^c$}}
\label{subsec:gibbssampler}

For the real-world data $\mathbf{X}_m\equiv\{X_1,X_2,\ldots,X_m\}$, each observation $X_i$ has the associated parameters $\pmb{\psi}_i$ with $i=1,2,\ldots,m$. 
For any set of cluster density  parameters $\pmb{\psi}$ generated by $G_0$, this component is called \textit{active} if it has at least one associated observed data point from $\mathbf{X}_m$. Otherwise, it is called \textit{inactive}.
According to Equation~\eqref{CondPrior}, parameters $\pmb{\psi}_i$ and $\pmb{\psi}_{i^\prime}$ with $i^\prime\neq i$ could take the same values. Let $K_0$ denote the number of \textit{distinct} values in $\{\pmb{\psi}_i\}_{i=1}^m$, and represent the distinct parameter values as $\{\pmb{\psi}^\star_1,\ldots,\pmb{\psi}^\star_{K_0}\}$. Notice that $K_0$ is bounded by $m$. Following \cite{Muller1996}, we introduce the latent indicator variables $\mathbf{c}=(c_1,c_2,\ldots,c_m)$ that associate the data $\{X_1,X_2,\ldots,X_m\}$ to $\{\pmb{\psi}^\star_1,\ldots,\pmb{\psi}^\star_{K_0}\}$, where $c_i = j$ if and only if $\pmb{\psi}_i =  \pmb{\psi}^\star_j$ for $i=1,2,\ldots,m$ and $j=1,2,\ldots,K_0$. Thus, the real-world data $\mathbf{X}_m$ are grouped to $K_0$ \textit{active} components with parameters $\{\pmb{\psi}^\star_1,\ldots,\pmb{\psi}^\star_{K_0}\}$.

Since the DPM model \eqref{DPM} does not have closed form distributions for analytical posterior analysis, we describe a Gibbs sampler to generate posterior samples of input model quantifying the input model estimation uncertainty. We first derive the conditional posteriors required in the sampling procedure.
According to \cite{Neal2000}, by setting $X_i$ to be the last observation, the conditional prior of $c_i$ can be derived based on Equation~(\ref{CondPrior}),
\begin{equation}
\label{eq.latent} \nonumber
\mbox{P}\big(c_i=j|\mathbf{c}^{-i})=\left\{
\begin{array}{ll}
\frac{m_j^{-i}}{m+\alpha-1} \mbox{  if  }\exists c_q=j \mbox{ for all } q\neq i \\
\frac{\alpha}{m+\alpha-1} \mbox{  otherwise }
\end{array}
\right.
\end{equation}
for $i=1,2,\ldots,m$, where $\mathbf{c}^{-i}$ denotes all the latent variables except $c_i$, and $m_j^{-i}$ is the number of latent variables with $c_q=j$ for all $q\in\{1,2,\ldots,m\}$ and $q\neq i$. 
Since the real-world data $\mathbf{X}_m=\{X_1,X_2,\ldots,X_m\}$ are i.i.d., when we classify $X_i$, we can suppose that all other observations have been classified.
Then, given the active component parameters $\Psi^\star \equiv \{\pmb{\psi}^\star_1,\ldots,\pmb{\psi}^\star_{K_0}\}$ and $X_i$, by applying the Bayes' rule, the conditional posterior is
\begin{equation}
\label{post_latent}
p(c_i=j|\mathbf{c}^{-i},\pmb{\psi}^\star_j,\alpha,X_i)=\left\{
\begin{array}{ll}
b_0\frac{m_j^{-i}}{m+\alpha-1}h(X_i|\pmb{\psi}^\star_j) \mbox{  if  }\exists c_q=j \mbox{ for all } q\neq i \\
b_0\frac{\alpha}{m+\alpha-1}\int h(X_i|\pmb{\psi})dG_0 \mbox{  otherwise }
\end{array}
\right.
\end{equation}
where $b_0$ is the normalizing constant.

The posterior for dispersion parameter $\alpha$ conditional on the number of active components is  $p(\alpha|K_0)\sim p(\alpha)p(K_0|\alpha)$. We impose a prior, $p(\alpha)=\mbox{Gamma}(\varsigma_1,\varsigma_2)$, on $\alpha$, with shape $\varsigma_1>0$ and scale $\varsigma_2>0$. Thus, the hyper-parameters for $\alpha$ are $\pmb{\theta}_\alpha=(\varsigma_1,\varsigma_2)^\top$. To simplify the sampling procedure for $p(\alpha|K_0)$, following \cite{Escobar_West_1995}, we introduce a new random variable $\nu$ and generate $\alpha$ from $p(\alpha|K_0)$ by
\begin{equation}
\begin{aligned}\label{alpha}
\nu|\alpha,K_0 &\sim \mbox{Beta}(\alpha+1,m)  \\
\alpha|\nu,K_0 &\sim \tau \mbox{Gamma}(\varsigma_1+K_0,\varsigma_2-\log(\nu))+(1-\tau)
\mbox{Gamma}(\varsigma_1+K_0-1,\varsigma_2-\log(\nu)),
\end{aligned}
\end{equation}
where $\tau$ is defined by $\tau/(1-\tau)=(\varsigma_1+K_0-1)/[m(\varsigma_2-\log(\nu))]$.

Therefore, given the real-world input data $\mathbf{X}_m$, we provide a sampling procedure in Algorithm~\ref{alg:procedure} to generate the samples of compound random variable  $U=\bar{Y}(\widetilde{F})$ and further build a percentile empirical CrI accounting for both input and simulation estimation uncertainties.
First, based on the support of input model $F^c$, choose an appropriate kernel density function $h(\cdot)$, and then
specify the hyper-parameters for both $G_0$ and $\alpha$ in Step~1; see Section~\ref{subsec:inputEstimation} for the values of hyper-parameters used in our empirical study. Then, motivated by \cite{Neal2000}, in Step~2, we propose a Gibbs sampling approach to generate $B$ posterior samples of input distribution,  $\widetilde{F}^{(b)}\sim p(F|\mathbf{X}_m)$ with $b=1,2,\ldots,B$, accounting for the input model estimation uncertainty.
At each $\widetilde{F}^{(b)}$, run the simulations with $n_b$ replications, obtain simulation outputs $\mathbf{Y}_b$, and record the sample mean $\bar{Y}_b$ in Step~3. The simulation uncertainty is characterized by the sampling distribution $\bar{Y}_b|\widetilde{F}^{(b)}$, with mean $\mu(\widetilde{F}^{(b)})$ and variance $\sigma^2_{e}(\widetilde{F}^{(b)})/n_b$. 
Thus, the samples $\{\bar{Y}_1,\bar{Y}_2,\ldots,\bar{Y}_B\}$ of $U=\bar{Y}(\widetilde{F})$ with $\widetilde{F}^{(b)}\sim p(F|\mathbf{X}_m)$ quantify both input and simulation uncertainties. We further construct a $(1-\alpha^*)100\%$ percentile empirical CrI, $\left[\bar{Y}_{\left(\lceil(\alpha^*/2)B\rceil\right)},
\bar{Y}_{\left(\lceil(1-\alpha^*/2)B\rceil\right)}\right]$, quantifying the overall uncertainty of system mean performance estimation in Step~4.
Notice that the approaches proposed to improve the Gibbs sampling efficiency for DPM through a collapse of the state space of the Markov chain in \cite{MacEachern_1994,MacEachern1998,MacEachern2000} could be incorporated into our nonparametric Bayesian framework.

\begin{algorithm}[hbt!]
	\caption{The Nonparametric Bayesian Framework for Uncertainty Quantification } \label{alg:procedure}
	\small
	\begin{algorithmic}
	\STATE
	1. Based on the support of $F^c$, choose an appropriate kernel density function $h(\cdot)$. Then, specify hyper-parameters $\pmb{\theta}_G$ and $\pmb{\theta}_\alpha$ for the base distribution $G_0$ and the dispersion parameter $\alpha$.\;
	
	\STATE
	2. Generate the posterior samples $\widetilde{F}^{(b)}\sim p(F|\mathbf{X}_m)$ through the Gibbs sampling for $b=1,2,\ldots,B$:\
	    
	    \INDSTATE 
	    (2.1) Initialization:\
	    
	        \INDSTATE[2] 
	        (a) Draw $\widetilde{\alpha} \sim \mbox{Gamma}(\varsigma_1,\varsigma_2)$;\;
	        
	        \INDSTATE[2] 
	        (b) Set $c_i = i$ for $i=1,2,\ldots,m$ and $\widetilde{K}_0 = m$;\;
	        
	        \INDSTATE[2]
	        (c) Generate $\widetilde{\pmb{\psi}}^\star_j\sim p({\pmb{\psi}}^\star_j|X_j)$ with $p({\pmb{\psi}}^\star_j|X_j) \propto p(\pmb{\psi}^\star_j) p(X_j|\pmb{\psi}^\star_j)$ for $j=1,2,\ldots,\widetilde{K}_0$ by using the sampling procedure described in {Supplement~\ref{subsec:gibbsSampling}}.\;
	        
	   \INDSTATE
	   (2.2) In each Gibbs sampling iteration, there are three main steps described as follows:\
            
            \INDSTATE[2] 
            (a) For $i = 1,2,\ldots,m$, generate the parameters associated with the inactive component as $\widetilde{\pmb{\psi}}^\star_{\widetilde{K}_0+1}\sim G_0$, and then draw a sample $\widetilde{c}_i$ from the conditional posterior $p(c_i=j|{\widetilde{\mathbf{c}}}^{-i},\widetilde{\pmb{\psi}}^\star_j,{\widetilde{\alpha}},X_i)$ for $j = 1,2,\ldots,\widetilde{K}_0+1$ by applying (\ref{post_latent}); see the detailed sampling procedure for DPM with Gamma, Gaussian, and Beta kernel densities in Supplement~\ref{subsec:gibbsSampling}.
            Then, remove inactive components and update the number of active components $\widetilde{K}_0$;\;
            
            \INDSTATE[2] 
            (b) For the $j$-th active component with $j=1,2,\ldots,\widetilde{K}_0$, generate the $r$-th parameter in $\pmb{\psi}^\star_j$, denoted by $\widetilde{\psi}^\star_{jr}$, from the conditional posterior $p(\psi^\star_{jr}|{\widetilde{\pmb{\psi}}^\star_{j,-r}},\mathbf{X}^j)$, where $\pmb{\psi}^\star_{j,-r}$ denotes the remaining parameters in $\pmb{\psi}^\star_j$ and $\mathbf{X}^j$ denotes all the data associated to the $j$-th component; see the sampling procedure in Supplement~\ref{subsec:gibbsSampling};\;
            
            \INDSTATE[2] 
            (c) Generate $\widetilde{\alpha}$ from the posterior $p(\alpha|\widetilde{K}_0)$ by using Equations~(\ref{alpha}).\;
            
3. At each $\widetilde{F}^{(b)}$ with $b=1,2,\ldots,B$, generate input variates by using Equation~(\ref{eq.generateX}), run simulations with $n_b$ replications, and obtain the outputs $\mathbf{Y}_b$. Then, record the sample mean $\bar{Y}_b=\sum_{j=1}^{n_b} Y_j(\widetilde{F}^{(b)})/n_b$.\;

4. Report a $(1-\alpha^*)100\%$ two-sided percentile empirical CrI for $\mu^c$,
\begin{equation}
	\mbox{CrI}=
	\left[\bar{Y}_{\left(\lceil(\alpha^*/2)B\rceil\right)},
	\bar{Y}_{\left(\lceil(1-\alpha^*/2)B\rceil\right)}\right],
	\label{eq.CrI}	
\end{equation}
with the order statistics $\bar{Y}_{(1)}\leq\bar{Y}_{(2)}\leq\ldots\leq\bar{Y}_{(B)}$. 
\end{algorithmic}
\end{algorithm}

Here, we present the detailed Gibbs sampling for generating posterior samples of indicator variables $\mathbf{c}=(c_1,c_2,\ldots,c_m)$, component parameters $\Psi^\star$, and dispersion parameter $\alpha$. 
For the initialization in Step~(2.1), we generate  $\widetilde{\alpha}$ from the prior $\mbox{Gamma}(\varsigma_1,\varsigma_2)$ and assign the observed data points $\mathbf{X}_m=\{X_1,X_2,\ldots,X_m\}$ to distinct components, which follows the idea of hierarchical agglomerative clustering; see Chapter 15 of \cite{maimon2005data}.
Then, each Gibbs sampling iteration in Step~2 includes three main parts.
In Step~(2.2.a), the conditional posterior in (\ref{post_latent}) is used to generate the sample of $\mathbf{c}$. For each observation $X_i$ with $i=1,2,\ldots,m$ in the real-world data $\mathbf{X}_m$, since the integration $\int h(X_i|\pmb{\psi})dG_0$ in Equation~(\ref{post_latent}) is often intractable, we
	first sample the parameters for inactive component $\widetilde{\pmb{\psi}}_{\widetilde{K}_0+1}\sim G_0$. Then, we update the latent indicator $c_i$ conditional on all the other parameters, remove the inactive components, and update the number of active distinct components $\widetilde{K}_0$. In Step~(2.2.b), for each active component, given the data associated to it, we update its kernel density parameters $\pmb{\psi}^\star_j$ with $j=1,2,\ldots,\widetilde{K}_0$. The conditional posteriors and the sampling procedure of $\mathbf{c}$ and $\pmb{\psi}^\star_j$ for Gamma, Gaussian, and Beta kernel densities are described in Supplement~\ref{subsec:gibbsSampling}. In Step~(2.2.c), we update the dispersion parameter $\alpha$ conditional on the number of active components and generate a posterior sample by applying Equations~(\ref{alpha}).

After the convergence of Gibbs sampling, we record $B$ posterior samples of input models,  $\{\widetilde{F}^{(1)},\widetilde{F}^{(2)},\ldots,\widetilde{F}^{(B)}\}$, quantifying the input model estimation uncertainty. 
Each sample $\widetilde{F}^{(b)}$ with $b=1,2,\ldots,B$ is specified by the dispersion parameter $\widetilde{\alpha}$ and the parameters of active clusters or components corresponding to real-world data, denoted by  $\widetilde{\Psi}\equiv \{\widetilde{\pmb{\psi}}_1,\widetilde{\pmb{\psi}}_2,\ldots,\widetilde{\pmb{\psi}}_m\}$ with $\widetilde{\pmb{\psi}}_i = \widetilde{\pmb{\psi}}^\star_{\widetilde{c}_i}$ for $i=1,2,\ldots,m$.
In our empirical study, we use 500 burn-in iterations to generate stable posterior samples.
{The density function of $\widetilde{F}^{(b)}$ can be represented as}
\begin{equation}
\widetilde{f}^{(b)}(x) = \frac{1}{m+\widetilde{\alpha}^{(b)}}\sum_{i=1}^{m}h(x|\widetilde{\pmb{\psi}}^{(b)}_i)+\frac{\widetilde{\alpha}^{(b)}}{m+\widetilde{\alpha}^{(b)}}\int h(x|\pmb{\psi})dG_0.
\label{eq.posteriorF}
\end{equation}
Notice that the number of active components, $\widetilde{K}_0$, can vary at different samples of input model.
In Step~3, to estimate the system mean response at the posterior sample of input model $\widetilde{F}^{(b)}$,
we can generate input variates,
\begin{equation}
\pmb{\psi}|\widetilde{\Psi}^{(b)}\sim \frac{1}{m+\widetilde{\alpha}^{(b)}}\sum_{i=1}^{m}\delta\Big(\widetilde{\pmb{\psi}}^{(b)}_i\Big)+\frac{\widetilde{\alpha}^{(b)}}{m+\widetilde{\alpha}^{(b)}}G_0 \quad \mbox{and} \quad
X|\pmb{\psi}\sim h(\cdot|\pmb{\psi}),
\label{eq.generateX}
\end{equation}
to drive the simulation and estimate the mean response by $\bar{Y}_b$.

We derive the computational complexity to generate $B$ posterior samples of input models, $\{\widetilde{F}^{(1)},\widetilde{F}^{(2)},\ldots,\widetilde{F}^{(B)}\}$, from the DPM posterior distribution. 
For Step (2.2.a), 
since we need to calculate the conditional posterior  $p(c_i=j|{\widetilde{\mathbf{c}}}^{-i},\widetilde{\pmb{\psi}}^\star_j,{\widetilde{\alpha}},X_i)$ for $i=1,2,\ldots,m$ and $j = 1,2,\ldots,\widetilde{K}_0$, 
the complexity is $\mathcal{O}(m\widetilde{K}_0)$. 
For Step (2.2.b), we need to calculate the conditional posteriors for $\pmb{\psi}^\star_j$ with $j=1,2,\ldots,\widetilde{K}_0$ (see the formula in Supplement~\ref{subsec:gibbsSampling}), which include
$\sum_{k=1}^{m_j}{\log(X_k^j)}$ in Equation (\ref{eq.posteriorGamma}) for Gamma kernel, $\sum_{k=1}^{m_j}(X^j_k-\bar{X}^j)^2$ in Equation (\ref{eq.PosteriorGauss_mu}) for Gaussian kernel, and $\sum_{k=1}^{m_j}{\log(1-X_k^j)}$ in Equation (\ref{eq.Beta2}) for Beta kernel, where $X_k^j$ denotes the $k$-th observation associated with the $j$-th component and $\bar{X}^j=\frac{1}{m_j}\sum_{k=1}^{m_j}X_k^j$.
For Gamma and Beta kernels, we develop  Metropolis-Hasting nested Gibbs samplers (see Supplement~\ref{subsec:gibbsSampling}) and run them for a fixed number of iterations independent of $m$ and $B$ in the proposed algorithm. 
Thus, the computational complexity for implementing Step~(2.2.b) is $\mathcal{O}(\widetilde{K}_0m)$.
In Step (2.2.c), the complexity for generating $\widetilde{\alpha}$ is $\mathcal{O}(1)$. 
Since the study in \cite{GhoVan17} (see Chapter 4 Proposition 4.8) shows that the number of active components $\widetilde K_0$ in Dirichlet process is at most $\mathcal{O}(\log m)$, the overall computational complexity to generate $B$ posterior samples of input model is $\mathcal{O}(B m\log m)$.
In addition, suppose that each detailed simulation output as a function of associated inputs has a fixed complexity $\mathcal{O}(1)$. Then, 
the computational complexity for running simulations at $B$ posterior samples of input models is $\mathcal{O}(BnR)$, where $n$ and $R$ represent the number of replications and the simulation run-length assigned to each posterior sample $\widetilde{F}^{(b)}$ with $b=1,2,\ldots,B$.

We need $B$ to be {reasonably} large to accurately estimate the percentile CrI. In the empirical study, we set $B=1000$ \cite{Xie_Bayesian_2014}. Without any prior information about the mean response $\mu(\cdot)$, in this paper, we assign equal replications to all samples of input distribution $\{\widetilde{F}^{(1)},\widetilde{F}^{(2)},\ldots,\widetilde{F}^{(B)}\}$. Since each simulation run can be computationally expensive, a sequential design of experiments could efficiently use the computational budget and reduce the impact of simulation estimation uncertainty on the system performance by finding the optimal setting for $(B,n_1,n_2,\ldots,n_B)$ \cite{Yi_Xie_2016}.


\subsection{Posterior Consistency of Input Models}
\label{subsec:inputSamplingConvergence}

In the Bayesian paradigm, a very basic requirement is the {\it posterior consistency} at the true input distribution \cite{Ghosal1999, Ghosal1995}. It means that as the amount of real-world data increases, the posterior becomes more and more concentrated around $F^c$ with probability approaching 1. The posterior consistency for DPM is studied in the statistics literature, such as \cite{Ghosal1999,Tok06,Wu_Ghosal_2008}, etc. Given the prior set as in Equations (\ref{DPM}), and the base measures $G_0$ for Dirichlet Process set in Equations~(\ref{eq.priorGamma}), (\ref{eq.GaussPrior}) or (\ref{eq.Betaprior}), Theorem~\ref{thm:consistency} summarizes posterior consistency results on DPM with Gamma, Gaussian, and Beta density functions as kernels for input distributions supported on $[0,\infty)$, $\Re$, and fixed interval $[a_1,a_2]$.

The posterior consistency in Theorem~\ref{thm:consistency} is stated in the sense of {\it weak consistency}. {The posterior distribution $p(\cdot\mid \mathbf{X}_m)$ is said to be {\it weakly consistent} at $F^c$ (or $f^c$), if with $P_{f^c}$-probability 1; this means $p(U\mid \mathbf{X}_m)\to 1$ for all weak neighborhoods $U$ of $f^c$. We defer the detailed introduction and definitions of weak neighborhood, weak consistency, and other related concepts in classic Bayesian nonparametric theory to Supplement \ref{subsec:proofThm1}. We also refer the readers to Chapter 4 of \cite{Ghosh_2003} and Chapter 7 of \cite{GhoVan17} for detailed discussions on posterior consistency.}

\begin{sloppypar}

\begin{theorem} \label{thm:consistency}
Let $\mathbf{X}_m\equiv \{X_1,X_2,\ldots,X_m\}$ with $X_i\stackrel{i.i.d.}\sim F^c$ for $i=1,2,\ldots,m$.
\begin{itemize}
		\item[(i)] {(Modified from \cite{Wu_Ghosal_2008} Theorem 14) Suppose the DPM with Gamma kernel has the prior specified as Equation~(\ref{eq.priorGamma}).  Let $f^c$ be a continuous and bounded density with support on $[0,\infty)$ satisfying the following conditions: (a) $f^c(x)>0$ for all $x\in (0,+\infty)$ and $f^c(x)\leq C_f<\infty$ for some finite constant $C_f$ for all $x\in [0,+\infty)$; 
		(b) $|\int_0^{\infty}f^c(x)\log f^c(x) dx | < \infty$; (c) $\int_0^{\infty} f^c(x)\log \frac{f^c(x)}{\phi_{\delta}(x)}dx <\infty$ for some $\delta>0$, where $\phi_{\delta}(x)=\inf_{[x,x+\delta)}f^c(t)$ if $0<x<1$ and $\phi_{\delta}(x)=\inf_{(x-\delta,x]}f^c(t)$ if $x\geq 1$; (d) there exists $\zeta>0$ such that $\int_0^{\infty} \max(x^{-\zeta-2},x^{\zeta+2})f^c(x)dx <\infty$.}
		Then, the posterior $p(F|\mathbf{X}_m)$ from DPM with Gamma kernel is weakly consistent at $F^c$.

		\item[(ii)] (\cite{Tok06} Theorem 3.3) Suppose the DPM with Gaussian kernel has the prior specified as Equation~(\ref{eq.GaussPrior}). Let $F^c$ (and the density $f^c$) be supported on $\Re$ and assume that it satisfies the following conditions: (a) $\left|\int_{-\infty}^{+\infty} f^c(x) \log f^c(x) dx\right|<+\infty$; (b) there exists an $\eta\in (0,1)$, such that $\int_{-\infty}^{+\infty} |x|^\eta f^c(x) dx <+\infty$; (c) there exist constants $\sigma_1>0$, $c_1\in (0,\eta)$,  $c_2>c_1$, $b_1, b_2>0$, such that for the base measure $G_0(u,\sigma)$ and for all large $x>0$:
		\begin{align*}
		&\max \left\{G_0\left([x-\sigma_1 x^{\eta/2},+\infty)\times [\sigma_1,+\infty) \right), G_0\left([0,+\infty)\times (x^{1-\eta/2},+\infty)\right)\right\} \geq b_1x^{-c_1};\\
		& \max \left\{G_0\left((-\infty, -x+\sigma_1 x^{\eta/2}] \times [\sigma_1,+\infty) \right), G_0\left((-\infty,0]\times (x^{1-\eta/2},+\infty)\right)\right\} \geq b_1x^{-c_1};\\
		& G_0\left((-\infty, x)\times (0,e^{x^\eta-1/2})\right) > 1- b_2x^{-c_2}; \quad  G_0\left((-x,+\infty)\times (0,e^{x^\eta-1/2})\right) > 1- b_2x^{-c_2}.
		\end{align*}
		Then, the posterior $p(F|\mathbf{X}_m)$ from DPM with Gaussian kernel is weakly consistent at $F^c$.
		
		\item[(iii)] {Suppose the DPM with Beta kernel has the prior specified as Equation~(\ref{eq.Betaprior}). Let $F^c$ (and the density $f^c$) be supported on $[a_1,a_2]$ and assume that  $f^c(x)$ is continuous density on $[a_1,a_2]$.} Then, the posterior $p(F|\mathbf{X}_m)$ from DPM with Beta kernel is weakly consistent at $F^c$.
		
	\end{itemize}
\end{theorem}

\end{sloppypar}


The proof of Theorem \ref{thm:consistency} is given in Supplement \ref{subsec:proofThm1}. 
This theorem indicates that the posterior of DPM with Gamma, Gaussian, and Beta kernels can consistently estimate the true input distributions under some general sufficient conditions, including
the existence of moments and entropy of $F^c$ (or $f^c$), and the boundedness and continuity of $f^c$. In particular, for DPM with Gamma kernel, the posterior consistency holds if the true density $f^c(x)$ is upper bounded by constant, nonzero and continuous for all $x>0$, with a finite entropy and finite moments of certain order. For DPM with Gaussian kernel, the posterior consistency holds if the true density has a finite entropy and finite $\eta$-moment with $\eta\in (0,1)$, thus including heavy-tailed distributions like Cauchy. For DPM with Beta kernel, the posterior consistency holds as long as the true density is continuous on the finite interval $[a_1,a_2]$. Thus, there is no assumption on the analytic forms of $F^c$ and $f^c$ required for the posterior consistency. 

The posterior consistency in Theorem \ref{thm:consistency} for DPM with Gamma, Gaussian, and Beta kernels applies to a wide range of true distributions. In Supplement \ref{subsec:proofThm1}, we give examples of posterior consistency on the standard parametric distributions that are commonly used in simulations,  such as normal, logistic, Student's $t$, Cauchy, uniform, triangular, power function, Beta, truncated normal, log-normal, Gamma with shape parameter greater than 2, Weibull with shape parameter greater than 3, log-logistic with shape parameter greater than 2, Pearson Type V and Type VI, Johnson $S_B$, Johnson $S_L$, Johnson $S_U$, by applying Theorem \ref{thm:consistency}, though one of the main motivations for using DPM is to flexibilly model the underlying distribution that does not belong to any standard distribution families.




Like any other statistical models, the proposed DPM-based input modeling cannot cover all situations. Some distributions, such as standard (unshifted) versions of the gamma, Pearson VI, Pearson V, and log-logistic with shape parameter less than or equal to 2, and Weibull distributions with shape parameter less than or equal to 3, do not satisfy the finite moment condition or bounded density requirement in Theorem \ref{thm:consistency}. This does not necessarily mean that the proposed nonparametric approach cannot consistently estimate such true distributions. Theorem \ref{thm:consistency} provides only \textit{sufficient} conditions on which posterior consistency holds.
We provide the empirical study in Section~\ref{sec:EmpiricalStudy} to  demonstrate the robustness of DPM finite-sample performance even when the \textit{sufficient conditions} for the posterior consistency listed in Theorem~\ref{thm:consistency}
does not hold.
		 
		 
Theoretically, by choosing different kernel functions, the DPM can lead to consistent posterior on the distributions that are not covered by Theorem~\ref{thm:consistency}.  For example, if the true density is completely monotone on $[0,\infty)$, such as Gamma,  Pearson Type V, log-logistic, and Weibull distributions  with shape parameters less than or equal to 2 or 3, the DPM priors with exponential density or scaled uniform density kernels have consistent posterior on them, as shown by Theorems~16 and 17 in \cite{Wu_Ghosal_2008}. 


{Part (iii) of Theorem~\ref{thm:consistency} is a completely  new result.  Existing Bayesian asymptotic results in the literature mainly focus on 
different versions of Beta mixtures, such as the finite mixtures of Bernstein polynomials (\cite{PetWas02,Wu_Ghosal_2008}), or the finite Beta mixtures in \cite{Rousseau_2010}. The study in \cite{GhoRoyTan08} contains partial results on the classes of distributions that can be expressed with an infinite mixture of Betas. We present a general result for posterior consistency of DPM with Beta kernel and its proof is given in Supplement \ref{subsec:proofThm1}. }
{Our empirical study also demonstrates the flexibility and adaptiveness of DPM with Beta kernel for fitting the continuous distributions with known bounded support.} 

{Theorem \ref{thm:consistency} directly applies to the situation that the support of the input distribution is known. As pointed out by \cite{Law_2015}, such assumption could be a limitation for some cases. These distributions include, but not limited to,  the shifted (scaled) versions of the distributions on $[0,\infty)$ and $[0,1]$. To deal with bounded or semi-bounded supported underlying input distributions with boundaries unknown,  
 we mildly extend the DPM prior as follows: Let $\xi$ denote the boundary value(s) of the support. For the bounded support case, we take $\xi=(a_1,a_2)$, while for the half-bounded case, we take $\xi=a_0$ and let the support be $[a_0,\infty)$ without loss of generality. We assign the prior $\pi$ on $\xi$ and the complete nonparametric prior on the input distribution is given by:
	\begin{equation}
	    	X_i|\pmb{\psi}_i\sim  h(\cdot|\pmb{\psi}_i),
	    	\quad
	\pmb{\psi}_i \mid G \sim G,  \quad
	G \mid \xi \sim DP(\alpha,G_{0})\times \Ind(\xi), \quad
	\xi \sim  \pi. \label{index}
	\end{equation}
The mixing distribution $G$ here is slightly different from the one defined in Model \eqref{DPM} and it assigns point mass $1$ on the boundary $\xi$ for given $\xi$, where $\xi$ is assigned a prior $\pi$ as an index parameter.}

{
With the prior defined as (\ref{index}), we have corollaries as follows. Please refer to Supplement \ref{subsec:proofThm1} Remark \ref{unknown} for the detailed discussion and proofs. 
\begin{corollary}\label{c-b}
Suppose that the true density $f^c$ is continuous and has bounded support with unknown boundary $\xi=(a_1,a_2)$. Let the DPM prior with location-scale transformed beta kernel as described above. Assume that the prior on the index parameter (boundary values) satisfies that for any $\delta>0$, $\pi([a_1-\delta, a_1]\times[a_2,a_2+\delta])>0$. Then the posterior $p(F|\mathbf{X}_m)$ from DPM with transformed Beta kernel is weakly consistent at $F^c$.
\end{corollary}	}

{
\begin{corollary} \label{C-G}
	Suppose the true density $f^c$ satisfies all the conditions in Theorem 1 Part (i), except that the support is $[a_0, \infty)$ with $a_0$ unknown. Let the DPM prior with shifted Gamma density kernel as described in (\ref{shiftGamma}). Assume the prior on the index parameter (boundary value) satisfies $\pi([a_0-\delta,a_0])>0$.Then the posterior $p(F|\mathbf{X}_m)$ from DPM with the shifted Gamma kernel is weakly consistent at $F^c$.
\end{corollary}	
}


\subsection{Asymptotic Properties of the CrI}
\label{subsubsec:asymptoticProperties}

In this section, we study the asymptotic properties of the empirical CrI
constructed from our framework in Section~\ref{subsec:gibbssampler}. In many situations, it could be hard or expensive to collect more real-world data when we make decisions.
Therefore, in Theorem~\ref{thm:cribound} Part~(i), we show that given finite real-world data $\mathbf{X}_m$, as the simulation budget increases, the interval constructed by our approach,  $\left[\bar{Y}_{\left(\lceil(\alpha^*/2)B\rceil\right)},
	\bar{Y}_{\left(\lceil(1-\alpha^*/2)B\rceil\right)}\right]$, converges to the $(1-\alpha^*)100\%$ percentile CrI induced by the input uncertainty with the true mean response surface $\mu(\cdot)$ known,
$\left[q_W(\alpha^*/2|\mathbf{X}_m),
	q_W(1-\alpha^*/2|\mathbf{X}_m)\right] $.
In Theorem~\ref{thm:cribound} part~(ii), we show that as the amount of real-world data and the simulation budget go to infinity, {the constructed CrI $\left[\bar{Y}_{\left(\lceil(\alpha^*/2)B\rceil\right)},
	\bar{Y}_{\left(\lceil(1-\alpha^*/2)B\rceil\right)}\right]$ shrinks to the true mean response $\mu^c$.}


\begin{sloppypar}
	
{The convergence between two credible intervals in Theorem~\ref{thm:cribound} is measured by} the {\it Hausdorff distance}, denoted by $d_H(\cdot,\cdot)$, which is widely used for measuring the distance between two sets. It has a simplified expression when $A_1$ and $A_2$ are both closed intervals: If $A_1=[a_1,b_1]$ and $A_2=[a_2,b_2]$, then $d_H(A_1,A_2)=\max(|a_1-a_2|,|b_1-b_2|)$. In this case, the convergence under Hausdoff distance is the same as the point-wise convergence for the two endpoints of CrIs.

For two generic distributions (and measures) $F_1$ and $F_2$ on $\Re$ with the Borel sigma algebra $\mathcal{B}(\Re)$, their {\it L\'evy-Prokhorov (L-P) distance} (\cite{billingsley1999}) is defined by $d_{LP}(F_1,F_2)\equiv \inf\{\eta>0~|~F_1(A)\leq F_2(A^\eta)+\eta \text{ and } F_2(A)\leq F_1(A^\eta)+\eta, \text{ for all } A\in \mathcal{B}(\Re)\}$, where $A^\eta\equiv \{a\in \Re~|~\exists b\in A, |a-b|<\eta\}$. The L-P distance, denoted by $d_{LP}$, is a metric under which the convergence is equivalent to the weak convergence of measures on $\Re$.

\begin{theorem} \label{thm:cribound}
{Let $n_{\min}=\min\{n_1,n_2,\ldots,n_B\}\geq 1$. Suppose that the following conditions hold:}
\begin{enumerate}
\item[(1)] {The posterior distribution function $F_W(\cdot|\mathbf{X}_m)$ is continuous with a positive density on its support; Furthermore, $\int_{\Re} w^2 d F_W(w|\mathbf{X}_m) <\infty$ almost surely for all $m$;}
\item[(2)] The simulation model satisfies
$\text{E}(\bar Y_b |\widetilde F^{(b)})=\mu(\widetilde F^{(b)})$ and $\text{Var}(\bar Y_b |\widetilde F^{(b)})=\sigma^2_{e}(\widetilde{F}^{(b)}) /n_b$ for $b=1,2,\ldots,B$. For almost surely all $\widetilde{F}\sim p(F|\mathbf{X}_m)$, there exists a finite constant $C_{\sigma}>0$ with $\sigma^2_{e}(\widetilde{F})\leq C^2_{\sigma}$;
\item[(3)] For any $\epsilon>0$, there exists a finite $\delta>0$ such that $|\mu(F)-\mu(F^c)|< \epsilon$ if $d_{LP}(F,F^c)<\delta$;
\item[(4)] The posterior distribution $p(F|\mathbf{X}_m)$ is weakly consistent at $F^c$.
\end{enumerate}
Then,
\begin{enumerate}
\item[(i)] {If Conditions~(1) and (2) hold, then for some absolute constant $C_1>0$ that does not depend on $\mathbf{X}_m$ and $\mu(\cdot)$, the CrI in Equation~\eqref{eq.CrI} satisfies
\begin{align}\label{eq.dhorder}
&\text{E}\left[ \int_0^{1} d_{H}\left(\left[\bar Y_{\left(\lceil(\alpha^*/2)B\rceil\right)},
\bar Y_{\left(\lceil(1-\alpha^*/2)B\rceil\right)}\right], \left[q_W\left(\alpha^*/2|\mathbf{X}_m\right) ,q_W\left(1-\alpha^*/2|\mathbf{X}_m\right)\right] \right) d\alpha^* ~\Big|~ \mathbf{X}_m\right] \nonumber\\
&\leq \frac{2C_{\sigma}}{\sqrt{n_{\min}}} + \frac{2C_1\int_{\mathbb{R}} w^2 dF_W(w|\mathbf{X}_m)}{\sqrt{B}}.
\end{align}
Furthermore, for any given $\epsilon>0$, $\delta>0$, and fixed $\alpha^*\in (0,1)$, there exist some integers $B_1$ and $n_{\min,1}$ that only depends on $\epsilon, \delta, \alpha^*, C_{\sigma}, \mathbf{X}_m$ and the function $\mu(\cdot)$, such that for all $B>B_1$ and $n_{\min}>n_{\min,1}$,
\begin{align}\label{eq.CrIconverge}
\text{P}\Big( &d_{H}\Big(\big[\bar Y_{\left(\lceil(\alpha^*/2)B\rceil\right)},
\bar Y_{\left(\lceil(1-\alpha^*/2)B\rceil\right)}\big],
\big[q_W\left(\alpha^*/2|\mathbf{X}_m\right) ,q_W\left(1-\alpha^*/2|\mathbf{X}_m\right)\big] \Big) < \delta ~\Big|~ \mathbf{X}_m \Big) > 1 - \epsilon.
\end{align}  }
\item[(ii)] If Conditions (1)--(4) hold, then for any given $\epsilon>0,\delta>0,\gamma\in(0,1)$, $\eta\in (0, 2\min\{\gamma,1-\gamma\})$, there exist sufficiently large integers $B_2$ and $n_{\min,2}$ that only depend on $\epsilon,\delta,\gamma,\eta,C_{\sigma},\mathbf{X}_m$ and the function $\mu(\cdot)$, and a sufficiently large integer $M_0$ that depends on $\epsilon,\delta,\eta$, such that for all $B>B_2$, $n_{\min}>n_{\min,2}$, and $m>M_0$,
\begin{align}\label{eq.Quantconverge}
& \text{P}_{F^c}\left[\text{P}\left( \left|\bar Y_{(\lceil\gamma B\rceil)}-\mu(F^c)\right| < \delta ~\Big|~ \mathbf{X}_m \right) > 1 - \eta \right] > 1 - \epsilon.
\end{align}
\end{enumerate}
\end{theorem}

\end{sloppypar}




{
Condition (1) is a mild condition that implies that the posterior distribution function of the system response $\mu(\widetilde{F})$ with $\widetilde{F}\sim p(F|\mathbf{X}_m)$ is continuous and strictly increasing, with finite second moment.  Condition~(2) requires that the simulation errors have a bounded variance.
Condition~(3) is about the continuity of the system response $\mu(F)$ with respect to $F$ around $F^c$ in terms of the L-P distance that can be used to characterize the weak posterior consistency.
The similar continuity assumption is commonly used in the literature on input uncertainty and the Gaussian process metamodel (an approximate input-output mean response surface)} when the parametric family of input model is known; see for example \cite{ankenman_nelson_staum_2010}, \cite{barton_nelson_xie_2011} and \cite{Xie_Bayesian_2014}. Condition~(3) generalizes it to the nonparametric situations.
Condition~(4) is a direct consequence from Theorem~\ref{thm:consistency} which only provides the asymptotic consistency for input models with support on $\Re^+$ and $\Re$.

Given finite real-world data $\mathbf{X}_m$, Part~(i) of Theorem \ref{thm:cribound} shows that as the simulation budget goes to infinity with $n_{\min},B\rightarrow \infty$, the empirical CrI obtained by our framework in Equation~\eqref{eq.CrI} converges to the true underlying CrI $[q_W(\alpha^*/2|\mathbf{X}_m),q_W(1-\alpha^*/2|\mathbf{X}_m)]$. { This convergence happens in the integrated sense, in which we take an average of the Hausdorff distance over the significance level $\alpha^*\in (0,1)$.} For finite $B$ and $n_{\min}$, the upper bound in \eqref{eq.dhorder} further provides a detailed breakdown of the approximation error from the simulation estimation uncertainty. The first error term on the right side of \eqref{eq.dhorder} comes from the finite replications ($n_{\min}$) allocated to the posterior samples of input model quantifying the input uncertainty. The second error term in \eqref{eq.dhorder} comes from using finite ($B$) posterior samples. The convergence of CrI in \eqref{eq.dhorder} is stated in the Bayesian setup conditional on the real-world data $\mathbf{X}_m$ and it does {\it not} require the sample size $m\rightarrow\infty$. Therefore, the bound in Part~(i) is {\it non-asymptotic} in $m$ and only asymptotic in the simulation budget
$(n_{\min},B)$. {The relation \eqref{eq.CrIconverge} further expresses this relation using the convergence in posterior probability for each fixed $\alpha^*\in (0,1)$.} Notice that Part~(i) only requires Conditions~(1) and (2).

{Part~(ii) shows that the convergence of quantile as the amount of real-world data $m$ increases to infinity, for which we have used the continuity of $\mu(\cdot)$ at $F^c$ from Condition~(3) and the weak consistency of the posterior distribution $p(F|\mathbf{X}_m)$ in Condition~(4). }
The detailed proof of Theorem~\ref{thm:cribound} is provided in Supplement~\ref{subsec:AppendixAsympticPropertiesProof}.


\subsection{Variance Decomposition}
\label{subsubsec:varianceDecomposition}

{Following the procedure in Section~\ref{subsec:gibbssampler}, we obtain samples of system mean response, $\bar{Y}_b$ with $b=1,2,\ldots,B$, quantifying the estimation uncertainty of $\mu^c$. We derive the variance decomposition in Theorem~\ref{thm:vardecomp}(i). The overall uncertainty can be written as the sum of two variance components, $\sigma^2_T=\sigma^2_I+\sigma^2_S$, quantifying the relative contributions from input and simulation uncertainties, which can guide how to improve the system performance estimation when the overall uncertainty is too large. If $\sigma_S^2$ is dominant, more simulation resource can be invested to reduce the estimation uncertainty of $\mu^c$.}

{We further study the asymptotic property of both variance components $\sigma^2_I$ and $\sigma_S^2$ in Theorem~\ref{thm:vardecomp}(ii), which depend on $\mathbf{X}_m$. The component $\sigma_I^2$ measuring the impact from input uncertainty decreases as the amount of real-world data increases. For the input model with support on $\Re^+$, $\Re$ or {$[a_1,a_2]$} satisfying the conditions in  Theorem~\ref{thm:consistency}, as $m\rightarrow\infty$,
the impact of input uncertainty disappears or $\sigma_I^2$ converges to zero in probability $\sigma_I^2\stackrel{p}\rightarrow 0$, since the posterior of input model $p(F|\mathbf{X}_m)$ converges to $F^c$ and $\mu(\cdot)$ is bounded and continuous around $F^c$ in terms of L-P distance.
As $n_{\min}\rightarrow\infty$, the variance component $\sigma_S^2$ measuring the impact from the simulation uncertainty disappears $\sigma_S^2\stackrel{p}\rightarrow 0$, if $\sigma^2_e(\cdot)$ is bounded.
The detailed derivation of Theorem~\ref{thm:vardecomp} is provided in Supplement~\ref{subsec:AppendixVarianceDecomposition}.}
{
\begin{theorem}\label{thm:vardecomp}
	At any $\widetilde{F}^{(b)}$ with $b=1,2,\ldots,B$, let $\mu_b=\mu(\widetilde{F}^{(b)})$ and $\sigma_b^2=\sigma^2_e(\widetilde{F}^{(b)})$.
	\begin{itemize}
	\item[(i)] Given $\mathbf{X}_m$, the total variance of $\bar{Y}(\widetilde{F})$ can be decomposed as 
	\begin{eqnarray} 
	\label{eq.VarDecomp}
\mbox{Var}(\bar{Y}(\widetilde{F})|\mathbf{X}_m) 
= \mbox{E}_{\widetilde{F}^{(b)}}\left[\left.\frac{\sigma_b^2}{n_b}\right|\mathbf{X}_m\right] + \mbox{Var}_{\widetilde{F}^{(b)}}[\mu_b|\mathbf{X}_m] 
	\end{eqnarray}
On the right side of equation, 
$\sigma_S^2\equiv\mbox{E}_{\widetilde{F}^{(b)}}\left[\left.\frac{\sigma_b^2}{n_b}\right|\mathbf{X}_m\right]$ and 
$\sigma_I^2\equiv \mbox{Var}_{\widetilde{F}^{(b)}}[\mu_b|\mathbf{X}_m] $
measure the impacts from simulation and input uncertainties. 
		Since the sample mean and variance $\bar{Y}_b$ and $S^2_b$ are the consistent estimators for $\mu_b$ and $\sigma^2_b$, we estimate $\sigma_S^2$ with $\widehat{\sigma}_S^2=\frac{1}{B} \sum_{b=1}^B \frac{S_b^2}{n_b}$ and estimate $\sigma_I^2$ with $ \widehat{\sigma}^2_I=\frac{1}{B} \sum_{b=1}^B (\bar{Y}_b-\bar{\bar{Y}})^2$, where $\bar{\bar{Y}}=\frac{1}{B}\sum_{b=1}^B \bar{Y}_b$.
	\item[(ii)] {Suppose that Conditions~(2)--(4) in Theorem~\ref{thm:cribound} hold. For almost surely all $\widetilde{F}\sim p(F|\mathbf{X}_m)$, there exists a finite constant $C_{\mu} >0$ such that $|\mu(\widetilde{F})| \leq C_{\mu}$.  Then, as} $m$ and $n_{\min}$ go to infinity, the variance components $\sigma_I^2$ and $\sigma_S^2$ converge to zero in probability: (a) $\sigma_I^2 \stackrel{p}\rightarrow 0$ as $m \rightarrow \infty$;  
		and (b) $\sigma_S^2 \stackrel{p}\rightarrow 0$ as $n_{\min}\rightarrow \infty$.	
	\end{itemize}	
\end{theorem}
}


\section{Empirical Study}
\label{sec:EmpiricalStudy}

We first compare the fitting performance of DPM models with the existing input modeling approaches by using simulated data in Section~\ref{subsec:inputEstimation}. The results demonstrate that DPM with appropriate kernel can capture the important properties in the simulated data. It has better and more robust finite-sample performance than existing approaches, including finite mixture, empirical distribution, kernel density estimation (KDE), and parametric distributions selected by using the Anderson-Darling (AD) and Kolmogorov-Smirnov (KS) tests.
Since some test examples in Section~\ref{subsec:inputEstimation} violate the conditions in Theorem~\ref{thm:consistency}, the results also indicate that the performance of DPM is robust to the violation of the sufficient conditions required for input model asymptotic consistency.
Then, we use the real raw material demand data collected from the biopharmaceutical manufacturing to show the robust performance of DPM model in Section~\ref{subsec:AppendixRealData}. 
Since the real-world data often represent many latent sources of uncertainties, we study the performance on identifying the underlying sources of uncertainties in Section~\ref{subsec:numberofCluster}. The results indicate that DPM model works better than existing finite mixture 
approach \cite{Cheng2003}.
After that, we use an $M/G/1$ queue to study the performance of our DPM-based Bayesian nonparametric framework in Sections~\ref{subsec:mm1Queue_sim}.
Results show that our approach has good and robust performance when there is no strong prior information on the input model and the mean response surface. 	
As the amount of real-world data and the simulation budget increase, the empirical CrI $\left[\bar{Y}_{\left(\lceil(\alpha^*/2)B\rceil\right)}, \bar{Y}_{\left(\lceil(1-\alpha^*/2)B\rceil\right)}\right]$ shrinks closer to $\mu^c$.
Further, the ratio $\widehat{\sigma}_I/\widehat{\sigma}_S$ provides a good indicator
of the relative contributions from both input and simulation uncertainty.

\subsection{Input Density Estimation}
\label{subsec:inputEstimation}

In the empirical study, a Gamma prior is used for the dispersion parameter 
$\alpha \sim \mbox{Gamma}(\varsigma_1,\varsigma_2)$.
Since our sensitivity study in Supplement~\ref{subsec:AppendixSensitivityAnalysis} indicates that the input model performance is not sensitive to the values of hyper-parameters $\pmb{\theta}_{\alpha}$, we set 
$\varsigma_1=1$ and $\varsigma_2=1$ in the empirical study. 
As for the hyper-parameters $\pmb{\theta}_G$ for the base distribution $G_0$, we use the noninformative prior. 
We set $\theta=0.01$, $r=2$ and $s=2$ for DPM with Gamma kernel density, and set $\mu_0=0$, $v_0=1.5$, $m_0=0.01$ and $\sigma_0 = 1$ for DPM with Gaussian kernel density. For DPM with Beta kernel density, we set $\lambda_0=1,\lambda_1=\lambda_2=0.01$.

We study the performance of the proposed nonparametric Bayesian input models by using simulated data generated from the test examples listed in Table~\ref{table:example}.
	Example~1 is an Pareto distribution with shape $1.1$ and the support on $\Re^+$, which violates Condition~(i) in Theorem~1. Examples~2 and 3 are shifted Gamma and shifted Weibull with shape less than $1$ and the support on $\Re^+$, which have unknown lower endpoints of their support, and violate Condition~(i) in Theorem~1. Example~4 is Log-Logistic with shape less than $1$ and the support on $\Re^+$, which violates Condition~(i) in Theorem~1.
	Example~5 is a mixture distribution of {log-normal} with the support on $\Re^+$. Example~6 is a mixture distribution of Gumbel with the support on $\Re$. Both {log-normal} and Gumbel mixtures have heavy tails.
	Example~7 is a mixture of Beta distributions, which has the support on $[0,1]$.

\begin{table}[!h]
	\centering
	\caption{Test examples to study the input distribution estimation}
	\label{table:example}
	\scalebox{0.8}{
		\begin{tabular}{|c|c|c|}
			\hline
			Example 1            & Pareto  &  Pareto(shape = 1.1, scale = 1)          \\ \hline
			Example 2            & Shifted Gamma  & Gamma(0.5,1) with shift = 1 \\ \hline
			Example 3            & Shifted Weibull  &  Weibull(shape = 0.5, scale = 1) with shift = 1          \\ \hline
			Example 4           & Log-logistic  &  Log-logistic(shape = 0.5, scale = 1)         \\ \hline
			Example 5            & {Log-normal} (L) Mixture  & 0.3L(0,0.1)+0.4L(1,0.1)+0.3L(2,0.1) \\ \hline
			Example 6            & Gumbel (Gum) Mixture     & 0.3Gum(1.5,0.1)+0.4Gum(2.5,0.3)+0.3Gum(5,0.5)           \\ \hline
			Example 7            & Beta (Be) Mixture     & 0.3Be(10,90)+0.4Be(20,60)+0.3Be(10,10)           \\ \hline
		\end{tabular}
	}
\end{table}

For Bayesian approaches, there exist various model selection criteria, including Bayes Factor \cite{Kass1995}, Posterior predictive density \cite{Gelman_2004}, and Deviance Information Criteria \cite{Spiegelhalter2002}. However, they are not suitable here since we consider both frequentist and Bayesian candidates.
As the KS and AD test statistics are commonly used to study the goodness of fit in the simulation community,
we use the KS and AD criteria to study the fitting performance obtained by various approaches. Since the underlying true input model $F^c$ for examples listed in Table~\ref{table:example} are known, we replace the hypothesized distribution in these test statistics with $F^c$ to obtain corresponding distance measures. The KS distance, defined as $D_m\equiv\underset{x\in \Re}\sup(|F^c(x)-\widehat{F}_m(x)|)$, records the largest vertical distance between $F^c(\cdot)$ and the distribution estimated by $m$ real-world data, denoted by $\widehat{F}_m(\cdot)$, which could be obtained by different approaches, including DPM with various kernel densities, empirical distribution, KDE and parametric approaches.
The KS distance assigns equal weight to all $x\in \Re$. Since it is typically more challenging to estimate the tail behavior compared to the central part, the AD distance, defined as $A_m^2\equiv m \int_{-\infty}^{\infty} |F^c(x)-\widehat{F}_m(x)|^2 w(x)dF^c(x)$, places more weight on the tails of $F^c$,
where the weight function is $w(x)=1/\left(F^c(x)(1-F^c(x))\right)$. Thus, the AD distance can better detect the discrepancies in the tails.

Table~$\ref{table:sizes}$ records the statistical behaviors of
KS and AD distances ($D_m$ and $A_m$) obtained by DPM with Gamma, Gaussian, and Beta kernel densities, {finite mixture \cite{Cheng2003}}, empirical distribution, KDE, and parametric distributions {selected based on KS and AD criteria}
when $m=50,100,500$. All results are based on $N=1000$ macro-replications. In the $i$-th macro-replication, we first draw $m$ samples, denoted by $\mathbf{X}_m^{(i)}$, from $F^c$ listed in Table~\ref{table:example} to mimic the procedure collecting $m$ ``real-world data". Then, various approaches are used to fit the real-world data, and calculate the KS and AD distances for the fitted distributions.
In the table, ``parametric (AD)" and ``parametric (KS)" refer to the parametric distributions selected based on the AD and KS statistics by using @Risk \cite{risk}.
KDE is obtained by using the R function, \texttt{kde}, and the bandwidth is selected to minimize the mean integrated squared error \cite{Sheather_Jones_1991}.
For empirical distribution, KDE and parametric approaches, we find the fitted distributions and then record the KS and AD distances for these fitted distributions.

\begin{table}[ht]
	\centering
	\caption{KS and AD distances obtained from DPM with Gamma, Gaussian and Beta kernel densities, the empirical distribution, KDE and parametric distributions selected based on KS and AD tests.}
	\label{table:sizes}
	\scalebox{0.64}{
		\begin{tabular}{|c|c|c|c|c|c|c|c|c|}
			\hline
			\multicolumn{2}{|c|}{$m = 50$}                  & Example 1        & Example 2        & Example 3        & Example 4        & Example 5        & Example 6        & Example 7        \\ \hline
			\multirow{2}{*}{DPM with Gamma}         & $D_m$ & \textbf{0.089$\pm$0.001}  & \textbf{0.116$\pm$0.001}  & \textbf{0.123$\pm$0.002}  & \textbf{0.072$\pm$0.001}  & \textbf{0.050$\pm$0.001}  & NA               & 0.075$\pm$0.001  \\ \cline{2-9}
			& $A_m$ & \textbf{12.079$\pm$0.153} & \textbf{15.649$\pm$0.210} & \textbf{13.172$\pm$0.231} & \textbf{11.253$\pm$0.144} & \textbf{7.093$\pm$0.098}  & NA               & \textbf{8.566$\pm$0.115}  \\ \hline
			\multirow{2}{*}{DPM with Gaussian}      & $D_m$ & 0.135$\pm$0.002  & 0.133$\pm$0.002  & 0.146$\pm$0.002  & 0.098$\pm$0.001  & 0.062$\pm$0.001  & \textbf{0.071$\pm$0.001}  & 0.084$\pm$0.001  \\ \cline{2-9}
			& $A_m$ & 16.894$\pm$0.250 & 17.759$\pm$0.288 & 18.658$\pm$0.319 & 14.599$\pm$0.221 & 8.861$\pm$0.105  & \textbf{5.940$\pm$0.074}  & 9.864$\pm$0.129  \\ \hline
			\multirow{2}{*}{DPM with Beta}          & $D_m$ & NA               & NA               & NA               & NA               & NA               & NA               & \textbf{0.068$\pm$0.001}  \\ \cline{2-9}
			& $A_m$ & NA               & NA               & NA               & NA               & NA               & NA               & 8.608$\pm$0.096  \\ \hline
			\multirow{2}{*}{Finite Mixture}         & $D_m$ & 0.113$\pm$0.001  & 0.230$\pm$0.003  & 0.206$\pm$0.002  & 0.142$\pm$0.002  & 0.143$\pm$0.002  & 0.094$\pm$0.001  & 0.082$\pm$0.001  \\ \cline{2-9}
			& $A_m$ & 14.324$\pm$0.218 & 21.658$\pm$0.293 & 25.023$\pm$0.324 & 18.549$\pm$0.287 & 16.127$\pm$0.251 & 9.387$\pm$0.133  & 10.528$\pm$0.146 \\ \hline
			\multirow{2}{*}{Empirical Distribution} & $D_m$ & 0.104$\pm$0.001  & 0.122$\pm$0.002  & \textbf{0.120$\pm$0.002}  & 0.115$\pm$0.002  & 0.082$\pm$0.001  & 0.081$\pm$0.001  & 0.084$\pm$0.001  \\ \cline{2-9}
			& $A_m$ & 12.224$\pm$0.163 & 16.181$\pm$0.188 & 13.769$\pm$0.169 & 13.934$\pm$0.174 & 9.734$\pm$0.118  & 6.592$\pm$0.087  & 9.809$\pm$0.120  \\ \hline
			\multirow{2}{*}{KDE}                    & $D_m$ & 0.161$\pm$0.002  & 0.174$\pm$0.002  & 0.207$\pm$0.002  & 0.085$\pm$0.001  & 0.141$\pm$0.002  & 0.130$\pm$0.002  & 0.084$\pm$0.001  \\ \cline{2-9}
			& $A_m$ & 19.086$\pm$0.266 & 20.654$\pm$0.291 & 23.996$\pm$0.325 & 13.404$\pm$0.190 & 18.951$\pm$0.283 & 11.245$\pm$0.148 & 10.408$\pm$0.142 \\ \hline
			\multirow{2}{*}{Parametric (KS)}        & $D_m$ & 0.153$\pm$0.002  & 0.180$\pm$0.002  & 0.191$\pm$0.002  & 0.076$\pm$0.001  & 0.147$\pm$0.002  & 0.128$\pm$0.002  & 0.110$\pm$0.001  \\ \cline{2-9}
			& $A_m$ & 22.045$\pm$0.305 & 25.541$\pm$0.328 & 26.410$\pm$0.346 & 12.679$\pm$0.170 & 19.653$\pm$0.281 & 12.273$\pm$0.168 & 11.506$\pm$0.157 \\ \hline
			\multirow{2}{*}{Parametric (AD)}        & $D_m$ & 0.154$\pm$0.002  & 0.184$\pm$0.002  & 0.191$\pm$0.002  & 0.078$\pm$0.001  & 0.148$\pm$0.002  & 0.131$\pm$0.002  & 0.113$\pm$0.001  \\ \cline{2-9}
			& $A_m$ & 21.922$\pm$0.292 & 25.528$\pm$0.303 & 26.207$\pm$0.352 & 12.100$\pm$0.174 & 19.744$\pm$0.283 & 11.924$\pm$0.159 & 11.233$\pm$0.152 \\ \hline
			\hline
			\multicolumn{2}{|c|}{$m=100$}                   & Example 1        & Example 2        & Example 3        & Example 4        & Example 5        & Example 6       & Example 7       \\ \hline
			\multirow{2}{*}{DPM with Gamma}         & $D_m$ & \textbf{0.058$\pm$0.001}  & \textbf{0.088$\pm$0.001}  & \textbf{0.084$\pm$0.001}  & \textbf{0.051$\pm$0.001}  & \textbf{0.056$\pm$0.001}  & NA              & 0.053$\pm$0.001 \\ \cline{2-9}
			& $A_m$ & \textbf{8.095$\pm$0.114}  & \textbf{9.093$\pm$0.142} & 10.516$\pm$0.166 & \textbf{7.160$\pm$0.109}  & \textbf{6.121$\pm$0.088}  & NA              & \textbf{6.208$\pm$0.090} \\ \hline
			\multirow{2}{*}{DPM with Gaussian}      & $D_m$ & 0.091$\pm$0.001  & 0.100$\pm$0.001  & 0.117$\pm$0.002  & 0.069$\pm$0.001  & 0.064$\pm$0.001  & \textbf{0.062$\pm$0.001} & 0.056$\pm$0.001 \\ \cline{2-9}
			& $A_m$ & 13.174$\pm$0.187 & 12.678$\pm$0.182 & 14.224$\pm$0.202 & 10.372$\pm$0.153 & 8.133$\pm$0.111  & \textbf{4.390$\pm$0.068} & 7.334$\pm$0.105 \\ \hline
			\multirow{2}{*}{DPM with Beta}          & $D_m$ & NA               & NA               & NA               & NA               & NA               & NA              & \textbf{0.049$\pm$0.001} \\ \cline{2-9}
			& $A_m$ & NA               & NA               & NA               & NA               & NA               & NA              & 6.416$\pm$0.092 \\ \hline
			\multirow{2}{*}{Finite Mixture}         & $D_m$ & 0.096$\pm$0.001  & 0.176$\pm$0.002  & 0.185$\pm$0.002  & 0.064$\pm$0.001  & 0.094$\pm$0.001  & 0.066$\pm$0.001 & 0.061$\pm$0.001 \\ \cline{2-9}
			& $A_m$ & 15.550$\pm$0.207 & 17.582$\pm$0.237 & 18.188$\pm$0.255 & 11.288$\pm$0.167 & 16.583$\pm$0.214 & 5.329$\pm$0.080 & 7.802$\pm$0.102 \\ \hline
			\multirow{2}{*}{Empirical Distribution} & $D_m$ & 0.071$\pm$0.001  & \textbf{0.086$\pm$0.001}  & 0.087$\pm$0.001  & 0.083$\pm$0.001  & 0.064$\pm$0.001  & 0.068$\pm$0.001 & 0.059$\pm$0.001 \\ \cline{2-9}
			& $A_m$ & 8.319$\pm$0.115  & 9.621$\pm$0.134  & \textbf{9.687$\pm$0.139}  & 8.860$\pm$0.128  & 6.874$\pm$0.096  & 5.443$\pm$0.075 & 7.230$\pm$0.098 \\ \hline
			\multirow{2}{*}{KDE}                    & $D_m$ & 0.130$\pm$0.002  & 0.176$\pm$0.002  & 0.211$\pm$0.001  & 0.062$\pm$0.001  & 0.132$\pm$0.002  & 0.078$\pm$0.001 & 0.060$\pm$0.001 \\ \cline{2-9}
			& $A_m$ & 15.431$\pm$0.204 & 18.544$\pm$0.239 & 22.895$\pm$0.315 & 10.623$\pm$0.161 & 18.037$\pm$0.243 & 6.386$\pm$0.094 & 8.095$\pm$0.112 \\ \hline
			\multirow{2}{*}{Parametric (KS)}        & $D_m$ & 0.141$\pm$0.002  & 0.174$\pm$0.002  & 0.182$\pm$0.002  & \textbf{0.049$\pm$0.001}  & 0.146$\pm$0.001  & 0.087$\pm$0.001 & 0.086$\pm$0.001 \\ \cline{2-9}
			& $A_m$ & 19.710$\pm$0.315 & 21.593$\pm$0.330 & 22.263$\pm$0.367 & 7.752$\pm$0.118  & 16.585$\pm$0.293 & 7.794$\pm$0.115 & 9.503$\pm$0.124 \\ \hline
			\multirow{2}{*}{Parametric (AD)}        & $D_m$ & 0.143$\pm$0.002  & 0.175$\pm$0.002  & 0.184$\pm$0.002  & 0.050$\pm$0.001  & 0.146$\pm$0.002  & 0.089$\pm$0.001 & 0.086$\pm$0.001 \\ \cline{2-9}
			& $A_m$ & 19.448$\pm$0.322 & 21.081$\pm$0.335 & 22.928$\pm$0.354 & 7.680$\pm$0.114  & 16.330$\pm$0.287 & 7.542$\pm$0.109 & 9.337$\pm$0.133 \\ \hline
				\hline
				\multicolumn{2}{|c|}{$m=500$}                   & Example 1        & Example 2        & Example 3        & Example 4       & Example 5        & Example 6       & Example 7       \\ \hline
				\multirow{2}{*}{DPM with Gamma}         & $D_m$ & \textbf{0.042$\pm$0.001}  & 0.065$\pm$0.001  & 0.061$\pm$0.001  & \textbf{0.034$\pm$0.001} & \textbf{0.028$\pm$0.001}  & NA              & 0.031$\pm$0.001 \\ \cline{2-9}
				& $A_m$ & \textbf{5.022$\pm$0.073}  & 8.577$\pm$0.118  & 7.397$\pm$0.109  & \textbf{3.820$\pm$0.057} & \textbf{3.147$\pm$0.054}  & NA              & 3.653$\pm$0.058 \\ \hline
				\multirow{2}{*}{DPM with Gaussian}      & $D_m$ & 0.050$\pm$0.001  & 0.053$\pm$0.001  & 0.059$\pm$0.001  & 0.033$\pm$0.001 & 0.031$\pm$0.001  & \textbf{0.026$\pm$0.001} & 0.035$\pm$0.001 \\ \cline{2-9}
				& $A_m$ & 6.880$\pm$0.095  & 6.558$\pm$0.093  & 7.056$\pm$0.102  & 5.045$\pm$0.074 & 3.599$\pm$0.038  & \textbf{2.744$\pm$0.044} & 4.490$\pm$0.069 \\ \hline
				\multirow{2}{*}{DPM with Beta}          & $D_m$ & NA               & NA               & NA               & NA              & NA               & NA              & \textbf{0.026$\pm$0.001} \\ \cline{2-9}
				& $A_m$ & NA               & NA               & NA               & NA              & NA               & NA              & \textbf{3.030$\pm$0.042} \\ \hline
				\multirow{2}{*}{Finite Mixture}         & $D_m$ & 0.082$\pm$0.001  & 0.105$\pm$0.001  & 0.134$\pm$0.002  & 0.070$\pm$0.001 & 0.114$\pm$0.001  & 0.041$\pm$0.001 & 0.040$\pm$0.001 \\ \cline{2-9}
				& $A_m$ & 12.828$\pm$0.174 & 14.302$\pm$0.195 & 16.899$\pm$0.212 & 8.393$\pm$0.124 & 13.676$\pm$0.176 & 4.568$\pm$0.065 & 5.559$\pm$0.072 \\ \hline
				\multirow{2}{*}{Empirical Distribution} & $D_m$ & 0.051$\pm$0.001  & \textbf{0.048$\pm$0.001}  & \textbf{0.049$\pm$0.001}  & 0.041$\pm$0.001 & 0.038$\pm$0.001  & 0.036$\pm$0.001 & 0.038$\pm$0.001 \\ \cline{2-9}
				& $A_m$ & 5.527$\pm$0.078  & \textbf{6.059$\pm$0.091}  & \textbf{6.038$\pm$0.088}  & 4.254$\pm$0.060 & 3.570$\pm$0.052  & 3.673$\pm$0.055 & 4.922$\pm$0.067 \\ \hline
				\multirow{2}{*}{KDE}                    & $D_m$ & 0.126$\pm$0.002  & 0.152$\pm$0.002  & 0.185$\pm$0.002  & 0.037$\pm$0.001 & 0.113$\pm$0.001  & 0.048$\pm$0.001 & 0.034$\pm$0.001 \\ \cline{2-9}
				& $A_m$ & 14.327$\pm$0.193 & 15.277$\pm$0.205 & 18.839$\pm$0.226 & 6.375$\pm$0.099 & 15.735$\pm$0.211 & 6.066$\pm$0.094 & 5.172$\pm$0.073 \\ \hline
				\multirow{2}{*}{Parametric (KS)}        & $D_m$ & 0.139$\pm$0.002  & 0.167$\pm$0.002  & 0.174$\pm$0.002  & 0.038$\pm$0.001 & 0.135$\pm$0.002  & 0.072$\pm$0.001 & 0.065$\pm$0.001 \\ \cline{2-9}
				& $A_m$ & 16.635$\pm$0.231 & 19.744$\pm$0.289 & 20.136$\pm$0.296 & 5.293$\pm$0.074 & 15.527$\pm$0.218 & 7.131$\pm$0.104 & 6.975$\pm$0.097 \\ \hline
				\multirow{2}{*}{Parametric (AD)}        & $D_m$ & 0.139$\pm$0.002  & 0.170$\pm$0.002  & 0.176$\pm$0.002  & 0.039$\pm$0.001 & 0.136$\pm$0.002  & 0.077$\pm$0.001 & 0.068$\pm$0.002 \\ \cline{2-9}
				& $A_m$ & 16.469$\pm$0.230 & 19.682$\pm$0.286 & 20.122$\pm$0.303 & 5.118$\pm$0.071 & 15.760$\pm$0.220 & 7.530$\pm$0.109 & 6.883$\pm$0.095 \\ \hline
				\multicolumn{9}{c}{}
	\end{tabular}}
\end{table}

Differing from these frequentist approaches that provide the point estimates of input distribution, {DPM and finite mixture are Bayesian approaches}.
According to \cite{Gelman_2004}, the posterior predictive distribution, defined by $f(X|\mathbf{X}_m)=\int f(X|F)dP(F|\mathbf{X}_m)$, is recommended for assessing the fitting performance of input model to the real-world data. Thus, the posterior predictive distribution is used to calculate the KS and AD distances. Specifically, we use the Gibbs samplers described in Section~\ref{subsec:gibbssampler} to generate 100 posterior samples of input models with the warmup equal to 500 and save the sample for each 10 draws. Then, we aggregate these posterior samples to obtain the posterior predictive distribution, $\widehat{f}(X|\mathbf{X}_m)=\sum_{b=1}^{B'} f(X|\widetilde{F}^{(b)})/B'$ with $\widetilde{F}^{(b)}\sim P(F|\mathbf{X}_m)$ for $b=1,2,\ldots,B'$ and $B'=100$. To calculate the KS and AD distances, the posterior predictive distribution is used to replace $\widehat{F}_m(x)$ in $D_m$ and $A_m^2$. 
We generate $10,000$ samples from $F^c$ and $10,000$ samples from posterior predictive distribution (i.e., draw 100 posterior input models $\widetilde{F}^{(b)}$ with $b=1,2,\ldots,B'$ and generate 100 samples of $X$ from each $\widetilde{F}^{(b)}$) to numerically estimate the KS and AD distances.

In the $i$-th macro-replication, we obtain the KS and AD distances, denoted by $D^{(i)}_m$ and $A^{(i)}_m$, with $i=1,2,\ldots,N$.
	Then, we record 95\% symmetric CIs for both KS and AD distances, denoted by $\bar{D}\pm 1.96 S_D/\sqrt{N}$ and $\bar{A}\pm 1.96 S_A/\sqrt{N}$, in Table~\ref{table:sizes}, and highlight the smallest values, where
	$\bar{D}=\sum_{i=1}^{N}D^{(i)}_m/N$, $\bar{A}=\sum_{i=1}^{N}A^{(i)}_m/N$, $S_D=\big[\sum_{i=1}^{N}(D^{(i)}_m-\bar{D})^2/(N-1)\big]^{1/2}$ and $S_A=\big[\sum_{i=1}^{N}(A^{(i)}_m-\bar{A})^2/(N-1)\big]^{1/2}$. As $m$ increases, the KS and AD distances obtained from all approaches decrease, and the DPM with appropriate kernel density typically has the best performance.
Notice that DPM with Gamma and Beta kernel densities performs better than DPM with Gaussian kernel, which is the main focus of study in both statistics and machine learning communities. Further, DPM with Gamma kernel fits different input models with support on $\Re^+$ well. Based on the results of AD distance, DPM tends to provide better estimation on the tail behavior compared with the finite mixture, empirical, KDE and parametric distributions, especially when $m$ is not large.


\subsection{Studying Input Model Performance by Using Real Raw Materials Demand Data}
	\label{subsec:AppendixRealData}
	
	Besides the simulated data in Section~\ref{subsec:inputEstimation}, we also assess the performance of our nonparametric input models by using the demand data of two representative raw materials (RM) collected from a real biopharmaceutical manufacturing system. The sample sizes are 101 and 142 respectively. Since the underlying true distributions are unknown, \textit{cross validation is applied for the density selection}; see more detailed description in \cite{Lian2009}. We perform a 5-folds cross validation. Table~\ref{table:cvResults} records the average log-likelihoods obtained by using different approaches. Specifically, we randomly divide all the data into 5 sets, select one set for validation and use the remaining sets as training data. For each combination of training and validation data sets, we first fit the input model by using the training data, apply it to the validation data and calculate the log-likelihood. After that, we record the average log-likelihood obtained from all combinations of training and validation data sets.
	
	Since the demand data have support on $\Re^+$, we use DPM with Gamma kernel density. The distribution family for the parametric approach is selected based on the KS test statistics by using @Risk since both this criteria and the likelihood are related to the overall fitting performance of input model. In addition, we skip the empirical distribution since it only has the information at the data points and does not return a density estimate.
	
	\begin{table}[]
		\centering
		\caption{Average log-likelihood results of cross validation for the distribution density function selection}
		\label{table:cvResults}
		\scalebox{0.75}{
			\begin{tabular}{|c|c|c|c|c|}
				\hline
				& DPM Gamma & Empirical Distribution & KDE      & Parametric \\ \hline
				Demand of raw material 1 & -218.731  & NA     & -226.928 & -393.239   \\ \hline
				Demand of raw material 2 & -233.971  & NA     & -270.577 & -605.704   \\ \hline
			\end{tabular}
		}
	\end{table}
	
	Since the posterior predictive distribution is recommended for the model selection \cite{Gelman_2004}, for DPM, the likelihood is calculated based on the posterior predictive distribution given by $f\big(\mathbf{X}_V^{(i)}|\mathbf{X}_T^{(i)}\big)=\int f\big(\mathbf{X}_V^{(i)}|F\big)dP\big(F|\mathbf{X}_T^{(i)}\big)$, where $\mathbf{X}_T^{(i)}$ and $\mathbf{X}_V^{(i)}$ denote the $i$-th combination of training and validation data with $i=1,2,\ldots,5$. Then, we record the average log-likelihood $\sum_{i=1}^5 \log \big[ f\big(\mathbf{X}_V^{(i)}|\mathbf{X}_T^{(i)}\big)\big]/5$. For the frequentist KDE and parametric approaches, we first find the fitted input density based on the training set, denoted by $\widehat{f}(\cdot|{\mathbf{X}_T^{(i)}})$, then apply it to the validation data and calculate the average log-likelihood $\sum_{i=1}^5 \log\big[\widehat{f}\big(\mathbf{X}_V^{(i)}|\mathbf{X}_T^{(i)}\big)\big]/5$. Table~\ref{table:cvResults} demonstrates that DPM with Gamma kernel maximizes the average log-likelihood and provides the best fit to the real RM demand data.

\subsection{Identifying the Underlying Sources of Uncertainty}
	\label{subsec:numberofCluster}
To study the number of components identified by DPM and the finite mixture \cite{Cheng2003}, we consider a mixture distribution as test example with $F^c$ equal to $0.3\mbox{Gum}(1,0.1)+0.3{L(2,0.1)}+0.4\mathcal{N}(4,0.5)$. It includes three components from different parametric families, where the Gumbel and log-normal components are asymmetric and the normal component is symmetric. Here, we use the DPM with Gaussian kernel, and record the marginal distribution for the number of active components,
	\begin{equation}
	\bar{p}(K_0=k)=\int p(K_0=k|\mathbf{X}_m) dF^c(\mathbf{X}_m).
	\label{eq.k0}  \nonumber
	\end{equation}
	To differentiate from the prior distribution of $K_0$, here we use the notation $\bar{p}(\cdot)$.
	The probability $\bar{p}(K_0=k)$ is estimated based on $N=100$ macro-replications and $B_0=100$ posterior samples of input model obtained in each macro-replication with results shown in Table~\ref{table:postK}. In the $i$-th macro-replication, we generate $\mathbf{X}_m^{(i)}\stackrel{i.i.d.} \sim F^c$ with $i=1,2,\ldots,N$. The marginal probability $\bar{p}(K_0=k)$ is estimated by using $\frac{1}{B_0N}\sum_{i=1}^{N} \sum_{b=1}^{B_0} \delta(K_0^{(b)}=k|\mathbf{X}_m^{(i)})$ for $k=1,2,\ldots,m$, where $\delta(\cdot|\mathbf{X}_m^{(i)})$ denotes an indicator function conditional on $\mathbf{X}_m^{(i)}$ for $i=1,2,\ldots,N$. The posterior samples of $K_0$ and input model can be obtained by following the procedure in Section~\ref{subsec:gibbssampler}. We compare the posterior of $K_0$ obtained by DPM with that obtained from the finite Gaussian mixture using Maximum A Posteriori Importance Sampling (MAPIS) described in \cite{Cheng2003}. We record the estimated marginal probability in Table~\ref{table:postK} when $m=50,100,500$.
	\textit{The DPM provides a better detection of the underlying number sources of uncertainty.}

\begin{table}[ht]
	\centering
	\caption{The Estimated Marginal Distribution for the Number of Active Components, $\bar{p}(K_0=k)$}
	\label{table:postK}
	\scalebox{0.75}{
		\begin{tabular}{|c|c|c|c|c|c|c|c|c|c|c|c|}
			\hline
			\multicolumn{2}{|c|}{$k$}      & 1     & 2     & 3     & 4     & 5     & 6     & 7     & 8     & 9     & \textgreater=10 \\ \hline
			\multirow{2}{*}{$m = 50$}  & DPM   & 0.014 & 0.165 & \textbf{0.453} & 0.239 & 0.084 & 0.026 & 0.013 & 0.004 & 0.002 & 0               \\ \cline{2-12}
			& MAPIS & 0.047 & 0.079 & 0.109 & 0.102 & 0.108 & 0.106 & 0.109 & \textbf{0.116} & 0.110 & 0.115           \\ \hline
			\multirow{2}{*}{$m = 100$} & DPM   & 0     & 0.138 & \textbf{0.503} & 0.255 & 0.072 & 0.023 & 0.008 & 0.001 & 0     & 0               \\ \cline{2-12}
			& MAPIS & 0.036 & 0.070 & 0.114 & 0.111 & \textbf{0.118} & 0.107 & 0.112 & 0.117 & 0.115 & 0.099           \\ \hline
			\multirow{2}{*}{$m = 500$} & DPM   & 0     & 0.097 & \textbf{0.558} & 0.246 & 0.077 & 0.016 & 0.006 & 0     & 0     & 0               \\ \cline{2-12}
			& MAPIS & 0.032 & 0.050 & \textbf{0.123} & 0.122 & 0.106 & 0.113 & \textbf{0.123} & 0.115 & 0.108 & 0.109           \\ \hline
		\end{tabular}
	}
\end{table}

\begin{figure}[ht]
	\begin{minipage}{.49\textwidth}
		\centering
		\includegraphics[width=.7\linewidth]{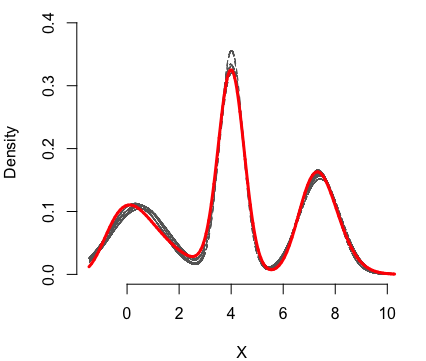}
		\caption{Posterior samples of input density obtained by DPM}
		\label{fig:densityDPM}
	\end{minipage}
	\begin{minipage}{.49\textwidth}
		\centering
		\includegraphics[width=.7\linewidth]{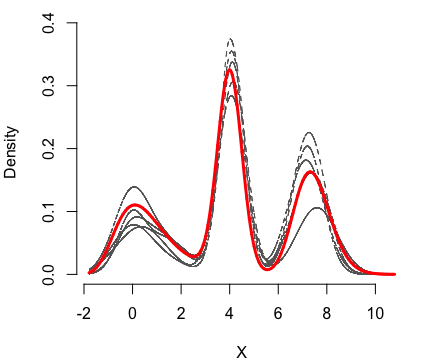}
		\caption{Posterior samples of input density obtained by finite mixture}
		\label{fig:densityMAPIS}
	\end{minipage}
\end{figure}

In addition, given the data $\mathbf{X}_m$, Figures~\ref{fig:densityDPM} and \ref{fig:densityMAPIS} give the representative posterior samples of input model obtained by DPM and finite mixture \cite{Cheng2003} when $m=500$. {The posterior density function of $\widetilde{F}^{(b)}$ is given in Equation~(\ref{eq.posteriorF}), where the integration can be estimated by $\frac{1}{N_G}\sum_{i=1}^{N_G}h(x|\pmb{\psi}_i)$ with $\pmb{\psi}_i\sim G_0$, and we use $N_G = 1000$ here.} In Figures~\ref{fig:densityDPM} and \ref{fig:densityMAPIS}, the solid line represents the true density function, and the dashed lines represent the posterior samples of input model. \textit{The figures show that DPM can deliver accurate density estimation.}

\subsection{An $M/G/1$ Queue}
\label{subsec:mm1Queue_sim}

An $M/G/1$ queue is used to study the performance of our DPM-based nonparametric Bayesian framework, and we compare it with the empirical distribution based direct bootstrap  \cite{Barton_Schruben_2001}.
	Suppose that the arrival process is known with the arrival rate equal to $\lambda$, and the true distribution of service time $F^c$ is unknown. We are interested in the probability of each customer having the time staying in the system greater than a threshold, denoted by $\tau$.
The unknown distribution for service time is estimated by using $m = 50, 500$ observations drawn from $F^c$. Among the seven examples in Table~1, Example~1 has infinite variance and Example~6 has support on $\Re$. Thus, we consider the remaining five examples as shown in Table~\ref{table:examplesOutput} as the underlying input distribution.
	For each case, the true mean system response $\mu^c$ is estimated by a side experiment with the runlength equal to $10^6$ customers.

\begin{table}[ht]
	\centering
	\caption{The settings for five $M/G/1$ test examples}
	\label{table:examplesOutput}
	\scalebox{0.75}{
		\begin{tabular}{|c|c|c|c|c|}
			\hline
			Example       & Distribution of Service Time                           & Inter-arrival Time   & Threshold   &  $\mu^c$ (\%) \\ \hline
			{Log-normal}     & $0.3L(0,0.1)+0.4L(1,0.1)+0.3L(2,0.1)$  & $\mbox{Exp}(\lambda = 0.2)$ & $\tau = 25$ & $8.43\pm 0.02$    \\ \hline
			{Log-logistic}  & $\mbox{Log-logistic}(\mbox{shape} = 0.5, \mbox{scale} = 1)$ & $\mbox{Exp}(\lambda = 0.1)$ & $\tau = 20$ & $11.23\pm 0.03$    \\ \hline
			Shifted-Gamma & $\mbox{Gamma}(0.5,1)$ with $\mbox{shift} = 1$          & $\mbox{Exp}(\lambda = 0.5)$ & $\tau = 8$  & $13.03\pm 0.03$    \\ \hline
			Shifted-Weibull & $\mbox{Weibull}(0.5,1)$ with $\mbox{shift} = 1$          & $\mbox{Exp}(\lambda = 0.25)$ & $\tau = 40$  & $11.98\pm 0.03$    \\ \hline
			Beta & $0.3\mbox{Be}(10,90)+0.4\mbox{Be}(20,60)+0.3\mbox{Be}(10,10)$           & $\mbox{Exp}(\lambda = 3)$ & $\tau = 3$  & $8.37\pm 0.02$    \\ \hline
	\end{tabular}}
\end{table}

\begin{sloppypar}
We use DPM with Gamma kernel to model the distribution of service time. By following the sampling procedure described in Section~\ref{subsec:gibbssampler}, we construct the $(1-\alpha^*)\times 100\%=90\%$ percentile empirical CrI,  $\left[\bar{Y}_{\left(\lceil(\alpha^*/2)B\rceil\right)}, \bar{Y}_{\left(\lceil(1-\alpha^*/2)B\rceil\right)}\right]$, accounting for both input and simulation uncertainties. We use $\bar{\bar{Y}} = \frac{1}{B}\sum_{b=1}^{B}\bar{Y}_b$ as the point estimator for the system mean response.
For the direct bootstrap, we generate $B$ sets of bootstrapped input data, build empirical distributions, $\widehat{F}^{(1)},\widehat{F}^{(2)},\ldots,\widehat{F}^{(B)}$, and record the output sample mean $\bar{Y}_b$ at each $\widehat{F}^{(b)}$ for $b=1,2,\ldots,B$. A percentile CI is constructed according to \cite{Barton_Schruben_2001}, and similarly, $\bar{\bar{Y}} = \frac{1}{B}\sum_{b=1}^{B}\bar{Y}_b$ is used as the point estimator.
	We set $B=1000$, and assign equal numbers of replications, $n=100,1000$, to each posterior or bootstrapped sample of input model. Each simulation run starts with the empty system with both warmup and runlength equal to $1000$ customers.
\end{sloppypar}

To compare the performance of our approach with direct bootstrap, we first estimate the mean and standard deviation (SD) of the deviation of the point estimator from $\mu^c$, defined by $\mbox{Err}=|\bar{\bar{Y}}-\mu^c|$, and then report $\mbox{mean}\pm1.96\cdot \mbox{SD}/\sqrt{N}$
of the CrI and CI width as shown in Table~\ref{table:Queue}, where ``DPM" and ``EMP" represent our approach and direct bootstrap, respectively. The results are based on $N=1000$ macro-replications. Our proposed nonparametric Bayesian framework provides smaller $\mbox{Err}$ and CrI's with shorter width.

Frequentist and Bayesian approaches have totally different philosophies in terms of uncertainty quantification and performance evaluation. In the frequentist approaches, we construct the random confidence interval (CI) for true mean response $\mu^c$, and then the percentage covering $\mu^c$ is used to assess the performance of this interval. Different from frequentist approaches, 
the belief of input uncertainty in Bayesian approaches is quantified by the posterior distribution of $W\equiv \mu(\widetilde{F})$ with $\widetilde{F}\sim p(F|\mathbf{X}_m)$.  
By following the studies in \cite{Xie_Bayesian_2014,akcay2017simulation}, we use the probability content (PC) covering the posterior distribution of $\mu(\widetilde{F})$ to evaluate the CrI constructed by our approach. Specifically, to estimate $\mbox{PC}(\widetilde{\mbox{CrI}})$, we draw $B=1000$ posterior samples of the input models $\widetilde{F}^{(b)}\sim p(F|\mathbf{X}_m)$ with $b=1,2,\ldots,B$. At each $\widetilde{F}^{(b)}$, the system mean response or the expected probability of each customer having the time staying in the system greater than the threshold is estimated with side simulation experiments with the runlength equal to $10^5$ customers. The PC is estimated by $\widehat{\mbox{PC}}(\widetilde{\mbox{CrI}}) = \frac{1}{B}\sum_{b=1}^{B} \b 1\left(\mu(\widetilde{F}^{(b)})\in \widetilde{\mbox{CrI}}\right)$, where $1(\cdot)$ represents the indicator function. The standard deviation (SD) of the PC estimator is obtained by using $N = 1000$ macro-replications.
For the frequentist bootstrap approach, the coverage probability (CP) of CI obtained by empirical distribution based bootstrap (EMP) is estimated with $\widehat{\mbox{CP}} = \frac{1}{N}\sum_{i=1}^{N} \b 1 \left( \mu^c \in {\mbox{CI}}^{(i)} \right)$, where ${\mbox{CI}}^{(i)}$ represents the 90\% CI constructed in the $i$-th macro-replication. The standard deviation is estimated by $\sqrt{\widehat{\mbox{CP}}(1-\widehat{\mbox{CP}})}$.
Based on the results in Table~\ref{table:Queue}, the PC of CrI delivered by the proposed Bayesian framework is close to the nominal value 90\%. Overall, the coverage performance of CrI is comparable with the CI obtained by the empirical distribution based bootstrap uncertainty quantification. Since bootstrap uncertainty quantification is built on asymptotic approximation, our Bayesian approach tends to work better when the amount of real-world data is limited, such as $m=50$.

\begin{table}[ht]
\centering
\caption{
The results obtained by using our Bayesian framework (DPM) and direct bootstrap (EMP)}
\label{table:Queue}
\scalebox{0.72}{
\begin{tabular}{|c|c|c|c|c|c|c|}
\hline
Log-normal      & Err\% (DPM)      & CrI Width\% (DPM) & $\widehat{\mbox{PC}}(\widetilde{\mbox{CrI}})\%$ (DPM) & Err\%  (EMP)     & CI Width\%  (EMP) & $\widehat{\mbox{CP}}$ \% (EMP) \\ \hline
$m=500,n=1000$  & 2.0 $\pm$ 0.012  & 6.8 $\pm$ 0.051    & 89.8 $\pm$ 0.086                                      & 2.6 $\pm$ 0.019  & 9.5 $\pm$ 0.068   & 89.6 $\pm$ 1.892             \\ \hline
$m=500,n=100$   & 2.4 $\pm$ 0.019  & 7.3 $\pm$ 0.056   & 89.7 $\pm$ 0.121                                      & 2.9 $\pm$ 0.019  & 10.2 $\pm$ 0.081  & 89.4 $\pm$ 1.908             \\ \hline
$m=50,n=1000$   & 5.1 $\pm$ 0.031  & 20.6 $\pm$ 0.149  & 89.7 $\pm$ 0.093                                      & 6.1 $\pm$ 0.037  & 25.8 $\pm$ 0.167  & 89.5 $\pm$ 1.901             \\ \hline
$m=50,n=100$    & 5.5 $\pm$ 0.037  & 22.3 $\pm$ 0.143  & 88.8 $\pm$ 0.112                                      & 6.9 $\pm$ 0.037  & 27.1 $\pm$ 0.186  & 86.2 $\pm$ 2.138             \\ \hline
Log-logistic    & Err\% (DPM)      & CrI Width\% (DPM) & $\widehat{\mbox{PC}}(\widetilde{\mbox{CrI}})\%$ (DPM) & Err\%  (EMP)     & CI Width\%  (EMP) & $\widehat{\mbox{CP}}$ \% (EMP) \\ \hline
$m=500,n=1000$  & 3.8 $\pm$ 0.025  & 7.4 $\pm$ 0.056   & 89.8 $\pm$ 0.049                                      & 4.4 $\pm$ 0.031  & 9.8 $\pm$ 0.068   & 81.2 $\pm$ 2.422             \\ \hline
$m=500,n=100$   & 4.5 $\pm$ 0.031  & 8.0 $\pm$ 0.062   & 89.8 $\pm$ 0.051                                      & 5.2 $\pm$ 0.037  & 10.5 $\pm$ 0.074  & 79.6 $\pm$ 2.497             \\ \hline
$m=50,n=1000$   & 10.3 $\pm$ 0.068 & 18.6 $\pm$ 0.143  & 89.6 $\pm$ 0.104                                      & 12.2 $\pm$ 0.081 & 25.4 $\pm$ 0.192  & 67.2 $\pm$ 2.909             \\ \hline
$m=50,n=100$    & 10.8 $\pm$ 0.074 & 19.8 $\pm$ 0.161  & 89.5 $\pm$ 0.128                                      & 12.8 $\pm$ 0.087 & 26.3 $\pm$ 0.198  & 62.4 $\pm$ 3.002             \\ \hline
Shifted-Gamma   & Err\% (DPM)      & CrI Width\% (DPM) & $\widehat{\mbox{PC}}(\widetilde{\mbox{CrI}})\%$ (DPM) & Err\%  (EMP)     & CI Width\%  (EMP) & $\widehat{\mbox{CP}}$ \% (EMP) \\ \hline
$m=500,n=1000$  & 4.3 $\pm$ 0.031  & 8.7 $\pm$ 0.068   & 89.8 $\pm$ 0.040                                      & 4.4 $\pm$ 0.031  & 9.6 $\pm$ 0.074   & 91.6 $\pm$ 1.719             \\ \hline
$m=500,n=100$   & 4.8 $\pm$ 0.031  & 9.5 $\pm$ 0.074   & 89.8 $\pm$ 0.041                                     & 4.7 $\pm$ 0.031  & 10.1 $\pm$ 0.074  & 87.9 $\pm$ 2.021             \\ \hline
$m=50,n=1000$   & 9.9 $\pm$ 0.068  & 28.2 $\pm$ 0.198  & 89.7 $\pm$ 0.041                                      & 10.5 $\pm$ 0.068 & 29.6 $\pm$ 0.211  & 84.2 $\pm$ 2.261             \\ \hline
$m=50,n=100$    & 10.4 $\pm$ 0.068 & 29.0 $\pm$ 0.211  & 89.5 $\pm$ 0.044                                      & 10.9 $\pm$ 0.074 & 30.7 $\pm$ 0.236  & 84.1 $\pm$ 2.266             \\ \hline
Shifted-Weibull & Err\% (DPM)      & CrI Width\% (DPM) & $\widehat{\mbox{PC}}(\widetilde{\mbox{CrI}})\%$ (DPM) & Err\%  (EMP)     & CI Width\%  (EMP) & $\widehat{\mbox{CP}}$ \% (EMP) \\ \hline
$m=500,n=1000$  & 6.4 $\pm$ 0.043  & 12.5 $\pm$ 0.112  & 89.4 $\pm$ 0.052                                      & 6.5 $\pm$ 0.043  & 13.3 $\pm$ 0.124  & 85.9 $\pm$ 2.157             \\ \hline
$m=500,n=100$   & 6.9 $\pm$ 0.05   & 13.1 $\pm$ 0.118  & 89.8 $\pm$ 0.059                                      & 7.1 $\pm$ 0.05   & 13.9 $\pm$ 0.136  & 84.5 $\pm$ 2.243             \\ \hline
$m=50,n=1000$   & 18.4 $\pm$ 0.136 & 40.9 $\pm$ 0.279  & 78.7 $\pm$ 0.354                                      & 22.0 $\pm$ 0.161 & 50.9 $\pm$ 0.341  & 75.6 $\pm$ 2.662             \\ \hline
$m=50,n=100$    & 19.1 $\pm$ 0.143 & 41.7 $\pm$ 0.291  & 77.3 $\pm$ 0.382                                      & 22.8 $\pm$ 0.167 & 52.2 $\pm$ 0.359  & 76.1 $\pm$ 2.643             \\ \hline
Beta            & Err\% (DPM)      & CrI Width\% (DPM) & $\widehat{\mbox{PC}}(\widetilde{\mbox{CrI}})\%$ (DPM) & Err\%  (EMP)     & CI Width\%  (EMP) & $\widehat{\mbox{CP}}$ \% (EMP) \\ \hline
$m=500,n=1000$  & 1.5 $\pm$ 0.012  & 4.1 $\pm$ 0.037   & 89.4 $\pm$ 0.091                                      & 1.9 $\pm$ 0.019  & 4.5 $\pm$ 0.037   & 90.2 $\pm$ 1.843             \\ \hline
$m=500,n=100$   & 1.8 $\pm$ 0.012  & 4.4 $\pm$ 0.037   & 88.7 $\pm$ 0.108                                      & 2.2 $\pm$ 0.019  & 4.9 $\pm$ 0.037   & 89.6 $\pm$ 1.892             \\ \hline
$m=50,n=1000$   & 2.9 $\pm$ 0.025  & 10.0 $\pm$ 0.074  & 88.4 $\pm$ 0.111                                      & 3.7 $\pm$ 0.031  & 12.1 $\pm$ 0.099  & 89.8 $\pm$ 1.876             \\ \hline
$m=50,n=100$    & 3.4 $\pm$ 0.031  & 10.5 $\pm$ 0.081  & 88.1 $\pm$ 0.120                                      & 4.1 $\pm$ 0.037  & 12.8 $\pm$ 0.099  & 86.8 $\pm$ 2.098             \\ \hline
\end{tabular}}
\end{table}

We also report the ratio ${\widehat{\sigma}_I^2}/{\widehat{\sigma}_S^2}$ in Table~\ref{table:Ratio} from our approach according to Section~\ref{subsubsec:varianceDecomposition}, which provide insights on the contributions of input and simulation estimation uncertainties.

\begin{table}[ht]
	\centering
	\caption{Ratio of Input and Simulation Uncertainties with 95\% Confidence Interval}
	\label{table:Ratio}
	\scalebox{0.8}{
		\begin{tabular}{|c|c|c|c|c|c|}
			\hline
			${\widehat{\sigma}_I^2}/{\widehat{\sigma}_S^2}$ & {Log-normal}    & Log-logistic   & Shifted-Gamma  & Shifted-Weibull & Beta          \\ \hline
			$m=500,n=1000$                                      & 4.957 $\pm$ 0.077 & 8.792 $\pm$ 0.218  & 6.183 $\pm$ 0.108 & 8.034 $\pm$ 0.152  & 5.870 $\pm$ 0.088 \\ \hline
			$m=500,n=100$                                       & 0.635 $\pm$ 0.009 & 1.455 $\pm$ 0.015  & 0.924 $\pm$ 0.017  & 1.252 $\pm$ 0.019 & 0.908 $\pm$ 0.015 \\ \hline
			$m=50,n=1000$                                       & 8.492 $\pm$ 0.189 & 22.560 $\pm$ 0.544 & 11.273 $\pm$ 0.250 & 16.813 $\pm$ 0.388 & 9.405 $\pm$ 0.230 \\ \hline
			$m=50,n=100$                                        & 1.280 $\pm$ 0.036 & 4.642 $\pm$ 0.114  & 1.765 $\pm$ 0.052 & 3.508 $\pm$ 0.093  & 2.072 $\pm$ 0.052 \\ \hline
	\end{tabular}}
\end{table}

\section{Conclusions}

\label{sec:Conclusion}

Without strong prior information on
the true input models and the system mean response surface, in this paper, a Bayesian nonparametric framework is proposed to quantify the overall uncertainty of system mean performance estimation.
The DPM can model the  mixture  of heterogeneous distributions and capture the important properties in the real-world data, including multi-modality, skewness, and tails. The posteriors of flexible input models can automatically account for both model selection and parameters value uncertainty. 
Then, direct simulation is used to propagate the input model estimation uncertainty to the outputs with the simulation uncertainty quantified by the sampling distribution of system mean responses. Therefore,
given the real-world input data, our framework leads to a sampling procedure
that can deliver a conditional distribution of the system mean response and provide a percentile empirical CrI accounting for both input and simulation uncertainties. A variance decomposition is further developed to quantify the relative contributions from both sources of uncertainty.
Our approach is supported by rigorous asymptotic study. Given a finite amount of real-world data, as the simulation budget increases, our CrI converges to the CrI accounting for input uncertainty with the true mean response surface known. As both real-world data and simulation budget go to infinity, our empirical CrI converges to the true system response.

The empirical study demonstrates obvious advantages of Bayesian nonparametric DPM for input density estimation compared to existing approaches, including empirical distribution, KDE and parametric approaches. 
The simulation results indicate that
our framework is robust to possible violation of the sufficient conditions required for the asymptotic consistency of DPM. Our approach demonstrates better empirical performance than the nonparametric bootstrap in uncertainty quantification.
The ratio $\sigma_I/\sigma_S$ provides a good measure of the relative contributions from input and simulation uncertainties.

\appendix

\section*{Acknowledgments}
The authors are grateful for constructive comments from Barry L. Nelson and help from Bo Wang and Keqi Wang on the empirical study. The research of Cheng Li is supported by the Singapore Ministry of Education Academic Research Funds Tier 1 grant R-155-000-201-114.

\bibliographystyle{plain}
\bibliography{arxiv}

\newpage
\section{Appendix: Gibbs Samplers for DPM with Gamma, Gaussian and Beta Kernels}\label{subsec:AppendixDPM}
	For DPM with Gamma, Gaussian and Beta kernels,
	we provide the posterior inference and sampling for the indicator variables $\mathbf{c}$ and component parameters $\pmb{\psi}^\star_j$ for $j=1,2,\ldots,K_0$ used in Steps~1 and 2 of the Gibbs samplers presented in Section~\ref{subsec:gibbssampler}.
	We describe the main results to support the Gibbs sampling in Section~\ref{subsec:gibbsSampling}. Then, in Section~\ref{subsec:deriveCondPost}, we provide the detailed derivation of the results used in the sampling procedure.

	\subsection{Gibbs Sampling for $\mathbf{c}$ and $\Psi^\star$}
	\label{subsec:gibbsSampling}
	
	\subsubsection{DPM with Gamma Kernel}
	\label{subsubsec:AppendixGammaDPM}
	Here, we present a posterior sampler for the DPM with Gamma kernel. Given the base distribution $G_0$ in Equation~(\ref{eq.priorGamma}), we first generate samples of latent variables $\mathbf{c}$ for Step~1 of the Gibbs sampler in Section~\ref{subsec:gibbssampler}. According to Equation~(\ref{post_latent}), the conditional posterior probabilities of $c_i$ in DPM with Gamma kernel is
	\begin{equation}
	\label{latent_gamma}
	p(c_i=j|\mathbf{c}^{-i},\pmb{\psi}^\star_j,\alpha,X_i)=\left\{
	\begin{array}{ll}
	b_0\frac{m_j^{-i}}{m+\alpha-1}X_i^{V_j-1}e^{-\frac{V_j}{u_j}X_i} \mbox{  if  }\exists c_q=j \mbox{ for all } q\neq i \\
	b_0\frac{\alpha}{m+\alpha-1}\int X_i^{V-1}e^{-\frac{V}{u}X_i}dG_0(V,u) \mbox{  otherwise }
	\end{array}
	\right.
	\end{equation}
	where $b_0$ denotes the normalizing constant.
	When $X_i$ comes from a new component, the conditional posterior for $c_i$ in Equation \eqref{latent_gamma} is not analytically tractable and a sampling approach is used to generate samples of $\mathbf{c}$ {by following} Algorithm~4 in the reference \cite{Neal2000}.
	
	Next we generate samples of the parameters $\pmb{\psi}^\star_j=\left(V_j,u_j\right)^\top$ for Step~2 of the Gibbs sampler. By the Bayes' rule, $p(V_j|u_j,\mathbf{X}^j) \propto p(V_j)f(\mathbf{X}^j|V_j,u_j)$ and
	$p(u_j|V_j,\mathbf{X}^j) \propto p(u_j)f(\mathbf{X}^j|V_j,u_j)$, the conditional posteriors of $V_j$ and $u_j$ are given by
	\begin{eqnarray}
	\label{eq.posteriorGamma}
	V_j|u_j,\mathbf{X}^j &\propto& \frac{V_j^{m_jV_j}}{\Gamma(V_j)^{m_j}}
	\exp\left[-V_j\left(\theta+\frac{\sum_{k=1}^{m_j}{X_k^j}}{u_j}+m_j\log(u_j)-
	\sum_{k=1}^{m_j}{\log(X_k^j)}\right)\right]  \\
	u_j|V_j,\mathbf{X}^j &\sim& \mbox{Inv-Gamma}\left(r+m_jV_j,s+V_j\sum_{k=1}^{m_j}X_k^j\right)
	\nonumber
	\end{eqnarray}
	where $X_k^j$ are the $k$th observation associated to the $j$-th component and $m_j$ is the size of $\mathbf{X}^j$.
	The detailed derivation for these posteriors can be found in Section~\ref{subsec:AppendixPostGamma}.


	The conditional posterior $p(V_j|u_j,\mathbf{X}^j)$ in Equation~(\ref{eq.posteriorGamma}) is not a standard distribution. A Metropolis-Hasting (M-H) nested Gibbs sampler is developed to generate samples of $V_j$ from the conditional posterior.
	Specifically, denote the sample from the previous iteration in the nested M-H sampling by $V_j^0$. We first generate a candidate sample $\widetilde{V}_j$ from a proposal distribution, denoted by $g(\cdot,V_j^0)$, and accept it with probability
	$$\min\left\{1,\frac{p(\widetilde{V}_j|u_j,\mathbf{X}^j)g(V_j^0,\widetilde{V}_j)}
	{p(V_j^0|u_j,\mathbf{X}^j)g(\widetilde{V}_j,V_j^0)}\right\},$$
	where $p(V_j^0|u_j,\mathbf{X}^j)$ and $p(\widetilde{V}_j|u_j,\mathbf{X}^j)$ are the conditional posteriors from Equation~(\ref{eq.posteriorGamma}). Otherwise, retain the value of $V_j^0$. The proposal distribution $g(\cdot,V_j^0)$ is chosen to be $\mbox{Gamma}(d,d/V_j^0)$ with mean located at $V_j^0$. This proposal distribution is determined by using the Stirling approximation so that it can capture the tail of the conditional posterior $p(V_j|u_j,\mathbf{X}^j)$ well. The detailed derivation can be found in Section~\ref{subsec:AppendixPostGamma}.
	To make the proposal distribution relatively flat, we recommend that the value of $d$ is set to be small, e.g., $d=2$ used in our empirical study.

	\subsubsection{DPM with Gaussian Kernel}
	\label{subsubsec:AppendixGaussianDPM}
	Given the base distribution $G_0$ in Equation~(\ref{eq.GaussPrior}), we first generate samples of the latent variables $\mathbf{c}$ for Step~1 of the Gibbs sampler. 
	If $c_i$ is associated with an existing $j$th component, then
	\begin{equation}
	p(c_i=j|\mathbf{c}^{-i},\pmb{\psi}^\star_j,\alpha,X_i)
	=b_0\frac{m_j^{-i}}{m+\alpha-1}\frac{1}{{\sqrt{2\pi}\sigma_j }}e^{{{-\left({X_i-u_j}\right)^2}\mathord{\left/{\vphantom {{-\left({x-u_j} \right)^2}{2\sigma_j ^2 }}}\right.\kern-\nulldelimiterspace}{2\sigma_j^2}}}. \nonumber
	\end{equation}
	If $c_i$ is associated with a new component, then
	\begin{equation} \label{eq.PosteriorGauss_c}
	p(c_i=j|\mathbf{c}^{-i},\pmb{\psi}^\star_j,\alpha,X_i)
	=b_0 \frac{\alpha}{m+\alpha-1}\frac{\left(v_0/2\right)^{v_0/2}}{\Gamma(v_0/2)}
	\sigma_0^{v_0}\sqrt{\frac{m_0}{2\pi (m_0+1)}}\frac{\Gamma(A)}{B^A} \nonumber
	\end{equation}
	where $A=(v_0+1)/2$, $B=[v_0\sigma_0^{2}+m_0(X_i-u_0)^2/(m_0+1)]/2$ and $b_0$ is the normalizing constant. The detailed derivation for this conditional posterior can be found in Section~\ref{subsec:AppendixPostGaussian}.
	
	Next we generate samples of the parameters $\pmb{\psi}^\star_j=(u_j,\sigma_j^2)^\top$ for Step~2 of the Gibbs sampler. The conditional posteriors for $u_j$ and $\sigma_j$ are derived by following Chapter~3 in the reference \cite{Gelman_2004}
	\begin{eqnarray}
	\label{eq.PosteriorGauss_mu}  
	u_j|\sigma_j,\mathbf{X}^j &\sim &\mathcal{N}\left(\frac{m_0}{m_0+m_j}u_0+\frac{m_j}{m_0+m_j}\bar{X}^j,
	\frac{\sigma_{0j}^2}{m_0+m_j}\right), \\
	\sigma_j^2/\sigma_0^2\Big|\mathbf{X}^j &\sim& \mbox{Inv-Gamma}\left(\frac{v_0+m_j}{2},\frac{1}{2}\right),
	\label{eq.PosteriorGauss_sigma}    \nonumber
	\end{eqnarray}
	where
	\[\sigma_{0j}^2=\frac{v_0\sigma_0^{2}+\sum_{k=1}^{m_j}(X^j_k-\bar{X}^j)^2+
		\frac{m_0 m_j(\bar{X}^j-u_0)^2}{m_0+m_j}}{v_0+m_j} \mbox{  with }  \bar{X}^j=\frac{1}{m_j}\sum_{k=1}^{m_j}X_k^j. \]

	\subsubsection{DPM with Beta Kernel}
	\label{subsubsec:AppendixBetaDPM}
	
	\begin{sloppypar}
		Here we develop a posterior sampler for DPM with the Beta kernel density to fit the input models with compact supports. We assume that $X_i|c_i=j,\omega_j,\beta_j\sim \mbox{Beta}(\omega_j,\beta_j)$ and denote the parameters for the $j$th component by $\pmb{\psi}^\star_j=\left(\omega_j,\beta_j\right)^\top$. Equation~(\ref{eq.Betaprior}) provides the base function $G_0(\omega,\beta)$. The derivation for this prior can be founded in Section~\ref{subsec:AppendixPostBeta}.
	\end{sloppypar}
	
	We first generate samples of the latent variable $c_i$ for Step~1 of the Gibbs sampler. According to Equation~(\ref{post_latent}), the conditional posterior probabilities of $c_i$ in DPM with Beta kernel is
	\begin{equation}
	p(c_i=j|\mathbf{c}^{-i},\pmb{\psi}^\star_j,\alpha,X_i)=\left\{
	\begin{array}{ll}
	b_0\frac{m_j^{-i}}{m+\alpha-1}X_i^{\omega_j-1}(1-X_i)^{\beta_j-1} \mbox{  if  }\exists c_q=j \mbox{ for all } q\neq i \\
	b_0\frac{\alpha}{m+\alpha-1}\int X_i^{\omega-1}(1-X_i)^{\beta-1}d G_0(\omega,\beta) \mbox{  otherwise }
	\end{array}
	\right. \nonumber
	\end{equation}
	where $b_0$ denotes the normalizing constant. Since the conditional posterior for $c_i$ associated with a new component does not have a closed form, we use the sampling approach by following Algorithm~4 in the reference \cite{Neal2000} to generate samples of $c_i$.
	
	Next we generate samples of the parameters $\pmb{\psi}^\star_j=(\omega_j,\beta_j)^\top$ for Step~2 of the Gibbs sampler. By applying the Bayes' rule, $p(\omega_j|\beta_j,\mathbf{X}^j)\propto p(\omega_j)p(\mathbf{X}^j|
	\omega_j,\beta_j)$ and $p(\beta_j|\omega_j,\mathbf{X}^j)
	\propto p(\beta_j)p(\mathbf{X}^j|\omega_j,\beta_j)$, the conditional posteriors of component parameters $\omega_j$ and $\beta_j$ are given by
	\begin{eqnarray}
	\label{eq.Beta1}
	\omega_j|\beta_j,\mathbf{X}^j &\propto &\exp\left\{\left[-\lambda_1+\sum_{k=1}^{m_j}{\log(X_k^j)}\right]\omega_j-(\lambda_0+m_j)
	\log\left[\frac{\Gamma(\omega_j)}{\Gamma(\omega_j+\beta_j)}\right]\right\}, \\
	\label{eq.Beta2}
	\beta_j|\omega_j,\mathbf{X}^j &\propto& \exp\left\{\left[-\lambda_2+\sum_{k=1}^{m_j}{\log(1-X_k^j)}\right]\beta_j-(\lambda_0+m_j)
	\log\left[\frac{\Gamma(\beta_j)}{\Gamma(\omega_j+\beta_j)}\right]\right\}.
	\end{eqnarray}
	The detailed derivation for these posteriors can be found in Section~\ref{subsec:AppendixPostBeta}.
	
	Since the conditional posteriors in Equations~(\ref{eq.Beta1}) and (\ref{eq.Beta2}) are not standard distributions, we again develop an M-H nested Gibbs sampler to generate samples for $\omega_j$ and $\beta_j$. Denote the samples from the previous iteration in the M-H sampling by $\omega_j^0$ and $\beta_j^0$. By using the Stirling approximation, we choose $\mbox{Gamma}(d,d/a)$ with relatively small $d$ and mean $a$ equal to $\omega_j^0$ or $\beta_j^0$
	as the proposal distribution; See the detailed derivation in Section~\ref{subsec:AppendixPostBeta}. Denote the proposal density by $g(\cdot,a)$. Specifically, for $\omega_j$, we randomly sample a candidate $\widetilde{\omega}_j$ from the proposal distribution $\mbox{Gamma}(d,d/\omega_j^0)$, and accept $\widetilde{\omega}_j$ with probability
	$$\min\left\{1,\frac{p(\widetilde{\omega}_j|\beta_j^0,\mathbf{X}^j)
		g(\omega_j^0,\widetilde{\omega}_j)}
	{p(\omega_j^0|\beta_j^0,\mathbf{X}^j)g(\widetilde{\omega}_j,\omega_j^0)}\right\},$$
	where $p(\widetilde{\omega}_j|\beta_j^0,\mathbf{X}^j)$ and $p(\omega_j^0|\beta_j^0,\mathbf{X}^j)$ are the conditional posterior in Equation~(\ref{eq.Beta1}). Otherwise, retain the value of $\omega_j^0$. Similarly, for $\beta_j$, we randomly sample a candidate $\widetilde{\beta}_j$ from the proposal distribution $\mbox{Gamma}(d,d/\beta_j^0)$, and accept $\widetilde{\beta}_j$ with probability
	$$\min\left\{1,\frac{p(\widetilde{\beta}_j|\omega_j^0,\mathbf{X}^j)
		g(\beta_j^0,\widetilde{\beta}_j)}
	{p(\beta_j^0|\omega_j^0,\mathbf{X}^j)g(\widetilde{\beta}_j,\beta_j^0)}\right\}, $$		
	where $p(\widetilde{\beta}_j|\omega_j^0,\mathbf{X}^j)$ and $p(\beta_j^0|\omega_j^0,\mathbf{X}^j)$ are the conditional posteriors in Equation~(\ref{eq.Beta2}). Otherwise, retain the value of $\beta_j^0$. In our empirical study, we set $d=2$ when we sample both $\omega_j$ and $\beta_j$.

	\subsection{Derivation of the Results Used in the Gibbs Sampling}
	\label{subsec:deriveCondPost}
	
	In this section, we provide the detailed derivation of priors, proposal distributions, and conditional posteriors used in the Gibbs samplers for DPM with Gamma, Gaussian and Beta kernel densities in Section~\ref{subsec:gibbsSampling}.
	
	\subsubsection{Conditional Posteriors of DPM with Gamma Kernel}
	\label{subsec:AppendixPostGamma}
	
	We derive the conditional posteriors of parameters $\pmb{\psi}^\star_j=(V_j,u_j)$ with $j=1,2,\ldots,K_0$ for DPM with Gamma kernel. Given the priors $V_j\sim \mbox{exp}(\theta)$, $u_j\sim \mbox{Inv-Gamma}(r,s)$ and the likelihood $X_i|c_i=j,\pmb{\psi}^\star_j \sim \mbox{Gamma}(V_j,V_j/u_j)$, by the Bayes' rule, we have the conditional posterior for $V_j$
	\begin{eqnarray}
	\lefteqn{ p(V_j|\mathbf{X}^j,u_j)\propto p(V_j)\prod_{k=1}^{m_j} p\left(X^j_k|V_j,u_j\right) } \nonumber \\
	&\propto& e^{-\theta V_j}\prod_{k=1}^{m_j}\frac{(V_j/u_j)^{V_j}}{\Gamma(V_j)}
	(X_k^j)^{V_j-1}e^{-(V_j/u_j)X_k^j} \nonumber \\
	&\propto& \frac{V_j^{m_jV_j}}{\Gamma(V_j)^{m_j}}
	\exp\left\{-V_j\left[\theta+\frac{\sum_{k=1}^{m_j}{X_k^j}}{u_j}
	+m_j\log(u_j)-\sum_{k=1}^{m_j}{\log\left(X_k^j\right)}\right]\right\} .
	\label{eq.Mid_V1}
	\end{eqnarray}
	
	Since the conditional posterior of $V_j$ in Equation~(\ref{eq.Mid_V1}) is not a standard distribution, we develop an M-H sampling algorithm to generate samples of $V_j$. We first find an appropriate proposal distribution for the M-H sampling. To get a fair degree of probability drawing samples from the tail part of the conditional posterior $p(V_j|\mathbf{X}^j,u_j)$, the Stirling approximation, $n!\approx \sqrt{2\pi n}(n/\mbox{e})^n$ for large n, is used to find an appropriate family for the proposal distribution. Since $\Gamma(n)=(n-1)!$,
	\begin{eqnarray}
	\lefteqn{p\left(V_j|\mathbf{X}^j,u_j\right)
		\propto\frac{V_j^{m_jV_j}}
		{\Gamma(V_j)^{m_j}}
		\exp\left\{-V_j\left[\theta+\frac{\sum_{k=1}^{m_j}{X_k^j}}{u_j}
		+m_j\log(u_j)-\sum_{k=1}^{m_j}{\log(X_k^j)}\right]\right\} } \nonumber \\
	&\approx & \left[\frac{V_j^{V_j}}{\sqrt{2\pi(V_j-1)}{\left(\frac{V_j-1}{e}\right)^{V_j-1}}}\right]^{m_j} e^{-V_jB},  \mbox{   if }V_j \mbox{  is large} \nonumber \\
	&\approx &\left[\frac{V_j^{V_j}}{(V_j-1)^{V_j}}\sqrt{\frac{V_j-1}{2\pi}}e^{V_j-1}\right]^{m_j}e^{-V_jB}
	\nonumber \\
	&\approx&
	\left(\sqrt{\frac{V_j-1}{2\pi}}e^{V_j}\right)^{m_j}e^{-V_jB}
	\approx
	\left(\frac{1}{2\pi}\right)^{m_j/2}(V_j)^{m_j/2}e^{-V_j(B-m_j)} \nonumber
	\end{eqnarray}
	where  $B=\theta+\sum_{k=1}^{m_j}{X_k^j}/u_j+m_j\log(u_j)-\sum_{k=1}^{m_j}{\log(X_k^j)}$.
	This approximation holds when $V_j$ is large and it returns a Gamma kernel function. Thus, we choose the proposal distribution to be $\mbox{Gamma}(d,d/V_j^0)$ with mean $V_j^0$ denoting the sample obtained from the previous M-H iteration. To have a non-negligible probability to draw samples far from $V_j^0$, the value of $d$ is recommended to be small, e.g., $d=2$ used in our empirical study.

	Next we derive the conditional posterior for parameter $u_j$. By applying the Bayes' rule, we have
	\begin{eqnarray}
	\lefteqn{p(u_j|\mathbf{X}^j,V_j)\propto p(u_j)\prod_{k=1}^{m_j} p(X^j_k|V_j,u_j) } \nonumber \\
	&\propto& u_j^{-(r+1)}e^{-s/u_j}\prod_{i=1}^{m_j}\frac{(V_j/u_j)^{V_j}}
	{\Gamma(V_j)}(X_k^j)^{V_j-1}e^{-(V_j/u_j)X_k^j} \nonumber \\
	&\propto&
	u_j^{-(r+1+m_jV_j)}
	\exp\left[-\frac{s+V_j\sum_{k=1}^{m_j}X_k^j}{u_j}\right] \nonumber \\
	&\sim & \mbox{Inv-Gamma}\left(r+m_jV_j,s+V_j\sum_{k=1}^{m_j}X_k^j\right). \nonumber
	\end{eqnarray}

	\subsubsection{Conditional Posteriors of DPM with Gaussian Kernel}
	\label{subsec:AppendixPostGaussian}

	For DPM with Gaussian kernel, we choose a conditional conjugate joint prior distribution for the component parameters $\pmb{\psi}^\star_j=(u_j,\sigma^2_j)$ with $j=1,2,\ldots,K_0$,
	\begin{equation}
	u_j|\sigma_j^2 \sim \mathcal{N}(u_0,\sigma_j^2/m_0 ) \mbox{  and  }
	\sigma_j^2/\sigma_0^2 \sim \mbox{Inv-Gamma}\left(\frac{v_0}{2},\frac{1}{2}\right) \nonumber
	\end{equation}
	which determines the base function $G_0(u,\sigma^2)$ with hyper-parameters $\pmb{\theta}_G=(u_0,m_0,v_0,\sigma_0)$. 

	Here, we derive the conditional posteriors of the latent variables $\mathbf{c}$. For $i=1,2,\ldots,m$, if $X_i$ is associated to an existing component, by applying the Bayes' rule,
	\[p(c_i=j|\mathbf{c}^{-i},\pmb{\psi}^\star_j,\alpha,X_i)=b_0\cdot p(c_i=j|\alpha,\mathbf{c}^{-i})p(X_i|c_i=j,\pmb{\psi}^\star_j)
	=b_0 \frac{m_j^{-i}}{m+\alpha-1}\frac{1}{{\sqrt{2\pi}\sigma_j }}
	e^{-(X_i-u_j)^2/2\sigma_j ^2}. \]
	If $X_i$ is associated to a new component,
	\begin{eqnarray}
	\lefteqn{ p(c_i=j|\mathbf{c}^{-i},\pmb{\psi}^\star_j,\alpha,X_i)
		=b_0\cdot p(c_i=j|\alpha,\mathbf{c}^{-i})p(X_i|c_i=j,\pmb{\psi}^\star_j) } \nonumber \\
	&=&b_0
	\frac{\alpha}{m+\alpha-1}\int_{0}^{\infty} \int_{-\infty}^{\infty} p(X_i|u_j,\sigma_j^2) p(u_j|\sigma_j^2)p(\sigma_j^2)d u_jd\sigma_j^2
	\nonumber \\
	&=& b_0\frac{\alpha}{m+\alpha-1}
	\int_{0}^{\infty} \int_{-\infty}^{\infty} (2\pi\sigma_j^2)^{-1/2} e^{-\frac{(X_i-u_j)^2}{2\sigma_j^2}}  \times
	\left(\frac{2\pi\sigma_j^2}{m_0}\right)^{-1/2} \exp\left[-\frac{m_0(u_j-u_0)^2}{2\sigma_j^2}\right]
	\nonumber \\
	&&\times
	\frac{(v_0/2)^{v_0/2}}{\Gamma(v_0/2)}\sigma_0^{v_0}(\sigma_j^2)^{-(v_0/2+1)} e^{\left(-\frac{v_0\sigma_0^2}{2\sigma_j^2}\right)}du_jd\sigma_j^2
	\nonumber \\
	&=&b_0\frac{\alpha}{m+\alpha-1}\frac{(v_0/2)^{v_0/2}}{\Gamma(v_0/2)}
	\sigma_0^{v_0}\sqrt{m_0/2\pi}  \int_{0}^{\infty} \int_{-\infty}^{\infty}(\sigma_j^2)^{-(\frac{v_0+3}{2})}(2\pi\sigma_j^2)^{-1/2}
	\nonumber \\
	&&
	\times \exp\left[-\left(\frac{(m_0+1)(u_j-\frac{X_i+m_0 u_0}{m_0+1})^2+\frac{m_0(X_i- u_0)^2}{m_0+1}
		+v_0\sigma_0^2}{2\sigma_j^2}\right)\right] d u_jd\sigma_j^2
	\nonumber \\
	&= & b_0\frac{\alpha}{m+\alpha-1}\frac{(v_0/2)^{v_0/2}}
	{\Gamma(v_0/2)}\sigma_0^{v_0}\sqrt{\frac{m_0}{2\pi(m_0+1)}}
	\int_{0}^{\infty}
	\int_{-\infty}^{\infty}\left(\frac{2\pi\sigma_j^2}{m_0+1}\right)^{-1/2}
	\nonumber \\
	&& \times\exp\left[-\left(\frac{\left( u_j-\frac{X_i+m_0 u_0}{m_0+1}\right)^2}
	{2\sigma_j^2/(m_0+1)}\right)
	\right]d u_j
	\exp\left[-\left(\frac{\frac{m_0(X_i-\mu_0)^2}{m_0+1}+v_0\sigma_0^2}
	{2\sigma_j^2}\right)\right](\sigma_j^2)^
	{-(\frac{v_0+1}{2}+1)}d\sigma_j^2
	\nonumber \\
	&=&b_0\frac{\alpha}{m+\alpha-1}\frac{(v_0/2)^{v_0/2}}{\Gamma(v_0/2)}
	\sigma_0^{v_0}\sqrt{\frac{m_0}{2\pi(m_0+1)}}\int_{0}^{\infty}
	\exp\left[-\left(\frac{\frac{m_0(X_i-u_0)^2}{m_0+1}+v_0\sigma_0^2}
	{2\sigma_j^2}\right)\right]
	(\sigma_j^2)^{-(\frac{v_0+1}{2}+1)}d\sigma_j^2
	\nonumber \\
	&=&b_0\frac{\alpha}{m+\alpha-1}\frac{(v_0/2)^{v_0/2}}
	{\Gamma(v_0/2)}\sigma_0^{v_0}\sqrt{\frac{m_0}{2\pi(m_0+1)}}\frac{\Gamma(A)}{B^A}
	\nonumber
	\end{eqnarray}
	where $b_0$ is a normalization constant, $A=\frac{v_0+1}{2}$ and $ B=\left[v_0\sigma_0^{2}+\frac{m_0(X_i-u_0)^2}{m_0+1}\right]/2$.
	
	\subsubsection{Conditional Posteriors of DPM with Beta Kernel}
	\label{subsec:AppendixPostBeta}
	
	In this section, we first find a conjugate joint prior and then
	derive the conditional posteriors of parameters $\pmb{\psi}^\star_j=(\omega_j,\beta_j)$ with $j=1,2,\ldots,K_0$ for DPM with Beta kernel density. The likelihood is $X_i\mid c_i=j, \pmb{\psi}^\star_j \sim \mbox{Beta}(\omega_j,\beta_j)$. Since Beta distribution belongs to the exponential family, we rewrite the Beta density into the general form
	\begin{equation}
	p(x|\omega_j,\beta_j)=\frac{\Gamma(\omega_j+\beta_j)}
	{\Gamma(\omega_j)\Gamma(\beta_j)}x^{\omega_j-1}(1-x)^{\beta_j-1}
	=\frac{\Gamma(\omega_j+\beta_j)}{\Gamma(\omega_j)
		\Gamma(\beta_j)}e^{(\omega_j-1)\log(x)+(\beta_j-1)\log(1-x)}.
	\nonumber
	\end{equation}
	Thus, we choose a conjugate joint prior for $(\omega_j,\beta_j)$
	with the hyper-parameters $\pmb{\theta}_G=(\lambda_0,\lambda_1,\lambda_2)$ \cite{Chick_2001}
	\begin{equation}
	\omega_j,\beta_j|\lambda_0,\lambda_1,\lambda_2 \propto \exp\left\{-\lambda_1\omega_j-\lambda_2\beta_j-\lambda_0
	\log\left[\frac{\Gamma(\omega_j)\Gamma(\beta_j)}{\Gamma(\omega_j+\beta_j)}\right]\right\}.
	\nonumber
	\end{equation}

	Then, we derive the conditional posteriors for parameters $(\omega_j,\beta_j)$ used in the Gibbs sampler in Section~\ref{subsubsec:AppendixBetaDPM}.
	By applying the Bayes' rule, the conditional posterior for $\omega_j$ is
	\begin{eqnarray}
	\lefteqn{p(\omega_j|\beta_j,\mathbf{X}^j)\propto p(\omega_j|\beta_j)p(\mathbf{X}^j|\omega_j,\beta_j) } \nonumber \\
	&\propto& \exp\left\{-\lambda_1\omega_j-\lambda_0
	\log\left[\frac{\Gamma(\omega_j)}{\Gamma(\omega_j+\beta_j)}\right]\right\}
	\prod_{k=1}^{m_j}\frac{\Gamma(\omega_j+\beta_j)}{\Gamma(\omega_j)}(X_k^j)^{\omega_j-1}
	\nonumber \\
	&\propto& \exp\left\{\left(-\lambda_1+\sum_{k=1}^{m_j}{\log(X_k^j)}\right)\omega_j-(\lambda_0+m_j)
	\log\left[\frac{\Gamma(\omega_j)}{\Gamma(\omega_j+\beta_j)}\right]\right\}.
	\label{eq.Mid_beta1}
	\end{eqnarray}
	
	Since the conditional posterior for $\omega_j$ in Equation~(\ref{eq.Mid_beta1}) is not a standard distribution, we develop an M-H sampling algorithm to draw
	samples of $\omega_j$ by following the similar procedure used in DPM with Gamma kernel density. The Stirling approximation is used to find an appropriate proposal distribution family. As $\omega_j$ is large, the conditional posterior distribution can be approximated by
	\begin{eqnarray}
	\lefteqn{ p(\omega_j|\beta_j,\mathbf{X}^j)\propto e^{\left(-\lambda_1+\sum_{k=1}^{m_j}{\log(X_k^j)}\right)\omega_j-(\lambda_0+m_j)
			\log\left[\frac{\Gamma(\omega_j)}{\Gamma(\omega_j+\beta_j)}\right]} }
	\nonumber \\
	&\approx& e^{\left(-\lambda_1+\sum_{k=1}^{m_j}{\log(X_k^j)}\right)\omega_j}\left[\frac
	{(\omega_j+\beta_j-1)!}{(\omega_j-1)!}\right]^{\lambda_0+m_j}  \nonumber \\
	&\approx& e^{-\left(\lambda_1-\sum_{k=1}^{m_j}{\log(X_k^j)}\right)\omega_j}\left(\omega_j^{\beta_j}\right)
	^{\lambda_0+m_j}
	,  \mbox{   if }\omega_j \mbox{  is large}
	\nonumber \\
	&\sim& \mbox{Gamma}\left(\beta_j(\lambda_0+m_j)+1,\lambda_1-\sum_{k=1}^{m_j}{\log(X_k^j)}\right). \label{eq.proposalGamma1} \nonumber
	\end{eqnarray}
	Thus, $\mbox{Gamma}(d,d/\omega_j^0)$ with small $d$, e.g., $d=2$ used in the empirical study, is used as the proposal distribution, where $\omega_j^0$ denotes the sample obtained from the previous M-H sampling iteration.

	Next, by applying the Bayes' rule, we derive the conditional posterior for $\beta_j$
	\begin{eqnarray}
	\lefteqn{ p(\beta_j|\omega_j,\mathbf{X}^j)\propto p(\beta_j|\omega_j)p(\mathbf{X}^j|\omega_j,\beta_j)} \nonumber \\
	&\propto&
	\exp\left\{-\lambda_2\beta_j-\lambda_0	\log\left[\frac{\Gamma(\beta_j)}{\Gamma(\omega_j+\beta_j)}\right]\right\}
	\prod_{k=1}^{m_j}\frac{\Gamma(\omega_j+\beta_j)}{\Gamma(\beta_j)}\left(1-X_k^j\right)^{\beta_j-1}
	\nonumber \\
	&\propto& \exp\left\{\left(-\lambda_2+\sum_{k=1}^{m_j}{\log(1-X_k^j)}\right)\beta_j-(\lambda_0+m_j)
	\log\left[\frac{\Gamma(\beta_j)}{\Gamma(\omega_j+\beta_j)}\right]\right\}. \label{eq.Mid_beta2}
	\end{eqnarray}
	Notice that Equations~(\ref{eq.Mid_beta1}) and (\ref{eq.Mid_beta2}) have the similar form, and they do not belong to any standard distribution. Thus, an M-H sampling approach is developed to generate samples for $\beta_j$. An appropriate
	proposal distribution family is found by applying the Stirling approximation,
	\begin{eqnarray}
	\lefteqn{ p(\beta_j|\omega_j,\mathbf{X}^j)\propto e^{\left(-\lambda_2+\sum_{k=1}^{m_j}{\log(1-X_k^j)}\right)\beta_j-(\lambda_0+m_j)
			\log\left[\frac{\Gamma(\beta_j)}{\Gamma(\omega_j+\beta_j)}\right]} }
	\nonumber \\
	&\approx& e^{\left(-\lambda_2+\sum_{k=1}^{m_j}{\log(1-X_k^j)}\right)\beta_j}\left[\frac
	{(\omega_j+\beta_j-1)!}{(\beta_j-1)!}\right]^{\lambda_0+m_j}  \nonumber \\
	&\approx& e^{-\left(\lambda_2-\sum_{k=1}^{m_j}{\log(1-X_k^j)}\right)\beta_j}\left(\beta_j^{\omega_j}\right)
	^{\lambda_0+m_j}
	,  \mbox{   if }\beta_j \mbox{  is large}
	\nonumber \\
	&\sim& \mbox{Gamma}\left(\omega_j(\lambda_0+m_j)+1,\lambda_2-\sum_{k=1}^{m_j}{\log(1-X_k^j)}\right). \label{eq.proposalGamma2}  \nonumber
	\end{eqnarray}
	In the M-H sampling, $\mbox{Gamma}(d,d/\beta_j^0)$ with small $d$ is used as the proposal distribution, where $\beta_j^0$ denotes the sample obtained from the previous iteration.


    \section{{Appendix: Posterior Consistency of DPM Nonparametric Input Models}} \label{subsec:proofThm1}
We first introduce a series of basic definitions and theorems related to posterior consistency from Bayesian nonparametrics theory. We refer the readers to the textbooks of \cite{Ghosh_2003} and \cite{GhoVan17} for technical details.
 
\begin{definition}\label{weak:nbr}
(\cite{Ghosal1999}) Let $\mathscr{F}$ be the set of all densities on $\Re$ with respect to the Lebesgue measure on $\Re$. Let $f^c\in \mathscr{F}$ denote the true probability density and $P_{f^c}$ be its associated probability measure. A \textit{weak neighborhood} $U$ of $f^c$ is a set containing a set of the form
$$V=\left\{f\in \mathscr{F}:~ \left|\int \phi_i(x) f(x)dx - \int \phi_i(x) f^c(x)dx \right| <\epsilon, ~ i=1,\ldots,k\right\},$$
where $\phi_i$'s are bounded continuous functions on $\Re$ and $k$ is a positive integer.
\end{definition}
The weak neighborhood is defined on the space of probability measures topologized by weak convergence (convergence in distribution). 
Refer to \cite{billingsley1999} Chapter 1 Section 2 for more details on weak convergence.

\begin{definition}\label{weak:consistency}
(\cite{Ghosal1999}) Let $\mathbf{X}_m=\{X_1, \ldots, X_m\}$ be an i.i.d. sample from $F^c$ (with density $f^c$). The posterior distribution $p(\cdot\mid\mathbf{X}_m)$ is said to be {\it weekly consistent} at $F^c$ or $f^c$, if with $P_{f^c}$-probability 1,
	\begin{equation}
	p(U\mid \mathbf{X}_m)\to 1, \quad \text{as } m\to\infty,
	\end{equation}
	 for all weak neighborhoods $U$ of $f^c$.
\end{definition}

Then we define the concept of \textit{Kullback-Leibler (K-L) support}: 
\begin{definition}\label{KLsupport}
(\cite{Ghosal1999}) Let $p$ be a prior distribution over the space $\mathscr{F}$, the set of all densities on $\Re$ with respect to the Lebesgue measure on $\Re$. A density $f^c$ is said to be in the {\it Kullback-Leibler (K-L) support} of the prior $p$ (denoted by $f^c\in KL(p)$), if for all $\epsilon>0$, $p(\mathcal{K}_{\epsilon}(f^c))>0$, where $\mathcal{K}_{\epsilon}(f^c)=\{g\in \mathscr{F}: \int f^c(x)\log\frac{f^c(x)}{g(x)} dx<\epsilon\}$ is the \textit{K-L neighborhood} of $f^c$. 
\end{definition}

Following this definition, we cite Theorem 4.4.2 in \cite{Ghosh_2003}, which is essentially derived from the Schwartz theorem (\cite{schwartz1965bayes}):
\begin{theorem}\label{thm:kl}
(\cite{Ghosh_2003} Theorem 4.4.2) Let $p$ be a prior distribution over the space $\mathscr{F}$, the set of all densities on $\Re$ with respect to the Lebesgue measure on $\Re$. If $f^c$ is in the Kullback-Leibler (K-L) support of $p$, then the posterior is weakly consistent at $f^c$.
\end{theorem}
Theorem \ref{thm:kl} shows that $f^c$ being in the K-L support of the prior $p$ implies the posterior weak consistency.  Therefore, to prove Theorem \ref{thm:consistency}, it is sufficient to show that $f^c$ is in the K-L support of those priors under the conditions of Theorem \ref{thm:consistency}.

In this paper, we define the support of a probability measure using the following standard definition on page 23 of \cite{billingsley1999}: 
\begin{definition}\label{support_billingsley}
(\cite{billingsley1999}) If $\mathcal{F}$ is a $\sigma$-field in $\Omega$ and $P$ is a probability measure on $\mathcal{F}$, the triple $(\Omega, \mathcal{F}, P)$ is called a probability measure space, or simply a probability space. \textit{A support of $P$} is any $\mathcal{F}$-measurable set $A$ for which $P(A) = 1$, denoted by $A=supp(P)$.
\end{definition}

To make our proof self-contained, we cite the original theorems and lemmas given in \cite{Wu_Ghosal_2008} that are used in the proof of Theorem \ref{thm:consistency}. In the citation, we changed the notation system in the original paper to the one used in this paper for easy understanding. 



\begin{theorem} \label{wu08t1}
(Theorem 1 of \cite{Wu_Ghosal_2008})
Let $f^c$ be the true density. Let $h(x;\psi,\phi)$ be a kernel density, where $\psi$ is the mixing parameter and $\phi$ is the hyper-parameter which lies in the parameter space $\Phi$. Consider the mixture distribution $f_{G,\phi} =\int h(x;\psi, \phi)dG(\psi)$, where $G$ lies in the space of all mixing distributions $\mathscr G$. Let $\mu$ and $\Pi$ be priors for the hyper parameter and the mixing distribution, and $p$ be the prior on $\mathscr F$ induced by $\mu\times \Pi$.  If for any $\epsilon>0$, there exists a mixing distribution $G_{\epsilon}$, a hyper-parameter $\phi_{\epsilon}$, a set $A\subset \Phi$ with $\mu(A)>0$, and a set $\mathscr W \subset \mathscr G$ with $\Pi(\mathscr W)>0$, such that
\begin{itemize}
	\item [A1.] $\int f^c \log \frac{f^c}{f_{G_{\epsilon}, \phi_{\epsilon}}}<\epsilon $,
	\item [A2.] $\int f^c \log \frac{f_{G_{\epsilon}, \phi_{\epsilon}}} {f_{G_{\epsilon}, \phi}}<\epsilon$ for every $\phi \in A$, and
	\item [A3.] $\int f^c \log \frac{f_{G_{\epsilon}, \phi}} {f_{G, \phi}}<\epsilon$ for every $G\in \mathscr W$, $\phi \in A$,
\end{itemize}
then $f^c \in KL(p)$.
\end{theorem}
This is Theorem 1 in \cite{Wu_Ghosal_2008}, which was constructed for Bayesian nonparametric kernel mixture models that are more general than the model \eqref{DPM}.  The kernel function for DPM considered in the cited theorem contains two of parameters $\psi$ and $\phi$, while the prior being focused in this paper as presented in \eqref{DPM} has only one parameter $\psi$. As same as presented in \eqref{DPM}, $\phi$ in Theorem \ref{wu08t1} is mixed over by the mixing distribution $G$, which is further given a prior $DP(\alpha,G_0)$. The additional parameter $\phi$ in Theorem \ref{wu08t1} is known as the hyper-parameter and is  given a separate prior $\mu$ directly. We use $\Psi$ to denote the sample space of $\psi$, and $\mathscr M(\Psi)$ to denote the space of mixing distributions on $\Psi$. Notice that $\mathscr W\subset \mathscr M(\Psi)$.

Prior \eqref{DPM} does not involve the hyper-parameter $\phi$, and hence it is equivalent to the prior induced by $\Pi$ only, instead of by $\mu \times \Pi$. Specifically, $\Pi$ is the Dirichlet Process (DP) with parameters $\alpha$ and base measure $G_0$ on $\mathscr G$, the space of mixing distribution $G$; and $p$ is the induced Dirichlet Process Mixtrure (DPM)  on $\mathscr F$, the space of density functions. Therefore,  to prove Theorem \ref{thm:consistency}, we only need to verify Conditions A1 and A3 in Theorem \ref{wu08t1}. 

Applying Theorem \ref{wu08t1} to prove the consistency is to verify that Conditions A1, A2 and A3 are satisfied for a given prior. Usually, Condition A1 is directly verified by construction, while A2 and A3 are verified through the following two lemmas.     

\begin{lemma}\label{wg08lemma2}
(Lemma 2 of \cite{Wu_Ghosal_2008})
Let $f^c$, $\Pi$, $\mu$ and $p$ be the same as in Theorem \ref{wu08t1}. If for any $\epsilon >0$, there exist a mixing distribution $G_{\epsilon}$, a set $D\supseteq supp(G_{\epsilon})$, and $\phi_{\epsilon}\in supp(\mu)$ such that Condition A1 holds and the kernel density function $h$ satisfies
	\begin{itemize}
		\item [A4.] for any given $x$ and $\psi$, the map $\phi \mapsto h(x;\psi, \phi)$ is continuous on the interior of $supp(\mu)$;
		\item [A5.] $\int _{\mathfrak X} f^c(x) \left\{
		\left| \log \frac{\sup_{\psi\in D} h(x;\psi,\phi_{\epsilon})}{\inf_{\psi\in D} h(x;\psi,\phi)} \right|+
		\left| \log \frac{\sup_{\psi\in D} h(x;\psi,\phi)}{\inf_{\psi\in D} h(x;\psi,\phi_{\epsilon})} \right|
		\right\}dx<\infty$ for every $\phi\in N(\phi_{\epsilon})$, where $N(\phi_{\epsilon})$ is an open neighborhood of $\phi_{\epsilon}$;
		\item [A6.] for any given $x\in \mathfrak X$, $\psi\in D$ and $\phi\in N(\phi_{\epsilon})$, there exists a function $k(x,\psi)$ such that $k(x,\psi)\geq h(x;\psi, \phi)$, and $\int k(x,\psi)dG_{\epsilon}(\psi)<\infty$;
	\end{itemize}
	then there exists a set $A \subset \Phi$ such that Condition A2 holds.
\end{lemma}

\begin{lemma} \label{wg08lemma3}
(Lemma 3 of \cite{Wu_Ghosal_2008})
	Let $f^c$, $\Pi$, $\mu$ and $p$ be the same as in Theorem \ref{wu08t1}. If for any $\epsilon >0$, there exist a mixing distribution $G_{\epsilon} \in supp(\Pi)$, a hyper-parameter $\phi_{\epsilon}\in supp(\mu)$, and a set $A\in \Phi$ with $\mu(A)>0$, such that Conditions A1 and A2 hold and for some $D\supseteq supp(G_{\epsilon})$, the kernel density function $h$ and the prior $\Pi$ satisfy
	\begin{itemize}
		\item [A7.] for any $\phi \in A$, $\int_{\mathfrak X} f^c(x) \log \frac{f_{G_{\epsilon},\phi}(x)}{\inf_{\psi\in D}h(x;\psi,\phi)}dx<\infty$;
		\item [A8.] for any $\epsilon>0$, there exists a compact set $C\subset \mathfrak X$ with the complement set denoted by $C^c$, such that
		\begin{equation}\label{w08.4}
		\int_{C^c} f^c(x) \log \frac{f_{G_{\epsilon},\phi}(x)}{\inf_{\psi\in D} h(x;\psi,\phi)}dx<\epsilon/4,
		\end{equation}
		and $P_{f^c}(C^c)<\epsilon/(4\log2)$, we have that $c:=\inf_{x\in C}\inf_{\psi\in D} h(x;\psi, \phi)>0$;
		\item [A9.] for any given $\phi\in A$ and some compact $C\subset \mathfrak X$ as required in A8, such that the family of maps $\{\psi \mapsto h(x;\psi,\phi), x\in C\}$ is uniformly equicontinuous on $D$;
	\end{itemize}
	then there exists a set $\mathscr W \subset \mathscr M(\Psi)$ such that Condition A3 holds and $\Pi(\mathscr W)>0$.
\end{lemma}
Notice that, Conditions A8 and A9 are a little different from their original form in \cite{Wu_Ghosal_2008}. The modification of Condition A9 follows \cite{wu2009correction}, while the modification on A8 is justified by the fact that in detailed proof of this lemma in \cite{Wu_Ghosal_2008}. We only need the existence of $C$ that satisfies (\ref{w08.4}) and $P_{f^c}(C^c)<\epsilon/(4\log2)$, where the existence is implied by Condition A7. Since the conditions of the lemma are different from the original version in \cite{Wu_Ghosal_2008}, we include the proof of this lemma below, which reflexes the change in the conditions. 

{{\noindent \bf Proof of Lemma \ref{wg08lemma3}:}}

For any $\phi\in A$, write
\begin{eqnarray}
\label{eq:x3}
\int_{\mathfrak{X}}f^c(x)\log\frac{f_{G_{\epsilon},\phi}
(x)}{f_{G,\phi}(x)}dx &=& \int_{C^c}f^c(x)\log\frac{f_{G_{\epsilon},\phi}
(x)}{f_{G,\phi}(x)}dx\nn\\
&&+\int_Cf^c(x)\log\frac{f_{G_{\epsilon},\phi}(x)}{f_{G,\phi}(x)}dx.
\end{eqnarray}
Now, since $G_\epsilon(D)=1>\frac12$, $\mathscr{V}=\{G:\ G({D})>\frac{1}{2} \}$
 is an open neighborhood of $G_{\epsilon}$ by the Portmanteau Theorem. For any $G\in \mathscr{V}$
and $\phi \in A$,
\begin{eqnarray*}
\lefteqn{\int_{C^c}
f^c(x)\log\frac{f_{G_{\epsilon},\phi}(x)}{f_{G,\phi}(x)}dx} \\
&&\leq \int_{C^c}f^c(x)\log\frac{f_{G_{\epsilon},\phi}(x)}{\int_{\psi\in D}\inf_{\psi\in D}h(x;\psi,\phi)dG(\psi)}dx\\
&&\leq \int_{C^c}f^c(x)\log\frac{f_{G_{\epsilon},\phi}(x)}{\inf_{\psi\in D}h(x;\psi,\phi)\int_{\psi\in D}dG(\psi)}dx\\
&& < 
{\int_{C^c}f^c(x)\log\frac{2 f_{G_{\epsilon},\phi}(x)}{\inf_{\psi\in D}h(x;\psi,\phi)}dx }
\end{eqnarray*}
The last term in the above inequality is due to the fact that $G(D)>1/2$ as defined in the definition of $\mathscr V$. Now we have that 
\begin{eqnarray*}
\lefteqn{\int_{C^c}f^c(x)\log\frac{2 f_{G_{\epsilon},\phi}(x)}{\inf_{\psi\in D}h(x;\psi,\phi)}dx }\\
&&= \int_{C^c}f^c(x)\log\frac{f_{G_{\epsilon},\phi}(x)}{\inf_{\psi\in
D}h(x;\psi,\phi)}dx+(\log2)P_{f^c}(C^c);
\end{eqnarray*}
where $P_{f^c}(C^c)=\int _{C^c}f^c(x)dx$,
 and $P_{f^c}$ denotes the  probability measure corresponding to $f^c$.    
By Condition A7, there exists compact $C \subset \mathfrak{X}$,
such that
\begin{equation}\label{eq:setC}
\int
_{C^c}f^c(x)\log\frac{f_{G_{\epsilon},\phi}(x)}{\inf_{\psi\in D}h(x;\psi,\phi)}dx<\epsilon/4
.
\end{equation}
We can further ensure that $P_{f^c}(C^c)<\epsilon/(4\log 2)$, so the bound for $\int_{C^c}f^c\log\frac{f_{G_{\epsilon,\phi}}}{f_{G,\phi}}$ is less than $\epsilon/2$.
 Now, if we can show that for the given $\epsilon>0$, there
exists a weak  neighborhood $\mathscr{U}$ of $G_{\epsilon}$,
such that
$\int_Cf^c(x)\log\frac{f_{G_{\epsilon},\phi}(x)}{f_{G,\phi}(x)}dx<\epsilon/2$
for any $G \in \mathscr{U}$ and $\phi \in A$, then Lemma
\ref{wg08lemma3} is proved by letting $\mathscr{W}=\mathscr{U}\cap
\mathscr{V}$.

Observing that for any given $\phi\in A$, the family of maps $\{\psi\mapsto
h(x;\psi,\phi):x\in C\}$ is uniformly equicontinuous on $D\subset
\Psi$. By the Arzela-Ascoli theorem (see \cite{royden1988} [pp. 169]), for any $\delta>0$, there exist
$ x_1,x_2,\ldots,x_m$, such that, for any $x\in C$,
\begin{equation}\label{eq:cdelta}
\sup_{\psi\in D}|h(x;\psi,\phi)-h(x_i;\psi,\phi)|<c\delta.
\end{equation}
for some $i=1,2,\ldots,k$.

Let $\mathscr{U}=\{
G:|\int_D h(x_i;\psi,\phi)dG_{\epsilon}(\psi)-\int_D
h(x;\psi,\phi)dG(\psi)|<c \delta, \ \ x \in C, i=1,2,\ldots,k\}.$ Then $\mathscr{U}$ is an open weak neighborhoods of $G_{\epsilon}$ since $G_{\epsilon}\in
\mathrm{supp}(\Pi)$ and $G_{\epsilon}(\partial D)=0$.
Hence, given $\phi\in A$, for any $G\in \mathscr U \cap \mathscr V$ and $x\in C$, by applying Condition A8, 
\begin{eqnarray*}
\left|\frac{\int_{\Psi}
h(x;\psi,\phi)dG_{\epsilon}(\psi)}{\int_{\Psi}
h(x;\psi,\phi)dG(\psi)}-1\right|
&\leq &\left|\frac{\int_{D}
h(x;\psi,\phi)dG_{\epsilon}(\psi)}{\int_{D}
h(x;\psi,\phi)dG(\psi)}-1\right| \\
&=&\left|\frac{\int_{D}
h(x;\psi,\phi)dG_{\epsilon}(\psi)-\int_{D}
h(x;\psi,\phi)dG(\psi)}{\int_{D}
h(x;\psi,\phi)dG(\psi)}\right|\\
&{\leq}&\frac{c\delta}{c/2}=2\delta,
\end{eqnarray*}
since $|\int_{D}
h(x;\psi,\phi)dG_{\epsilon}(\psi)-\int_{D}
h(x;\psi,\phi)dG(\psi)|<c\delta$ for any $G\in \mathscr U$, and $\int_{D}
h(x;\psi,\phi)dG(\psi)\leq \inf_{x\in C}\inf_{\psi\in D} h(x;\psi, \phi) G(D)\geq c/2$.
By choosing  $\delta$ small enough, we can ensure that the right hand side (RHS) of the last display is less than $\epsilon/2$.  Hence, for any given
$\phi\in A$
\[
\scalebox{0.93}{$\int_C f^c(x)\log\frac{f_{G_{\epsilon},\phi}(x)}{f_{G,\phi}(x)}dx
\leq \sup_{x\in C}\log\frac{f_{G_{\epsilon},\phi}(x)}{f_{G,\phi}(x)}
\leq \sup_{x\in C}\left | \frac{f_{G_{\epsilon},\phi}(x)}{f_{G,\phi}(x)}-1
\right |
\leq 
\sup_{x\in C}\left|\frac{\int_{\Psi} h(x;\psi,\phi)dG_{\epsilon}(\psi)}{\int_{\Psi} h(x;\psi,\phi)dG(\psi)}-1\right| < \epsilon/2$}\]
for any $G\in
\mathscr{U}\cap \mathscr V$.
$\mbox{}\Box$




Next, we present several lemmas which will be used for proving Theorem 1 Part (i), together with their proofs when necessary. Differing with Theorem~14 in \cite{Wu_Ghosal_2008}, the restriction of $f^c(0)> 0$  is not necessary here. In the proofs of the Lemmas and Theorem \ref{thm:consistency} (i), we will provide the detailed explanations.

 
\begin{lemma}\label{lemma:gamma}
(Lemma 7 in \cite{Wu_Ghosal_2008}) Let $l$ be a positive integer and  $A_l=[a_l,b_l]\subset\mathfrak{X}$, and let $h_l(x,t)$ be a sequence of continuous functions for $x\in\mathfrak{X}$ and $t\in
A_l$. Define $f_l(x)=\int_{A_l}h_l(x,t)f(t)dt$, $m=1,2,\ldots$, where
$f$ is bounded, uniformly continuous and integrable on $\mathfrak{X}$. If
$h_l$ satisfies that for any $a_l,b_l\in \mathfrak{X}$,
\begin{itemize}
\item[C1.]$\int_{A_l}h_l(x,t)dt\rightarrow 1$ as
$l\rightarrow \infty$,
\item[C2.] for each $\delta>0$,  $\int_{|x-t|>\delta,t\in
A_l}|h_l(x,t)|dt\rightarrow0$  as
$l\rightarrow\infty$,
\item[C3.]$\int_{A_l}|h_l(x,t)|dt\leq M(x)<\infty$ for each
$x\in\mathfrak{X}$, $l=1,2\ldots$, where the bound $M(x)$ may depend on $x$ but not on $l$,
\end{itemize}
then  $f_l(x)\rightarrow f(x)$ for each $x\in\mathfrak{X}$ as $l\to\infty$.
\end{lemma}

{\noindent \bf Proof of Lemma \ref{lemma:gamma}:} 

For any $\epsilon>0$, there is  $\delta>0$ so small that
that $|f(t)-f(x)|<\epsilon$ for $|x-t|\leq\delta$. By Condition C1,
\begin{equation}
f_l(x)-f(x)=\int_{A_l}[f(t)-f(x)]h_l(x,t)dt+o(1),
\label{eq.mid101}
\end{equation}
where the last term goes to 0 as $l\rightarrow\infty$, for each $x\in \mathfrak{X}$. 
To complete the proof, we will show that $\int_{A_l}[f(t)-f(x)]h_l(x,t)dt\to 0 $ as $l\to \infty$ for all $x\in \mathfrak{X}$. 
By Condition C3, we have that
\begin{equation}
\Big|\int_{|x-t|\leq\delta,\,t\in
A_l}[f(t)-f(x)]h_l(x,t)dt\Big|\leq\epsilon\int_{|x-t|\leq\delta,\,t\in
A_l}|h_l(x,t)|dt\leq\epsilon M(x).
\label{eq.mid102}
\end{equation}
We also have that
\begin{equation}
\int_{|x-t|>\delta,\,t\in
A_l}[f(t)-f(x)]h_l(x,t)dt\rightarrow0 \mbox{    as } l\rightarrow\infty,
\label{eq.mid103}
\end{equation}
by Condition C2, and the condition that function $f$ is bounded.
Combining (\ref{eq.mid101})-(\ref{eq.mid103}), we have
 $ |f_l(x)-f(x)|\leq\epsilon M(x)+o(1)$, and  hence the result follows. $\Box$
\vspace{1cm}

In the following, we first fix some notation for proving the posterior consistency of DPM of Gamma. Let $h_l(x;V)\equiv h(x;V, V/l)$ be the Gamma density kernel with shape parameter $V$ and mean equals to $V/l$, where $V>0$ and $l>0$. Define
\begin{equation} \label{eq:hm}
h_l(x;V)=h(x;V, V/l)=\frac{l^V}{\Gamma(V)}x^{V-1}e^{-x/l}   
\end{equation}
Let
\begin{equation}\label{eq:fm}
f_l(x)= t_l\int_{2}^{1+l^2} h_l(x;V)l^{-1} f^c((V-1)/l)dV
\end{equation}
where  $t_l = (\int_{l^{-1}}^l f^c(s)ds)^{-1}$.

\begin{lemma}\label{lemma:gammac} (Lemma 8 in \cite{Wu_Ghosal_2008}) 
Let $h_l(x;V)$ be defined as \eqref{eq:hm}. If Condition (d) in Theorem \ref{thm:consistency} is satisfied, then there exists a function $0<C(x)<1$ such that for all sufficiently large $l>0$,
\begin{equation}\label{eq:c0}
C(x)\leq \left\{
\begin{array}{ll}
\int_{l^{-1}\vee x}^{x+\delta} h_l(x;lv+1)dv,& \ l^{-1}<x<1,\\
\mbox{} & \mbox{} \\
\int_{x-\delta}^{l \wedge x} h_l(x;lv+1)dv,& \ 1\leq x\leq l+l^{-1},
\end{array}
\right.
\end{equation}
and $\int_{\mathfrak{X}} f^c(x)\log \frac{1}{C(x)}dx<\infty$.
\end{lemma}


{\noindent \bf Proof of Lemma \ref{lemma:gammac}:} 

For $l^{-1}<x<1$, applying Stirling's inequality and noting that $v<x+\delta<1+\delta$ (with $\delta>0$ chosen later) in the following integral, it follows that
{
\setlength\arraycolsep{0pt}
\begin{eqnarray}\label{eq:83}
\lefteqn{\int_{{l}^{-1}\vee x}^{x+\delta}h_l(x;lv+1)dv}\nn\\
&=&\int_{l^{-1}\vee x}^{x+\delta}\frac{l^{lv+1}x^{lv}e^{-lx}}{\Gamma(lv+1)}dv\nonumber\\
&\geq& \int_{l^{-1}\vee x}^{x+\delta}\frac{l^{lv+1}x^{lv}e^{-lx}}{\sqrt{2\pi}(lv+1)^{lv+1/2}\exp\{-(lv+1)+(12x)^{-1}\}}dv\nonumber\\
&=&\sqrt{\frac{l}{2\pi}} \exp(1-(12x)^{-1})\int_{l^{-1}\vee x}^{x+\delta}\frac{x^{lv}e^{l(v-x)}}{(v+l^{-1})^{lv+{1}/{2}}}dv\nonumber\\
&\geq&\frac{\sqrt{l}}{\sqrt{2\pi(1+\delta+l^{-1})}} \exp(1-(12x)^{-1})\int_{l^{-1}\vee x}^{x+\delta}\frac{x^{lv}e^{l(v-x)}}{(v+l^{-1})^{lv}}dv.
\end{eqnarray}
}
Note that
\begin{eqnarray*}
\lefteqn{\int_{l^{-1}\vee x}^{x+\delta}\frac{x^{lv}e^{l(v-x)}}{(v+l^{-1})^{lv}}dv}\\
&=&\int_{l^{-1}\vee x}^{x+\delta}\exp\left[lv\left \{\log\frac{x}{v+l^{-1}}-(\frac{x}{v}-1)\right \}\right]dv\\
&=&\int_{l^{-1}\vee x}^{x+\delta}\exp\Big[lv\Big \{\log\frac{x}{v+l^{-1}}-\Big(\frac{x}{v+l^{-1}}-1\Big)\\
&&\qquad\qquad\qquad\qquad\qquad\qquad\quad\,\,\,\ +\Big(\frac{x}{v+l^{-1}}-1\Big)-\Big(\frac{x}{v}-1\Big)\Big \}\Big]dv\\
&>&\int_{l^{-1}\vee x}^{x+\delta}\exp\left[lv\left\{-\frac{1}{2\frac{x}{v+l^{-1}}}\left(\frac{x}{v+l^{-1}}-1\right)^2+\frac{-{x}/{l}}{v(v+l^{-1})}\right\}\right]dv\\
&=&\int_{l^{-1}\vee x}^{x+\delta}\exp\left[\frac{-lv(x-v-l^{-1})^2-2x^2}{2x(v+l^{-1})}\right]dv.
\end{eqnarray*}
The above inequality holds, because of that, for $0<u<1$,
\begin{eqnarray*}
\log u-(u-1)&=&-(1-u)^2
\left\{\frac{1}{2}+\frac{(1-u)}{3}+\frac{(1-u)^2}{4}+\cdots\right\}\\
&\geq& -\frac{(1-u)^2}{2}\left\{1+(1-u)+(1-u)^2+\cdots\right\}=-\frac{(1-u)^2}{2u}.
\end{eqnarray*}
Since $1+\delta>x+\delta>v>x$ in the following integral, we have that
\begin{eqnarray}\label{eq:82}
\lefteqn{\int_{l^{-1}\vee x}^{x+\delta}\exp\left(\frac{-lv(x-v-l^{-1})^2-2x^2}{2x(v+l^{-1})}\right)dv}\nn\\
&\geq& \int_{(l^{-1}\vee x)+l^{-1}}^{x+\delta+l^{-1}}\exp\left(\frac{-l(1+\delta)(x-\tilde v)^2-2x^2}{2x^2}\right)d \tilde v\nonumber\\
&
 =&\sqrt{\frac{2\pi}{l}}\frac{x}{\sqrt{1+\delta}}e^{-1}\left\{\Phi\left(\frac{\delta+l^{-1}}{x/\sqrt{l(1+\delta)}}\right)-\Phi\left(\frac{l^{-1}}{x/\sqrt{l(1+\delta)}}\right)\right\}\nonumber\\
 &\geq&\sqrt{\frac{2\pi}{l}}\frac{x}{\sqrt{1+\delta}}e^{-1}\left\{\Phi\left(\frac{\delta+l^{-1}}{x/\sqrt{l(1+\delta)}}\right)-\Phi\left(\frac{l^{-1}}{x/\sqrt{l(1+\delta)}}\right)\right\},
\end{eqnarray}
where $\tilde v =v+l^{-1}$ and $\Phi(\cdot)$ is the cdf of the standard normal distribution.  For $l$ large, such that $\delta>l^{-1/2}$,
\begin{eqnarray}\label{eq:b1}
\lefteqn{\Phi\left(\frac{\delta+l^{-1}}{x/\sqrt{l(1+\delta)}}\right)-\Phi\left (\frac{l^{-1}}{x/\sqrt{l(1+\delta)}}\right)}\nn\\
&=&\Phi\left(\sqrt{1+\delta}\ \frac{l^{1/2}\delta+l^{-1/2}}{x}\right)-\Phi\left(\sqrt{1+\delta}\ \frac{l^{-1/2}}{x}\right)\nn\\
&\geq&\Phi(2\sqrt{1+\delta}\ \sqrt{\delta}/x)-\Phi(\sqrt{1+\delta}\ \delta/x)\nn\\
&\geq&\Phi(2\sqrt{1+\delta}\ \delta/x)-\Phi(\sqrt{1+\delta}\ \delta/x).
\end{eqnarray}
The last inequality holds since we chose $\delta<1$. Now for $u>0$,
\begin{equation*}
\frac {\frac{1+u^2}{u}\phi(u)}{\frac{1}{2u}\phi(2u)}=2(1+u^2)e^{3u^2/2}\geq 2,
\end{equation*}
where $\phi(x)=(2\pi)^{-1/2}e^{-x^2/2}$ is the standard normal pdf. By the fact that
\begin{equation}\label{eq:bound}
\frac{x}{1+x^2}\phi(x)<1-\Phi(x)<\frac{\phi(x)}{x},
\end{equation}
we have that
$$
\Phi(2u)-\Phi(u)\geq \frac{1+u^2}{u}\phi(u)-\frac{1}{2u}\phi(2u)\geq \frac{1}{2u}\phi(2u).
$$
Hence,  the the right hand side (RHS) of (\ref{eq:b1}) is greater than
\begin{equation}\label{eq:81}
 \frac{x}{2\delta\sqrt{2\pi(1+\delta)}}\exp\left(-\frac{2(1+\delta)\delta^2}{x^2}\right).
 \end{equation}

Now, combining the expressions (\ref{eq:83}), (\ref{eq:82}) and (\ref{eq:81}), it follows that
\begin{equation}\label{eq:c1}
C(x)=\frac{x^2}{2\delta(1+\delta)\sqrt{2+\delta}}\exp\left({-\frac{1}{12x}-\frac{2(1+\delta)\delta^2}{x^2}}\right),\qquad 0<x<1,
\end{equation}
satisfies (\ref{eq:c0}) for $l^{-1}<x<1$.

Now let $l+l^{-1}>x\geq 1$. Applying Stirling's inequality, we have that
\begin{eqnarray}\label{eq:84}
\lefteqn{\int_{(x-\delta)}^{l \wedge x}h_l(x;lv+1)dv}\nn\\
&=& \int_{(x-\delta)}^{l \wedge x} \frac{l^{lv+1}x^{lv}e^{-lx}}{\Gamma(lv+1)}dv\nonumber\\
&\geq&\int_{(x-\delta)}^{l \wedge x}\frac{l^{lv+1}x^{lv}e^{-lx}}{\sqrt{2\pi}(lv+1)^{lv+1/2}\exp[-(lv+1)+(12x)^{-1}]}dv\nonumber\\
&=&\sqrt{\frac{l}{2\pi}}\ e^{1-(12x)^{-1}}\int_{(x-\delta)}^{l \wedge x}\frac{x^{lv}e^{l(v-x)}}{(v+l^{-1})^{lv+1/2}}dv\nonumber\\
&\geq&\frac{\sqrt{l}\ e^{1-(12x)^{-1}}}{\sqrt{2\pi(x+\delta)}}\int^{x\wedge l}_{x-\delta}\exp\left[lv\left\{\log\left(\frac{x}{v+l^{-1}}\right)+(1-\frac{x}{v})\right\}\right]dv,
\end{eqnarray}
since   $v+l^{-1}<x+\delta$, when  $l>\delta^{-1}$.  Note that
\begin{eqnarray*}
\log u-(u-1)&=&(u-1)^2
\left\{-\frac{1}{2}+\frac{(u-1)}{3}-\frac{(u-1)^2}{4}+\cdots\right\}\\
&\geq&(u-1)^2\left\{-\frac{1}{2}+(u-1)-(u-1)^2+\cdots\right\}\\
&=&(u-1)^2\left\{-\frac{1}{2}-\left[(1-u)+(1-u)^2+(1-u)^3+\cdots\right]\right\}\\
&=&(u-1)^2\left(-\frac{1}{2}-\frac{1-u}{u}\right),
\end{eqnarray*}
for $0<u<1$. Further,
$
\log u-(u-1)
\geq-(u-1)^2/2,
$
for $1\leq u<2$, since $\frac{(u-1)}{3}-\frac{(u-1)^2}{4}+\frac{(u-1)^3}{5}-\cdots\geq0$. Note that $0< \frac{x}{v+l^{-1}}\leq \frac{1}{1-\delta}$, where $\delta<\frac{1}{2}$ without loss of generality. Now it follows that
\begin{eqnarray}\label{equation:x1}
\lefteqn{\log\left(\frac{x}{v+l^{-1}}\right)+1-\frac{x}{v}}\nn\\
&=&\log\left(\frac{x}{v+l^{-1}}\right)-\left(\frac{x}{v+l^{-1}}-1\right)+\left(\frac{x}{v+l^{-1}}-1\right)-\left(\frac{x}{v}-1\right)\nonumber\\
&\geq&\left(-\frac{1}{2}-\frac{v+l^{-1}-x}{x}\right)\left(\frac{x}{v+l^{-1}}-1\right)^2+\frac{x}{(v+l^{-1})lv}.
\end{eqnarray}
Letting $\tilde v$  denote $v+l^{-1}$,  RHS of (\ref{equation:x1}) is equal to
\begin{eqnarray*}
\frac{(x-2\tilde v)(x-\tilde v)^2}{2x\tilde v}+\frac{x}{lv\tilde v}&\geq&-\frac{(x-\tilde v)^2}{2\tilde v^2}+\frac{x}{lv\tilde v},
\end{eqnarray*}
since $\frac{x-2\tilde v}{x}\geq -1$ for $\tilde v<x+l^{-1}$ (i.e. $v<x$) and $x>1$. Now,
\begin{eqnarray}\label{eq:85}
\lefteqn{\int^{x\wedge l}_{x-\delta}\exp\left [lv\left\{\log\left(\frac{x}{v+l^{-1}}\right)+\left(1-\frac{x}{v}\right)\right\}\right ]dv\nonumber}\\
&=&\int_{x-\delta+l^{-1}}^{(x\wedge l)+l^{-1}}\exp\left[-\frac{lv(x-\tilde v)^2}{2(x-\delta)^2}+\frac{x}{\tilde v}\right]d \tilde v\nonumber\\
&\geq&e^{1/2}\int_{x-\delta+l^{-1}}^{(x\wedge l)+l^{-1}}\exp\left[-\frac{lx(x-\tilde v)^2}{2(x-\delta)^2}\right]d \tilde v\nonumber\\
&\geq&e^{1/2}\sqrt{\frac{2\pi}{lx}}\ (x-\delta)\left\{\Phi\left(\frac{\sqrt{lx}\ (\delta-l^{-1})}{x-\delta}\right)-\frac{1}{2}\right\}\nonumber\\
&\geq&e^{1/2}\sqrt{\frac{2\pi}{lx}}(x-\delta)\left\{\Phi\left(\frac{\sqrt x\ \delta}{2(x-\delta)}\right)-\frac{1}{2}\right\}\nonumber\\
&\geq& \frac{1}{2}\delta\sqrt{\frac{e}{l}}\exp\left(-\frac{x \delta^2}{8(x-\delta)^2}\right),
\end{eqnarray}
for $l/2>{\delta}^{-1}$, since $\Phi(z)-1/2>\phi(z)z$ for any $z>0$ and since $x>v>1-\delta$ for all $x>1$.
Combining expressions (\ref{eq:84}) and (\ref{eq:85}) and simplifying, we conclude that
\begin{equation}\label{eq:c2}
C(x)=\frac{\delta\exp(3/2-12x^{-1})}{2\sqrt{2\pi(x+\delta)}}\exp\left(-\frac{x \delta^2}{8(x-\delta)^2}\right),\qquad x\geq1,
\end{equation}
satisfies (\ref{eq:c0}) for $1\leq x<l+l^{-1}$.

Now for $C(x)$ defined by (\ref{eq:c1}) and (\ref{eq:c2}) satisfies (\ref{eq:c0}). Further, by straightforward calculations,  $\int_{\mathfrak{X}} f^c(x)\log \frac{1}{C(x)}dx<\infty$  under Condition (d) in Theorem \ref{thm:consistency}.
$\Box$

\vspace{1cm}

\begin{lemma}\label{thm:gamma12} (Lemma 6 in \cite{Wu_Ghosal_2008}) 
For $f_l(x)$ defined in \eqref{eq:fm}, $f_l(x)\to f^c(x)$ as $l\to \infty$ for each given $x>0$, where
$f^c$ is bounded, uniformly continuous and integrable on $\mathfrak{X}$.
\end{lemma}

 
{\noindent \bf Proof of  Lemma  \ref{thm:gamma12}:} 

To use Lemma \ref{lemma:gamma}, we re-parameterize the kernel function by the following tranformations. Let $v=(V-1)l^{-1}$ and $u=l^{-1}$. Let
$$
h(x;v,u)=\frac{x^{v/u}e^{-x/u}}{\Gamma(v/u+1)u^{v/u+1}},
$$
and let $h_l(x;v)=h(x;v,l^{-1})$,
where $v\in A_l$, $A_l=[l^{-1},l]$, for $l>0$. Now  we have that 
$$ f_l(x) = t_l \int_{2}^{1+l^2} h_l(x;V)l^{-1} f^c((V-1)/l)dV =\int_{l^{-1}}^l h_l(x;v)f^c(v)dv,$$ 
and we show that such $h_l(x;v)$ satisfies Conditions
C1--C3 in Lemma \ref{lemma:gamma}.

Observe that 
\begin{equation}\label{eq:gamdev}
\frac{d}{dV}\log (h_l(x;V))=\log l +\log
x-{\Psi_0}(V),
\end{equation}
where  $\Psi_0(z)=\frac{d}{dz}\log(\Gamma(z))$ is the digamma
function. Also $\Psi_0(z)$  is continuous and monotone increasing for
$z\in(0,\infty)$,
$
\Psi_0(z+1)=\Psi_0(z)+\frac{1}{z},  \mbox{ and } \Psi_0(z)-\log(z-1)\to 0
$;
see \cite{arfken1985method}[pp. 549--555]  for details.

Given $x>0$, consider 
{expression (\ref{eq:gamdev}). }
For $l$ sufficient large, such that $l^{-1}<x<l+l^{-1}$, we have
 \begin{equation}\label{eq:gamdev1}
\frac{d}{dv} h_l(x;v)\left \{
\begin{array}{l}
> 0\qquad l^{-1}\leq v < x-l^{-1},\\
\\
< 0 \qquad l \geq v> x-l^{-1}+\rho,
\end{array}
\right.
\end{equation}
where $\rho$ is some small positive number.
Also, note that $\frac{d^2}{dv^2}h_l(x;v)<0$ for all $x>0$ and $l^{-1}\leq v\leq l$.  Thus, the first order derivative
changes from  positive to negative as $v$ changes from $l^{-1}$ to
$l$ for given $x$ and sufficient large $l$. Hence, there exists $l_0$ such that $h(x;v,l^{-1})$ is increasing
as a function of $v$ when $v\leq l_0$ and decreasing when $v>l_0$.
For sufficient  large $l$,
\begin{eqnarray}\label{eq:91}
\lefteqn{e^{-xl}
\left[\sum_{t=0}^{[l^{2}]}\frac{(xl)^t}{t!}-1-\frac{(xl)^{[l_0]+1}}{([l_0]+1)!}\right]}\nonumber\\
 &&\leq
e^{-xl} \left[\int_{1}^{l^{2}}\frac{(xl)^{vl}}{\Gamma(vl+1)}d(vl)\right]
\leq e^{-xl}\int _1^{l^2}\frac{(xl)^{vl}}{\Gamma(vl+1)}d(vl)
=\int_{l^{-1}}^l h_l(x;v)dv
 \nonumber\\
&&\leq e^{-xl}\left[\sum_{t=0}^{[l^{2}]}\frac{(xl)^t}{t!}-1+\frac{(xl)^{l_0}}{([l_0]-1)!}\right],
\end{eqnarray}
 where $[z]$ stands for the largest integer less than or equal to $z$. Expression (\ref{eq:91}) is obtained by discretizing the integral term in the mid line of the expression.  Notice that $\sum_{t=0}^{[l^{2}]}\frac{(xl)^t}{t!}$ is the Taylor's expension of $e^{xl}$,  using the expression for the remainder of Taylor's series, we have that 
\begin{equation}\label{eq:92}
e^{-xl}
\left[\sum_{t=0}^{[l^{2}]}\frac{(xl)^t}{t!}-1-\frac{(xl)^{[l_0]+1}}{([l_0]+1)!}\right]
\geq 
1-\frac{\frac{(xl)^{[l^{2}]+1}}{([l^{2}]+1)!}e^{x^*l}}{e^{xl}}
-\frac{1}{e^{xl}}
-\frac{\frac{(xl)^{[l_0]+1}}{([l_0]+1)!}}{e^{xl}},
\end{equation}
where $x^*\in(0,x)$. It is obvious that the expression in (\ref{eq:92}) tends to 1 as
$l\rightarrow \infty$. Similarly, we have that the
 RHS of (\ref{eq:91}) tends to 1 as $l\rightarrow\infty$. Hence,
\[
\int_{{l}^{-1}}^l h(x;v,u)dv={e^{-xl}}{
 \int_{1}^{l^{2}}\frac{(xl)^{vl}}{\Gamma(vl+1)}d(vl)}
 \rightarrow 1 \quad \mbox{as } l\rightarrow\infty,
\]
that is, Condition C1 is satisfied.

From above, we also know that Condition C3 is satisfied, since
$h_l(x;v)>0$ for all $v\in A_l$ and $x\in\mathfrak{X}$.

To verify Condition C2, for any $\delta>0$ and $x\in \mathfrak{X}$, we want
\begin{eqnarray*}
&&\int_{|x-v|>\delta\,, v\in A_m}\Big|h_l(x,v)\Big|dv
=\int_{|x-v|>\delta,\, v\in
A_l}\frac{e^{-xl}(xm)^{vl}}{\Gamma(vl+1)}dv\rightarrow
0,
\end{eqnarray*}
as $l\rightarrow\infty$. We show that for any $\delta>0$,
\[l\sup_{|x-v|>\delta, v\in
A_l}\frac{e^{-xl}(xl)^{vl}}{\Gamma(vl+1)}\rightarrow
0 \quad\mbox{as } l \rightarrow \infty,\] which is equivalent to showing that
\[
\log
l + \log\frac{e^{-xl}(xl)^{vl}}{\Gamma(vl+1)}\rightarrow
-\infty\quad\mbox{
for all }v\in A_l,|x-v|>\delta.\]
 For any $v$ such that
$v\in A_l,|x-v|>\delta$, we have by Stirling's inequality for factorials,
\begin{eqnarray*}
\lefteqn{\log l+\log\frac{e^{-xl}(xl)^{vl}}{\Gamma(vl+1)}}\\
&&\leq\log l+\log\frac{e^{-xl}(xl)^{vl}}{[vl]!}\\
&&\leq\log l+vl\log(xl)-xl-vl\log vl+vl\\
&&=\log l+ \{1+\log (x/v)-x/v\}vl\rightarrow -\infty,
\end{eqnarray*}
as $l\rightarrow \infty$, since for any given $x$ and $\delta$, there exists $q<0$ such that $1+\log (x/v)-x/v<q$ for all
the $v\in A_l$, $|x-v|>\delta$.

Thus Conditions C1--C3 in Lemma \ref{lemma:gamma}
are all satisfied and we have that $f_l(x)\rightarrow f^c(x)$ as
$l\rightarrow \infty$ for each $x>0$. $\Box$

\vspace{1cm}

 \begin{lemma}\label{lemma:ga1}
(Lemma 5 in \cite{Wu_Ghosal_2008}) Let $f_l(x)$ be defined as in \eqref{eq:fm}. If the conditions of Theorem \ref{thm:consistency} (i) are satisfied, then  $\int_{\mathfrak{X}} f^c(x)\log\frac{f^c(x)}{f_l(x)}dx\to 0$ as $l\to \infty$.
\end{lemma}

{\noindent \bf Proof of Lemma \ref{lemma:ga1}:} 

First, we derive the lower bound of $f_l(x)$ for $x$ in different intervals. 

Following expression (\ref{eq:gamdev}), for $x<l^{-1}$, $\log(lx)<0$, and $\Psi_0(V)\geq
\Psi_0(2)=0.42$ for $V\in[2,1+l^2]$, and hence $\frac{d}{d V}\log (h_l(x;V))<0$. For
$x>l+l^{-1}$ and $V \in [2,1+l^2]$, $\log(lx)\geq \log (l^2)\geq \Psi_0(1+l^2)\geq\Psi_0(V)$, and hence $\frac{d}{dV}\log (h_l(x;V))>0$. Thus replacing  $V$ by $1+l^2$ in the integrand, we obtain a lower bound for $f_l(x)$ with $x<l^{-1}$ as,
\begin{equation}\label{eq:gamma1}
f_l(x)
\geq
t_l\int_{2}^{1+l^2}\frac{x^{l^{2}}e^{-lx}l^{l^{2}+1}}{\Gamma(l^{2}+1)}f^c(\alpha)d\alpha
=
\frac{x^{l^{2}}e^{-lx}l^{l^{2}+1}}{\Gamma(l^{2}+1)}. 
\end{equation}
Similarly, replacing $\alpha$ by $2$ in the integrand, we obtain that for $x>l+l^{-1}$,
\begin{equation}\label{eq:gamma2}
f_l(x)\geq{x^{}e^{-lx}l^{2}}.
\end{equation}
Consider the RHS of equation (\ref{eq:gamma1}). For $x<{l}^{-1}$, we have
$$
\frac{d}{dl}\log\left(\frac{x^{l^{2}}e^{-xl}l^{l^{2}+1}}{\Gamma(l^{2}+1)}\right)
= 2l[\,\log(xl)-\Psi_0(l^2+1)]+\frac{l^2+1}{l}-x
<0,
$$
for all $l$ sufficiently large, where $c_1>0$ is some
constant.
Consider the RHS of equation  (\ref{eq:gamma2}), for $x>l+l^{-1}$, we have
$
\frac{d}{dl}\left ({xe^{-xl}l^2}\right )
=xle^{-xl}
(2-xl)<0
$.

Hence, replacing $l$ by $x^{-1}$ on the RHS of (\ref{eq:gamma1}), we obtain a lower bound of $f_l(x)$ for $x<{l}^{-1}$ as below,
\begin{equation}
f_l(x)\geq\frac{x^{l^{2}}e^{-xl^{}}l^{l^{2}+1}}{\Gamma(l^{2}+1)}
\geq\frac{x^{x^{-2}}e^{-1}x^{-x^{-2}-1}}{\Gamma(x^{-2}+1)}=\frac{1}{ex\Gamma(x^{-2}+1)};
\end{equation}
and similarly, replacing $l$ by $x$ on the RHS of (\ref{eq:gamma2}), we obtain that  for $x>l+l^{-1}$,
\begin{equation}
f_l(x)\geq{xe^{-xl^{}}l^{2}}
\geq{e^{-x^{2}}x^{3}}.
\end{equation}

Now, we consider $f_l(x)$ for ${l}^{-1}\leq x\leq l+l^{-1}$. Let $\delta>0 $
be fixed and $v=(V-1)/l$. For $l$ large,
\begin{eqnarray*}
 f_l(x)&\geq& \int_{x-\delta}^{x+\delta} h_l(x;lv+1)t_lf^c(v)dv\\
&\geq& \left\{
\begin{array}{ll}
\phi_{\delta}(x)t_l\int_{{l}^{-1}\vee x}^{x+\delta
}h_l(x;lv+1)dv,& \ x<1\\
&\\
\phi_{\delta}(x)t_l\int_{x-\delta}^{l \wedge x}h_l(x;lv+1)dv,& \ x\geq1
\end{array}
\right .\\
&\geq& C(x)\phi_{\delta}(x),
\end{eqnarray*}
where $C(x)$ is given in Lemma \ref{lemma:gammac}.

Now we have the lower bound of function $f_l(x)$,
\begin{equation}\label{eq:gamma4}
f_l(x)\geq \left\{
\begin{array}{ll}
C(x)\phi_{\delta}(x),&R^{-1}\leq x\leq R,\\
&\\
\min(C(x)\phi_{\delta}(x),\frac{1}{ex\Gamma(x^{-2}+1)}),&
0<x<R^{-1},\\
&\\
\min(C(x)\phi_{\delta}(x),{ e^{-x^{2}}x^{3}}),&
R<x,
\end{array}
\right .
\end{equation}
where $0<R<l$. Hence, we have that
{\setlength\arraycolsep{3pt}
\begin{eqnarray*}
\log\frac{f^c(x)}{f_l(x)}&\leq& \xi(x)\\
 &:=&\left\{
\begin{array}{ll}
\log\frac{f^c(x)}{C(x)\phi_{\delta}(x)},&R^{-1}\leq x\leq R,\\
&\\
\max\left\{\log\frac{f^c(x)}{C(x)\phi_{\delta}(x)},\log([ex\Gamma(x^{-2}+1)]^{-1}f^c(x))\right\},&
0<x<R^{-1},\\
&\\
\max\left\{\log\frac{f^c(x)}{C(x)\phi_{\delta}(x)},\log\frac{f^c(x)}{{ e^{-x^{2}}x^{3}}}\right\},&
R<x.
\end{array}
\right .
\end{eqnarray*}
}Since $f^c(x)<C_f<\infty$, we also have that
$\log\frac{f^c}{f_l}\geq\log\frac{f^c(x)}{C_ft_2}$ for $l>2$, where $t_2$ is the $t_l$ defined in (\ref{eq:fm}) with $l=2$. Further, as
$\log\frac{f^c(x)}{C_ft_2}<0$, we have
$|\log\frac{f^c(x)}{f_l(x)}|\leq\max\{\xi(x),|\log\frac{f^c(x)}{C_ft_2}|\}$.

By Condition (b) in Theorem \ref{thm:consistency},
$\int|\log\frac{f^c(x)}{C_ft_2}|f^c(x)dx
{ = }\log C_ft_2-\int
f^c\log(f^c)dx<\infty$, since $f^c(x)<C_f$ for all $x\in C$ and $t_2\leq 1$. Now, consider $\int\xi(x)f^c(x)dx$, which equals to
{\setlength\arraycolsep{1pt}
\begin{eqnarray}\label{eq:gamma3}
\lefteqn{\int_{R^{-1}}^R f^c(x)\log\frac{f^c(x)}{C(x)\phi_{\delta}(x)}dx}\nonumber\\
&+&
\int_0^{R^{-1}}f^c(x)\max\left\{\log\frac{f^c(x)}{C(x)\phi_{\delta}(x)},{\log(f^c(x))
-
\log (ex\Gamma(x^{-2}+1))
}
\right\}
dx\nonumber\\
&+&
\int_R^{\infty}f^c(x)\max\left\{\log\frac{f^c(x)}{C(x)\phi_{\delta}(x)},
{
\log(f^c(x))
-\log({ e^{-x^{2}}x^{3}})
}\right\}dx\nonumber\\
\leq &&  \int_0^{\infty}f^c(x)\log\frac{f^c(x)}{\phi_{\delta}(x)}dx  +\int_0^{\infty}f^c(x)\log\frac{1}{C(x)}dx\\
&&\quad +
\int_{(0,R^{-1}]\cap A}
f^c(x)\Big[\log([ex\Gamma(x^{-2}+1)]^{-1}f^c(x))\Big]dx\nonumber\\
&&\quad +\int_{(R,\infty)\cap B}
f^c(x)\Big[\log\frac{f^c(x)}{ e^{-x^{2}}x^{3}}\Big]dx,\nn
\end{eqnarray}
}where $A=\{x:f^c(x)\ge [{ex\Gamma(x^{-2}+1)}]^{-1}\},$ and
$B=\{x:f^c(x)\ge { e^{-x^{2}}x^{3}}\}$. 
The above relation (\ref{eq:gamma3}) holds since  $C(x)<1$ by Lemma \ref{lemma:gammac} and $\max(x_1,x_2)\leq x_1+x_2^+$ if $x_1>0$.

The
first term on the RHS of (\ref{eq:gamma3}) is less than
infinity by Condition (c) in Theorem \ref{thm:consistency}. By Lemma \ref{lemma:gammac}, the second terms on the RHS of (\ref{eq:gamma3}) is also less than infinity. Note that, by Stirling's inequality, (see \cite{feller1957introduction} [vol. I. pp. 50-53])
\begin{eqnarray*}
\lefteqn{\Big|\log \frac{1}{ex\Gamma(x^{-2}+1)}\Big|}\\
&\leq& |\log x|+1+\log(2\pi)
+(x^{-2}+1)\log(x^{-2}+1)+\frac{(x^{-2}+1)^2+1}{12(x^{-2}+1)},
\end{eqnarray*}
for $0<x<1$. Hence, the third term on the RHS of
(\ref{eq:gamma3}) is less than infinity by Condition (d) in Theorem \ref{thm:consistency}.
Similarly, so is the fourth term. 

Since $\int _0^{\infty}f^c(x)\log \frac{f^c(x)}{f_l(x)}dx\leq \int_0^{\infty}f^c(x)\xi(x)dx<\infty$, we have that for any $\epsilon >0$, there exists $l_0$, such that for $l\geq l_0$, $\int _{l_0}^{\infty}f^c(x)\log\frac{f^c(x)}{f_l(x)}<\epsilon/2$.  By Lemma \ref{thm:gamma12},  we have that $f_l(x)\to f^c(x)$ for each $x\in [0,l_0]$ when $l\to \infty$, since $f^c(x)$ is uniformly continuous, bounded and integrable on $[0,l_0]$. Thus, by the Dominated Convergence Theorem (DCT), for any $\epsilon>0$, there exists $l_1$, such that for $l>l_1$, $\int_0^{l_0} f^c(x)\log\frac{f^c(x)}{f_l(x)}dx<\epsilon/2$. Therefore, for any $\epsilon>0$,  there exists $l_0$ and $l_1$, such that for $l>\max(l_0,l_1)$ we have that  $$\int_{\mathfrak X} f^c(x)\log\frac{f^c(x)}{f_l(x)}dx=\int_0^{l_0}f^c(x)\log\frac{f^c(x)}{f_l(x)}dx+\int_{l_0}^{\infty}f^c(x)\log\frac{f^c(x)}{f_l(x)}dx<\epsilon.\qquad \Box$$
 
\vspace{1cm}



Notice that the condition of of Lemma \ref{lemma:ga1} in this paper is different from its original form corresponding to the change in condition in Theorem \ref{thm:consistency} (i), which removed the restriction $f^c(0)\neq 0$. Lemma \ref{lemma:ga1} is applicable for Gamma kernel, and proved below by applying Lemma \ref{thm:gamma12}, DCT, and the fact that that $\log\frac{f^c(x)}{f_l(x)}$ is bounded from below by $\xi(x)$ and $f^c(x)\xi(x)$ is integerable on $[0,\infty)$. Lemma \ref{thm:gamma12} is the same as Lemma 6 in \cite{Wu_Ghosal_2008}, which provides the point-wise convergence, $f_l(x)\to f^c(x)$ as $l\to \infty$, for each $x>0$.  Lemma \ref{thm:gamma12} is then proved by verifying that Conditions C1-C3 in Lemma \ref{lemma:gamma} (Lemma 7 in \cite{Wu_Ghosal_2008}) are satisfied by the Gamma kernel. As pointed out in \cite{Wu_Ghosal_2008} page 324, Lemma \ref{lemma:gamma} is applicable when the support of function $h_l$ and $f^c$ is possibly non-compact. To apply DCT and show that the two facts mentioned above are true with respect to the Gamma kernel, we used the Conditions (b) and (c) in Theorem \ref{thm:consistency}, and result of Lemma \ref{lemma:gammac}.  Lemma \ref{lemma:gammac} is the same as Lemma 8 in in \cite{Wu_Ghosal_2008}, which is about the property of the kernel function, and does not involve the true density function $f^c$. Its proof depends on the Condition (d) in Theorem \ref{thm:consistency}.  Refer to the lemmas and their proofs above for details. 
\vspace{1cm}



\noindent {\bf Proof. of Theorem \ref{thm:consistency} (i):} 

Part (i) is a modified version of Theorem 14 in \cite{Wu_Ghosal_2008}, where the original condition:\\
B4. for some $0<M<\infty$ , $0<f^c(x)\leq M$ for all $x$;
is changed to:\\
(a) $f^c$ is nowhere zero, except at $x=0$ and bounded above by $C_f<\infty$.

Since no hyper-parameter involved in \eqref{DPM}, to complete the proof, we only need to verify Conditions A1 and A3 in Theorem \ref{wu08t1}. 
More specifically, we are going to show:
\begin{itemize}
	\item [a1.] $\int f^c \log\frac{f^c}{f_{G_{\epsilon}}}<\epsilon$,
	\item [a2.] $\int f^c\log \frac{f_{G_{\epsilon}}}{f_G}<\epsilon$ for every $G\in \mathscr W$,
\end{itemize}
where $f_G:=\int h(x;\psi)dG(\psi)$, to complete the proof.

For any $l>0$, let $G_l$ denote $F_l^*\times \delta (V/l)$, where $F_l^*$ is the probability measure corresponding to $t_l l^{-1}f^c((V-1)/l)\Ind(V\in[2,1+l^2])$ as a density function for $V$, and $\Ind$ is the indicator function. Obviously, $G_l$ is compactly supported, and we let $f_l(x)=f_{G_l}(x)$. Let $F_l$ be the probability measure corresponding to $f_l$. By Lemma \ref{lemma:ga1}, $\int f^c(x)\log\frac {f^c(x)}{f_l(x)}dx \to 0$ as $l\to \infty$, which implies that Condition a1 is satisfied.

Now we need to verify Condition A3 (equivalently Condition~a2) to complete the proof for part (i).  This is shown by verifying the Conditions A7, A8, and A9 without considering $\phi$ and $A$ in the statement of Lemma \ref{wg08lemma3}. For any given $\epsilon>0$, let $D=\{(V,V/l_{\epsilon}): V \in [2,1+l_{\epsilon}^2]\}$, where $l_{\epsilon}$ is such that $\int f^c(x)\log\frac{f^c(x)}{f_{l_{\epsilon}}(x)}<\epsilon$. By the verification of Condition A1, we have that $l_{\epsilon}$ exists.
To verify Condition A7,  it is sufficient to  show that
$\int f^c(x)|\log f_{l_{\epsilon}}(x)|dx<\infty$ and $\int f^c(x)| \log \inf_{(V,u)\in D}h(x;V,u) |dx<\infty$. 
Observe that 
\begin{equation*}
\frac{d}{dV}\log (h_l(x;V))=\log l +\log
x-{\Psi_0}(V),
\end{equation*}
where  $\Psi_0(z)=\frac{d}{dz}\log(\Gamma(z))$, is the digamma
function. Also $\Psi_0(z)$  is continuous and monotone increasing for
$z\in(0,\infty)$,
$
\Psi_0(z+1)=\Psi_0(z)+\frac{1}{z},  \mbox{ and } \Psi_0(z)-\log(z-1)\to 0
$;
see \cite{arfken1985method}[pp. 549--555]  for details.
We have that for $x\leq l^{-1}$, $\log(lx)<0$, and $\Psi_0(V)\geq
\Psi_0(2)=0.42$ for $V\in[2,1+l^2]$, and hence $\frac{d}{d V}\log (h_l(x;V))<0$. For
$x\geq l$ and $V \in [2,1+l^2]$, $\log(lx) \geq \log (l^2)\geq \Psi_0(1+l^2)\geq\Psi_0(V)$, and hence $\frac{d}{dV}\log (h_l(x;V))>0$. Then, for $l$ sufficient large, such that $l^{-1}<x<l+l^{-1}$, we have
 \begin{equation*}
\frac{d}{dV} h_l(x;V)\left \{
\begin{array}{l}
> 0\qquad 2\leq V < x-2,\\
\\
< 0 \qquad l^2+1 \geq V> x-2+\rho,
\end{array}
\right.
\end{equation*}
where $\rho$ is some small positive number.
Also, note that $\frac{d^2}{dv^2}h_l(x;v)<0$ for all $x>0$ and $2\leq v\leq l^2+1$.  Thus, the first order derivative
changes from  positive to negative as $v$ changes from $l^{-1}$ to
$l$ for given $x$ and sufficient large $l$. Hence, for any given $l$ sufficiently large, there exists $l_0$,  such that $h(x;V,Vl^{-1})$ is increasing
as a function of $v$ when $v\leq l_0$ and decreasing when $v>l_0$. Therefore, we have that
$$
\log \inf_{(V,u)\in D}h(x;V,u)=\log(\min\{h(x;1+l^2_{\epsilon},l_{\epsilon}^{-1}),h(x;2,l_{\epsilon}^{-1})\}),
$$
for any $0<x<\infty$. Hence,
\begin{equation}
|\log \inf_{(V,u)\in D}h(x;V,u)|
< xl_{\epsilon}  +(l_{\epsilon}^2)|\log x| + |\log (\Gamma(l_{\epsilon}^2+1)l_{\epsilon}^{-(l^2_{\epsilon}+1)})| +|\log(l^{-2}_{\epsilon})|.\nn
\end{equation}

By Condition (c) of Theorem \ref{thm:consistency} (i), we have that $\int |\log \inf_{(V,u)\in D}h(x;V,u)|f^c(x)dx<\infty$. Further, $\log f_{l_{\epsilon}}(x)$ is also $f^c$-integrable by a similar argument. 

To see that Condition A8 is satisfied, let $D=\{(V,V/l_{\epsilon}): V \in [2,1+l_{\epsilon}^2]\}$, the same as we used for verifying condition A7. Since Condition A7 is satisfied, for $C:=[l^{-1}, l]$, we can always have a large enough $l$ such that inequality (\ref{w08.4}) and $P_{f^c}(C^c)<\epsilon/(4\log2)$ both satisfied. For such $C$ and $D$, we have that $\inf_{x\in C}\inf_{\psi\in D}h(x,\psi)>0$, where $h(x,\psi)=\frac{l_{\epsilon}^V}{\Gamma(V)}x^{V-1}e^{-x/l_{\epsilon}}$ is the Gamma density with parameters in $D$.

Condition A9 is also satisfied for the $C$ and $D$ defined as above. Due to the setting of $D$, we need to show that $\frac{l_{\epsilon}^V}{\Gamma(V)}x^{V-1}e^{-x/l_{\epsilon}}$ is uniformly equicontinuous on $[2, 1+l_{\epsilon}]$ as a function of $V$ while $x\in [l^{-1}, l]$. By direct calculation, we have that
\begin{equation}\label{equicon}
   \frac{d\frac{l_{\epsilon}^V}{\Gamma(V)}x^{V-1}e^{-x/l_{\epsilon}}}{dV}=\frac{e^{-x/l_{\epsilon}}}{x}\left [ \log (l_{\epsilon})xe^{V\log(l_{\epsilon})x}\Gamma(V)+e^{V\log(l_{\epsilon})x}\Psi_0(V)
\right ], 
\end{equation}
where $\Psi_0(\cdot)$ is the digamma function. Notice that (\ref{equicon}) is bounded as a function of $V$ for any given $x\in [1/l,l]$, which implies that $\frac{l_{\epsilon}^V}{\Gamma(V)}x^{V-1}e^{-x/l_{\epsilon}}$ is pointwise equicontinuous on $[1/l,l]$, and it is also bounded as a function of $V\in [l_{\epsilon}^{-1}, l_{\epsilon}]$ and $x\in [l^{-1}, l]$, which implies that it is uniformly equicontinuous. $\Box$

\vspace{1cm}

\noindent {\bf Proof. of Theorem \ref{thm:consistency} (ii):} Part (ii) of this theorem has been proved by Theorem 3.3 of \cite{Tok06}.
\vspace{2mm}

Before proving Part (iii) of Theorem \ref{thm:consistency}, we cite the following lemma from \cite{Wu_Ghosal_2008}.

\begin{lemma} \label{wg08lemma4}
	For any density $f^c$ on $[0,1]$ and any $\epsilon>0$, there exist $m>0$ and $f_1(x)\geq m>0$, such that $p({\mathcal K}_{\epsilon}(f_1))>0$ implies that $p({\mathcal K}_{2\epsilon+\sqrt{\epsilon}}(f^c))>0$, where ${\mathcal K}_{\epsilon}(\cdot)$ is defined in Definition \ref{KLsupport}.
\end{lemma}

\noindent {\bf Proof. of Theorem \ref{thm:consistency} (iii):}
Based on Lemma \ref{wg08lemma4}, we only need to consider the densities that bounded  away from 0.  By Theorem 1 of \cite{diaconis1985quantifying}, Bernstein polynomials uniformly approximate any continuous density. Hence, for any $\epsilon>0$ and continuous density $f^c(x)$, there exists
\begin{equation}\label{eq:k}
f(x)=\sum_{j=0}^k w_j {k \choose j} x^j(1-x)^{k-j} ,
\end{equation} where $\sum_0^k w_j=1$,
such that $|f(x)-f^c(x)|<\epsilon$ for any $x\in [0,1]$.
It follows that Condition A1 is satisfied, when $G_{\epsilon}$ is defined as $G_{\epsilon}(\alpha=j+1, \beta=k-j+1)=w_j$ for $j=0.\ldots, k$, and $0$ otherwise.

We use Lemma \ref{wg08lemma3} to verify that Condition A3. Let $$D=\cup_{j=0,\ldots, k}[j+1-\delta,j+1+\delta]\times[k-j+1-\delta,k-j+1+\delta]$$ be a set that contains support of $G_{\epsilon}$, for some $0<\delta<1$. It is sufficient to show that Condition A7. holds by showing that $\int |\log f_{G_{\epsilon}}|f^c(x)dx<\infty$ and $\int \log \inf_{(\alpha,\beta)\in D} h(x;\alpha, \beta)f^c(x)dx<\infty$. Note that the $h(x;\alpha, \beta)$ is the Beta probability density function with parameter $\alpha$ and $\beta$ here.  The first inequality holds, since we only considering $f^c(x)$ bounded away from 0 and any continuous function on $[0,1]$ is bounded. Since $f^c(x)$ is bounded on $[0,1]$, to show  the second inequality holds is equivalent to  show $\int_0^1 |\log \inf_{(\alpha,\beta)\in D} h(x;\alpha, \beta)|dx<\infty$.  By the definition of $D$, We have that $B(\alpha, \beta)$ is bounded, where $B(\cdot)$ is beta function. Hence, we only need to show that
\begin{equation}\label{eq:betaineq}
\int_0^1 |\log \inf_{(\alpha,\beta)\in D} x^{\alpha-1}(1-x)^{\beta-1}|dx<\infty.
\end{equation}
 We have that   $x^{\alpha+\beta-2}\leq x^{\alpha-1}(1-x)^{\beta-1}$ for $0\leq x\leq 0.5$, and  $(1-x)^{\alpha+\beta-2}\leq x^{\alpha-1}(1-x)^{\beta-1}$ for $1\geq x >0.5$. For any given $\epsilon>0$, let $k$ is chosen by (\ref{eq:k}), and $0<\delta<1$ as defined above, we have 
    $$\int_0^1 |\log \inf_{(\alpha,\beta)\in D} x^{\alpha-1}(1-x)^{\beta-1}|dx
    {\stackrel{(*)}{\leq} }
    (k+\delta)\int_0^1 (|\log x|+|\log (1-x)|)dx<2(k+1)<\infty,$$  since $\int_0^1 |\log x|dx=1$. 
    Step (*) holds since $\alpha-1+\beta-1\leq k+\delta$, when $(\alpha,\beta)\in D$.
    Therefore,  Condition A7 holds.

 We have the  equicontinuity of Beta density family since $x^{\alpha-1}(1-x)^{\beta-1}$ is continuous function on $\alpha$ and $\beta$ for any $x\in C \subset (0,1)$, where $C$ is a compact set, and hence Condition~A9 is satisfied. 
Condition A8  is satisfied, since by Condition A7, the compact $C$ with $P_{f^c}(C^c)<\epsilon/(4\log 2)$ and satisfies (\ref{w08.4}) exists, and Beta density always greater than $0$ on $C\subset (0,1)$.
Therefore, Condition A3 is satisfied, and  the proof is completed. $\Box$


\begin{remark}
It is worth pointing out that all the priors we have chosen, with more consideration of the computational convenience, satisfy the conditions in Theorem \ref{thm:consistency}, which means that the weak consistency holds true as long as $f^c$ satisfies the conditions in Theorem \ref{thm:consistency}. For Gamma and Beta kernel cases, there is no explicit condition required on the choice of the ``parameters'' of the Dirichlet process, the $\alpha$ and $G_0$. Therefore, we only need to show that the $G_0$ we chose satisfies Condition (c) in Part (ii) of Theorem \ref{thm:consistency} for Gaussian kernel case. The prior chosen in this paper for $G_0$ is the conjugate normal-inverse gamma distribution in \eqref{eq.GaussPrior} with hyperparameters $\mu_0, v_0, m_0, \sigma_0$. We show that this prior satisfies Condition (c) in Theorem \ref{thm:consistency} Part (ii).
\end{remark}

\begin{lemma}\label{lemma:tokdar}
The normal-inverse gamma prior specified in \eqref{eq.GaussPrior} with $\mu_0=0$, $v_0\in(1,2)$, $m_0>0$ and $\sigma_0>0$ satisfies the conditions in Theorem \ref{thm:consistency} Part (ii).
\end{lemma}

\noindent {\bf Proof of Lemma \ref{lemma:tokdar}:}


We proceed with the same argument as in Remark 3.4 of \cite{Tok06}.  We let $\eta\in  (v_0/(1+v_0 /2), 1)$, $c_1=v_0(2-\eta)/2$, $c_2=v_0$ in Condition (c). Then $c_1>0$ and $c_1-\eta = v_0-(1+v_0/2)\eta<0$, satisfying $c_1\in (0,\eta)$; $c_2-c_1=v_0\eta/2>0$, satisfying $c_2>c_1$. Such choices are possible since $v_0\in (1,2)$. We show that all the inequalities in Condition (c) hold.

\begin{sloppypar}
\underline{The first inequality in Condition (c):} Since $\sigma^2/\sigma_0^2\sim \text{Inv-Gamma}(v_0/2,1/2)$, we have that $\sigma_0^2/\sigma^2\sim \text{Gamma}(v_0/2,1/2)$. For any $x>\max\{\sigma_0^{2/(2-\eta)},1\}$,
\begin{align*}
& G_0\left([0,+\infty)\times (x^{1-\eta/2},+\infty)\right) = G_0\left(u\in [0,+\infty)\mid \sigma\right)\cdot G_0\left( \sigma\in (x^{1-\eta/2},+\infty)\right) \\
\stackrel{(i)}{=}& \frac{1}{2}\text{P} \left(\sigma >x^{1-\eta/2} \right) = \frac{1}{2}\text{P} \left(\sigma_0^2/\sigma^{2} < \sigma_0^2 x^{-(2-\eta)} \right) \\
=& \frac{1}{2} \int_0^{\sigma_0^2 x^{-(2-\eta)}}\frac{1}{2^{v_0/2}\Gamma(v_0/2)} t^{v_0/2-1}e^{-t/2} dt \\
\stackrel{(ii)}{\geq} & \frac{1}{2^{v_0/2 +1}\Gamma(v_0/2)} (\sigma_0^2 x^{-(2-\eta)})^{v_0/2-1}\int_0^{\sigma_0^2 x^{-(2-\eta)}} e^{-t/2} dt\\
=& \frac{1}{2^{v_0/2 +1}\Gamma(v_0/2)} (\sigma_0^2 x^{-(2-\eta)})^{v_0/2-1} \cdot 2\left[1- \exp(-\sigma_0^2 x^{-(2-\eta)}/2)\right] \\
\stackrel{(iii)}{\geq} & \frac{1}{2^{v_0/2}\Gamma(v_0/2)} (\sigma_0^2 x^{-(2-\eta)})^{v_0/2-1} \cdot \frac{1}{4}\sigma_0^2 x^{-(2-\eta)} \\
=& \frac{1}{2^{v_0/2+2}\Gamma(v_0/2)} (\sigma_0^2)^{v_0/2} \cdot x^{-(2-\eta)v_0/2}
\end{align*}
where in (i) we use the fact that the prior of $u\mid \sigma^2$ is the symmetric distribution $N(0,\sigma^2/m_0)$, in (ii) we use the fact that $v_0/2-1<0$ and the function $t^{v_0/2-1}$ decreases with $t$, and in (iii) we use the inequality $1-\exp(-t)>t/2$ on $t\in (0,1/2)$ and the fact that $x>\sigma_0^{2/(2-\eta)}$ (so that $\sigma_0^2 x^{-(2-\eta)}/2<1/2$). Hence we can take $b_1=\frac{1}{2^{v_0/2+2}\Gamma(v_0/2)} (\sigma_0^2)^{v_0/2}$ and $c_1=v_0(2-\eta)/2$ as specified before, and the first inequality of Condition (c) is proved.
\end{sloppypar}

\vspace{5mm}

\underline{The second inequality in Condition (c):} Since the prior of $u\mid \sigma^2$ is the symmetric distribution $N(0,\sigma^2/m_0)$,
\begin{align*}
\scalebox{0.99}{$
G_0\left((-\infty,0]\times (x^{1-\eta/2},+\infty)\right) = G_0\left(u\in (-\infty,0]\mid \sigma\right)\cdot G_0\left( \sigma\in (x^{1-\eta/2},+\infty)\right)
= \frac{1}{2}\text{P} \left(\sigma >x^{1-\eta/2} \right).$}
\end{align*}
The rest is exactly the same as the proof for the first inequality of Condition (c).
\vspace{5mm}

\underline{The third and fourth inequalities in Condition (c):} Since the prior of $u\mid \sigma^2$ is the symmetric distribution $N(0,\sigma^2/m_0)$, by symmetry we only prove the third inequality, and the fourth inequality follows the same proof. Since
\begin{align}\label{tokinq30}
1-G_0\left((-\infty,x)\times (0,e^{x^\eta-1/2})\right)\leq G_0(u\in[x,+\infty)) + G_0(\sigma\in[e^{x^\eta-1/2},+\infty)),
\end{align}
it suffices to upper bound both terms $G_0(u\in(x,+\infty))$ and $G_0(\sigma\in (0,e^{x^\eta-1/2}))$, respectively. For the second term in \eqref{tokinq30}, we have that for all $x>1$,
\begin{align}\label{tokinq32}
&G_0(\sigma\in [e^{x^\eta-1/2},+\infty)) = \text{P} \left(\sigma_0^2/\sigma^{2} \leq \sigma_0^2 e^{-2x^{\eta}+1} \right) \nonumber \\
=& \int_0^{\sigma_0^2 e^{-2x^{\eta}+1}} \frac{1}{2^{v_0/2}\Gamma(v_0/2)} t^{v_0/2-1}e^{-t/2} dt \leq \int_0^{\sigma_0^2 e^{-2x^{\eta}+1}} \frac{1}{2^{v_0/2}\Gamma(v_0/2)} t^{v_0/2-1}\cdot 1 dt \nonumber \\
=& \frac{1}{2^{v_0/2-1}v_0\Gamma(v_0/2)}  \sigma_0^{v_0} e^{-v_0x^{\eta}+v_0/2} < C_1 x^{-c_2},
\end{align}
for sufficiently large constant $C_1>0$, where the last step follows since $\lim_{x\to+\infty} e^{-v_0x^\eta}/x^{-c_2}=0$. Now for the first term in \eqref{tokinq30}, we have that for all sufficiently large $x>1$,
\begin{align}\label{tokinq31}
& G_0(u\in([x,+\infty)) = \int_0^{\infty} G_0(u\in[x,+\infty)\mid \sigma) \frac{1}{2^{v_0/2}\Gamma(v_0/2)} t^{v_0/2-1}e^{-t/2} dt \nonumber \\
\stackrel{(i)}{=} & \int_0^{\infty} \left[1-\Phi(x/\sqrt{1/(m_0t)})\right] \frac{1}{2^{v_0/2}\Gamma(v_0/2)} t^{v_0/2-1}e^{-t/2} dt \nonumber \\
\stackrel{(ii)}{\leq} & \frac{1}{2^{v_0/2}\Gamma(v_0/2)}\int_0^{\infty} \frac{\phi(\sqrt{m_0t}x)}{\sqrt{m_0t}x} t^{v_0/2-1}e^{-t/2} dt \nonumber \\
=& \frac{1}{\sqrt{2\pi m_0}2^{v_0/2}\Gamma(v_0/2)} \cdot \frac{1}{x} \int_0^{\infty} t^{\frac{v_0-1}{2}-1}e^{-\frac{m_0x^2+1}{2}t} dt \nonumber \\
\stackrel{(iii)}{=}& \frac{1}{\sqrt{2\pi m_0}2^{v_0/2}\Gamma(v_0/2)} \cdot \frac{1}{x\left(\frac{m_0x^2+1}{2}\right)^{\frac{v_0-1}{2}}} \int_0^{\infty} s^{\frac{v_0-1}{2}-1}e^{-s} ds \nonumber \\
=& \frac{\Gamma((v_0-1)/2)}{2\sqrt{\pi m_0}2^{v_0/2}\Gamma(v_0/2)}  \frac{1}{x\left(m_0x^2+1\right)^{\frac{v_0-1}{2}}} \nonumber \\
\leq & \frac{\Gamma((v_0-1)/2)}{2\sqrt{\pi}2^{v_0/2}m_0^{v_0/2}\Gamma(v_0/2)} x^{-v_0}.
\end{align}
where in (i) we use the prior $u\mid \sigma^2\sim N(0,\sigma^2/m_0)$, in (ii) we use the inequality $1-\Phi(z)\leq \phi(z)/z$ for all sufficiently large $z>0$, in (iii) we use the change of variable $s=\frac{m_0x^2+1}{2}t$. Since we have set $c_2=v_0$, now we combine \eqref{tokinq30}, \eqref{tokinq32}, and \eqref{tokinq32} to conclude that
\begin{align}\label{tokinq34}
1-G_0\left((-\infty,x)\times (0,e^{x^\eta-1/2})\right)\leq \left(\frac{\Gamma((v_0-1)/2)}{2\sqrt{\pi}2^{v_0/2}m_0^{v_0/2}\Gamma(v_0/2)} +C_1\right) x^{-c_2}.
\end{align}
Then we set $b_2=\frac{\Gamma((v_0-1)/2)}{2\sqrt{\pi}2^{v_0/2}m_0^{v_0/2}\Gamma(v_0/2)} +C_1$ and this proves the third inequality of Condition (c). The fourth inequality follows similarly. \hfill $\Box$

\begin{remark}
 Theorem \ref{thm:consistency} is applicable to the following distributions that belong to commonly used distribution families.  
 \vspace{3mm}
 
\begin{enumerate}
	\item {\bf Distributions on $[a_1,a_2]$}	
	
	For distributions defined on $[a_1,a_2]$ with $a_1$ and $a_2$ known, all the continuous densities on $[a_1,a_2]$ satisfy the conditions in Theorem \ref{thm:consistency} Part (iii). Therefore it is easy to see that  when $f^c$ is the density function of one of the following distributions, the consistency holds, since $f^c$ is continuous on $[a_1,a_2]$:
	\begin{enumerate}
		\item {\bf Uniform distribution on $[a_1, a_2]$}: observed input $Y\sim \text{Uniform}(a_1,a_2)$. We apply DPM with Beta kernel on transformed $X=(Y-a_1)/(a_2-a_1)$. It is easy to see that $X$ has continuous density function, which is defined on $[0,1]$, and hence the posterior  consistency holds;
		\item {\bf Power function distribution on $[a_1,a_2]$}:  Let $Y$ denote the observed input that follows the power function distribution on $[a_1,a_2]$. We apply DPM with Beta kernel on transformed data $X=(Y-a_1)/(a_2-a_1)$. It is easy to see that $X$ has continuous density function, which is  defined on $[0,1]$, and hence the posterior  consistency holds;
		\item {\bf Triagular Distribution on $[a_1,a_2]$}:  Let $Y$ denote the observed input that follows the triangular distribution on $[a_1,a_2]$.We apply DPM with Beta kernel on transformed $X=(Y-a_1)/(a_2-a_1)$. It is easy to see that $X$ has continuous density function, which is defined on $[0,1]$, and hence the posterior consistency holds;
		\item {\bf Beta distributions (with location-scale transformation)}: Observed input can be modeled as $Y=a_1+(a_2-a_1)X$, where $X\sim \text{Beta}(\alpha, \beta)$, apply the DPM with Beta kernel on $X$. We have that $X$ has continuous density function, which is defined on $[0,1]$, and hence the posterior consistency holds;
		\item {\bf Truncated Normal distribution on $[a_1,a_2]$}: The observed input variable $Y\sim \text{TruncNorm}(\mu,\sigma^2, a_1, a_2)$, which is the normal distribution $\mathcal{N}(\mu,\sigma^2)$ truncated to the interval $[a_1,a_2]$. We apply the DPM with Beta kernel on the transformed variable $X=(Y-a_1)/(a_2-a_1)$. It is easy to see that $X$ has continuous density function, which is defined on $[0,1]$, and hence the posterior consistency holds;
		\item {\bf Johnson's $S_B$ distribution on $[a_1, a_2]$}: By definition of Johnson's $S_B$, the observed input can be presented as $X=\sigma g((Y-\gamma)/\delta)+\mu$, where $Y\sim \mathcal{N}(0,1)$, and $g(x)=1/(1+\exp(-x))$.  Notice that $a_1=\mu$ and $a_2=\mu+\sigma$. We apply the DPM with Beta kernel on the transformed variable $Z=(Y-a_1)/(a_2-a_1)$. It is obvious that $Z$ has continuous density function and defined on $[0,1]$, and hence the posterior consistency holds.
	\end{enumerate}
 \vspace{3mm}
	
	\item {\bf Distributions on $\Re$}
	
	We apply the DPM with Gaussian kernel to model input distributions on $\Re$. We show that the consistency holds for the following distributions by verifying the conditions in Part (ii) of Theorem \ref{thm:consistency}. Since only conditions 
	
    (a): $|\int^{\infty}_{-\infty}f^c(x)\log f^c(x)dx|<\infty$, and 

	(b) there exists an $\eta\in (0,1)$, such that $\int^{\infty}_{-\infty}|x|^{\eta}f^c(x)dx<\infty$,\\
	are related to the properties of $f^c$, we verify these two conditions for the following examples:
	\begin{enumerate}
		\item {\bf Normal distribution:} The true density $f^c(x)$ is $\phi_{\mu,\sigma}(x)$, the normal density function with mean $\mu$ and standard deviation $\sigma$. 
		
		For Condition (a), we have that $\log f^c(x)=-\log(2\pi\sigma)-(x-\mu)^2/(2\sigma^2)$, and $\left |\int^{\infty}_{-\infty} 
		\left( -\log(2\pi\sigma)-(x-\mu)^2/(2\sigma^2) \right) \phi_{\mu,\sigma}(x)dx\right |=|-\log(2\pi\sigma)-1/2|<\infty
		 $.
		
		For Condition (b), we have that, for any given $\eta\in (0,1)$,  
		\begin{eqnarray*}
		\int^{\infty}_{-\infty}|x|^{\eta}\phi_{\mu,\sigma}(x) dx & \leq & \int^{\infty}_{-\infty}|x|^{\eta}\phi_{0,\sigma}(x) dx \leq 2\int^{1}_{-1}|x|^{\eta}\phi_{0,\sigma}(x) dx+2\int^{\infty}_{1}|x|^{\eta}\phi_{0,\sigma}(x) dx\\
		&\leq&  1-2\left(1-\Phi(1/\delta)\right) + \int^{\infty}_{-\infty}|x|\phi_{0,\sigma}(x) dx \leq {1}/{(\pi\sigma)}+\sqrt {2/\pi}\sigma<\infty
		\end{eqnarray*}
		
		\item {\bf Logistic distribution:} The true density function is 
		$$f^c(x)= e^{-\frac{x-\mu}{s}}/(s(1+e^{-\frac{x-\mu}{s}})^2),$$ 
		
		For Condition (a), we have that $\log f^c(x)=-(x-\mu)/s-2\log(1+e^{-(x-\mu)/s}) -\log s$.
	To verify Condition (a), it is sufficient to show that $|\int_{-\infty}^{\infty} x f^c(x)dx|<\infty$ and $|\int_{-\infty}^{\infty} \log (1+e^{-(x-\mu)/s})f^c(x)dx|<\infty$. The first inequality holds since the logistic distribution has finite first moment. The second one holds, since
		\begin{eqnarray*}
		\left |\int_{-\infty}^{\infty} \log (1+e^{-(x-\mu)/s})f^c(x)dx \right |& \leq &
	\left	|\int_{-\infty}^{\mu}  \log (2e^{-(x-\mu)/s}) f^c(x)dx\right |  \\
	&+& \left |\int_{\mu}^{\infty} \log (1+1) f^c(x)dx\right | \\
	&\leq& \left  |\int_{-\infty}^{\mu}  -(x-\mu)/s f^c(x)dx\right | + \log 2= \log 8<\infty,
	\end{eqnarray*}
	where the last equality is based on the expectation of half logistic distribution is $\log 4$.  
	
	For Condition (b), we have that, for any $0<\eta<1$,
	\begin{eqnarray*}
	{\int^{\infty}_{-\infty}|x|^{\eta}f^c(x)dx}
&=&\int^{\infty}_{1}|x|^{\eta}f^c(x)dx+\int^{-1}_{-\infty}|x|^{\eta}f^c(x)dx+\int^{1}_{1}|x^{\eta}|f^c(x)dx\\
&\leq & 2 \int^{\infty}_{0}|x|f^c(x)dx+\int^{1}_{-1}1 \cdot f^c(x)dx\leq 2\log4 +1 <\infty.
	\end{eqnarray*}

	\item {\bf Student's t distribution: }
	The true density function is given by
	$$
		f^c(x)=\frac{\Gamma(\frac{\nu+1}{2})}{\sqrt{\nu\pi}\Gamma(\nu/2)}\left( 1+\frac{x^2}{\nu}\right)^{-\frac{\nu+1}{2}}.
		$$
	The entropy of Student's t distribution is
	\begin{align*}
	\left|\int^{\infty}_{-\infty}f^c(x)\log f^c(x)dx \right | & = \frac{\nu+1}{2}[\Psi_0((\nu+1)/2)-\Psi_0(\nu/2)] + \log [\sqrt{\nu}\text{B}(\nu/2,1/2)],
	\end{align*}
	where $\Psi_0(x)=\Gamma'(x)/\Gamma(x)$ is the digamma function for all $x>0$ and is monotonely increasing in $x$, and $\text{B}(a,b)$ is the Beta function with parameters $a>0$ and $b>0$. This entropy is always finite for all $\nu>0$. Therefore, Condition (a) is satisfied. 
		
	For Condition (b), since
	\begin{align*}
	\int^{\infty}_{-\infty}|x|^{\eta}f^c(x)dx & = \int^{\infty}_{-\infty}\frac{\Gamma(\frac{\nu+1}{2})}{\sqrt{\nu\pi}\Gamma(\nu/2)}\frac{|x|^{\eta}}{\left( 1+\frac{x^2}{\nu}\right)^{\frac{\nu+1}{2}}}dx,
	\end{align*}
	it is obvious that this integral is finite if $0<\eta<\nu$.



		
		\item {\bf Cauchy distribution:} The Cauchy distribution is the same as the Student's t distribution with $\nu=1$. Therefore Conditions (a) and (b) are verfied as above.

	\item {\bf Johnson's $S_U$ distribution:}	The random variable that follows Johnson's $S_U$ distribution can be represented as $X=\sigma \sinh ((Y-\gamma)/\delta)+\mu$, where $Y\sim \mathcal{N}(0,1)$ and $\sinh(x)= (e^x-e^{-x})/2$. The probability density function $f^c(x)$ of $X$ is 
	$$
	\frac{e^{-\frac{1}{2}(\gamma+\delta \sinh^{-1}(\frac{x-\mu}{\sigma}))^2}\delta}
	{\sqrt{2\pi}\sqrt {(x-\mu)^2+\sigma^2}}
	$$
	
	Since the Johnson's $S_U$ distribution has bounded density function and finite first moment, Condition (b) is obviously satisfied. 

    To check Condition (a), without loss of generality, we let $\mu=0$ and $\sigma=1$. Then it is sufficient to show that 
    \begin{equation}\label{s_u}
    \int_{_\infty}^{\infty} \log(1+x^2)f^c(x)dx+\left|  \int_{_\infty}^{\infty}\sinh^{-1}(x)f^c(x)dx\right |+ \int_{_\infty}^{\infty}\sinh^{-2}(x)f^c(x)dx<\infty
    \end{equation}
	By the facts that $f^c(x)$ for Johnson's $S_U$ is bounded, $\sinh^{-1}(x)=O \left (  \log (x)\right )$, and Johnson's $S_U$ distribution has finite first order moment, we have that inequality \eqref{s_u} holds.
	
	\item {\bf Gumbel distribution:} The Gumbel distribution with location parameter $\mu$ and scale parameter $\beta$ has the following density function:
	
	$$
	f^c(x) = \frac{1}{\beta} \exp\left(-\frac{x-\mu}{\beta} - e^{-\frac{x-\mu}{\beta}}\right), \quad x\in \Re,
	$$
    where $\mu\in\Re$ and $\beta>0$. 
	
	Because the entropy of this Gumbel distribution is
	$$\int_{-\infty}^{\infty} x f^c(x)dx = \ln \beta + C_{e}+1,$$
	where $C_e=0.5772\ldots$ is the Euler-Mascheroni constant, we know that Condition (a) is satisfied. Furthermore, for this Gumbel distribution, the mean is $\mu+C_e\beta$ and the variance is $\pi^2\beta^2/6$. To verify Condition (b), we have that
	\begin{align*}
	\int_{-\infty}^{\infty} |x| f^c(x)dx &= \text{E}|X| \leq \sqrt{\text{E}(X^2)} = \text{var}(X) + (\text{E}|X|)^2 \\
	&= \frac{\pi^2\beta^2}{6} + (\mu+C_e\beta)^2 < +\infty,
	\end{align*} 
	which means that Condition (b) is also satisfied by the Gumbel distribution with $\eta=1$.
	\end{enumerate}
 \vspace{3mm}

	\item {\bf Distributions on $[0,\infty)$}

	We use the DPM with Gamma kernel to model input distributions on $[0,\infty)$. We show that the posteriors are consistent for the following distributions by verifying that  the conditions in part (i) of Theorem \ref{thm:consistency}	 are satisfied. 
	
	Condition (a) is obviously true for all the following distributions. 
	
	Condition (c) requires that $\int_0^{\infty} f^c(x)\log \frac{f^c(x)}{\phi_{\delta}(x)}dx <\infty$ for some $\delta>0$, where $\phi_{\delta}(x)=\inf_{[x,x+\delta)}f^c(t)$ if $0<x<1$ and $\phi_{\delta}(x)=\inf_{(x-\delta,x]}f^c(t)$ if $x\geq 1$. Notice that all the following distributions are unimodal and have upper bounded densities. By the definition of $\phi_{\delta}(\cdot)$, $\log\frac{f^c(x)}{\phi_{\delta}(x)}=0$ in the following situations: (i)  $m\geq 1$ and $0<x< 1$; (ii) $x\geq m\geq 1$; (iii) $0<x\leq m<1$, where $m$ denotes the mode of the distribution. Therefore,  $\int_0^{\infty} f^c(x)\log \frac{f^c(x)}{\phi_{\delta}(x)}dx=\int_1^m f^c(x)\log \frac{f^c(x)}{\phi_{\delta}(x)}dx$ when $m\geq1$, and $\int_m^1 f^c(x)\log \frac{f^c(x)}{\phi_{\delta}(x)}dx$ when $0\leq m<1$. Due to the continuity of all the following $f^c(x)$ on the compact subset $[1,m]$ or $[m,1]$, we have that $\log \frac{f^c(x)}{\phi_{\delta}(x)}$ is bounded on $[1,m]$ or $[m,1]$, and hence $\int_m^1 f^c(x)\log \frac{f^c(x)}{\phi_{\delta}(x)}dx<\infty$ or $\int^m_1 f^c(x)\log \frac{f^c(x)}{\phi_{\delta}(x)}dx<\infty$, correspondingly, which implies that all the following listed distributions satisfies Condition (c).
	
	Below we show that Conditions (b) and (d) are satisfied for each of the following listed distributions respectively. Recall that Conditions (b) and (d) are:\\
	(b) $|\int _0^{\infty} f^c(x)\log f^c(x)dx|<\infty$,\\
	(d) there exists $\zeta>0$ such that $\int_0^{\infty} \max(x^{-\zeta-2},x^{\zeta+2})f^c(x)dx <\infty$.

	\begin{enumerate}
		
		\item {\bf	Inverse Gaussian distribution:} The density function 
		$$
		f^c(x)=\left[  \frac{\lambda}{2\pi x^3} \right]^{1/2}\exp\left\{ \frac{-\lambda(x-\mu)^2}
		{2\mu^2x} \right\}
		$$
		
	To show that Condition (b) holds,  we have that
		\begin{eqnarray*}
		\left|\int _0^{\infty} f^c(x)\log f^c(x)dx \right|&\leq & c_1+ \left| \int_0^{\infty}( -1.5 \log x - c_2 (x-\mu)^2/x )f^c(x)dx \right|\\
		&\leq& c_1+ c_3\int_0^1 \frac{1}{x}f^c(x)dx + c_4 \int_1^{\infty} xf^c(x)dx\\
		&\leq& c_1+ c^3 \int_0^1 x^{-2.5}\exp (-\lambda/(2x))dx+c_4\mu<\infty,
		\end{eqnarray*}
		since $\int_0^1 x^{-2.5}\exp (-\lambda/(2x))dx<\infty$, where $c_1, c_2, c_3$ and $c_4$ are some constants. 
		
For Condition (d), since the inverse Gaussian distribution has finite third moment, we only need to show that there exists $\zeta>0$ such that $\int _0^1 x^{-\zeta-2}f^c(x)dx<\infty$, which is true, since $\int_0^1 x^{-3.5-\zeta}\exp (-\lambda/(2x))dx<\infty$ for any $\zeta>0$.

		\item {\bf Log-Normal distribution:}  The  density function $f^c(x)=\frac{1}{x\sigma \sqrt {2\pi} }\exp(-\frac{(\log x-\mu)^2}{2\sigma^2} )$. 
		
		To  show that Condition (b) holds,  is to show that
		\begin{equation}\label{lognorm}
			\left |\int_0^{\infty} \left (-\log x- (\log x-\mu)^2/(2\sigma^2)\right )
			\frac{1}{x\sigma \sqrt {2\pi} }\exp(-\frac{(\log x-\mu)^2}{2\sigma^2} )
			dx\right |<\infty
		\end{equation} 
Consider a variable transformation $y=\log x$, we have that the integral in (\ref{lognorm}) is euqal to 
$$
\left|\int _{-\infty}^{\infty} (-y-y^2/\sigma^2)\frac{1}{\sigma \sqrt{2\pi}}\exp(-\frac{(y-\mu)^2}{2\sigma^2})dy\right |, 
$$
which is finite, since the normal distribution has finite mean and variance. 

For Condition (d), we know that the third moment of log-normal distribution is finite. Therefore, we only need to show that $\int _0^ 1 x^{-2-\zeta}f^c(x)dx<\infty$.  Substituting $x$ by $e^y$, the integration is then equal to $\int _{-\infty}^{\infty}e^{(-2-\zeta)y}\phi(y)dy$, where $\phi(y)$ denotes the density function of normal distribution with parameter $\mu$ and $\sigma$. The same as calculating the moment generating function for normal distribution, we have that $\int _{-\infty}^{\infty}e^{(-2-\zeta)y}\phi(y)dy= \exp ( \mu (-2-\zeta)+\sigma^2(-2-\zeta)^2/2)$, which is finite.

\item {\bf Log-logistic distribution with shape parameter $\beta> 2$:}	The density function for log-logistic distribution is $f^c(x)=\frac{(\beta/\alpha)(x/\alpha)^{\beta-1}}{(1+(x/\alpha)^{\beta})^2}$. 

To show that Condition (b) holds, it is sufficient to show that $\int_0^{\infty}[|\log x|+|\log(1+(x/\alpha)^{\beta})|]f^c(x)dx<\infty$. Since the half logistic distribution has finite mean, see \cite{kotz2004continuous} for more details, we have that $\int_0^{\infty}|\log x|f^c(x)dx<\infty$ by variable transformation. The rest is to show that 
\begin{eqnarray} \label{loglogistic}
\int_0^{\infty}|\log(1+(x/\alpha)^{\beta})|f^c(x)dx&=& \int_0^{\alpha}|\log(1+(x/\alpha)^{\beta})|f^c(x)dx \nn\\
&+& \int_{\alpha}^{\infty}|\log(1+(x/\alpha)^{\beta})|f^c(x)dx\nn\\
&\leq&  \int_0^{\alpha} \log 2 f^c(x)dx + \int_{\alpha}^{\infty} \beta|\log x|f^c(x)dx +c
\end{eqnarray}
is finite,  where $c$ is a constant, $\int_0^{\alpha}  f^c(x)\log 2dx<\log 2$, and  $\int_{\alpha}^{\infty} \beta|\log x|f^c(x)dx<\infty$ can be shown by variable transformation and the fact that  the half logistic distribution has finite mean. 

\begin{sloppypar}

 To verify Condition (d), we notice that the k-th moment of log-logistic distribution exists  when $k<\beta$. Therefore, what left to be shown is that $\int_0^{\infty}x^{-2-\zeta}f^c(x)dx<\infty$. By substituting $x$ by $e^y$,  we have that
 $
 \int_0^{\infty}x^{-2-\zeta}f^c(x)dx=
 \int_{-\infty}^{\infty} e^{(-2-\zeta)y} \frac{e^{-\beta (y-\log \alpha)}}{\left( 1+e^{-\beta (y-\log \alpha)} \right)^2/\beta}dy=\alpha^{-2-\zeta}\text{B}(1-\frac{2+\zeta}{\beta}, 1+\frac{2+\zeta}{\beta}),
 $
  which is finite. Hence, Condition (d) holds for  $\zeta\in (0, \beta-2)$.

	\item {\bf Pearson Type V (Inverse Gamma) distribution with shape parameter $\alpha>2$:} Pearson Type V distribution is also known as the inverse Gamma distribution and Wald distribution. For $Y\sim \text{Gamma}(\alpha,\beta)$, we have that $X=1/Y\sim \text{Inv-Gamma}(\alpha,1/\beta)$, whose density function $f^c(x)=\frac{\beta^{\alpha}}{\Gamma(\alpha)}x^{-\alpha-1}\exp(-\frac{\beta}{x})$.

	It is sufficient to show Condition (b) is satisfied by  showing that $\mbox{E}_{f^c(x)}(\log X)$ and $\mbox{E}_{f^c(x)} (1/X)$ are both finite. Since $1/X=Y\sim \text{Gamma} (\alpha,\beta)$, we have that $\mbox{E}_{f^c(x)}(1/X)=\mbox{E}(Y)=\alpha\beta<\infty$. We also have
	 \begin{equation}\label{gamma1}
	 \int_{1}^{\infty} |\log x| f^c(x)dx<\int_{1}^{\infty}  x f^c(x)dx<\int_{0}^{\infty}  x f^c(x)dx<\infty,
	 \end{equation}
	 and 
	 \begin{equation}\label{gamma2}
	 \int_{0}^1 |\log x| f^c(x)dx<\int_{0}^1 \frac{ f^c(x)}{x}dx<\int_{0}^{\infty}  \frac{ f^c(x)}{x}dx=\alpha\beta<\infty.
	 \end{equation}
Combining (\ref{gamma1}) and (\ref{gamma2}), we have that $|\int_{0}^{\infty} \log x f^c(x)dx|<\infty$, which means that the inverse gamma distribution with $\alpha>2$ satisfies Condition (b).

For Condition (d), we have that
\begin{eqnarray*}
\int_0^{\infty} \max(x^{-\zeta-2},x^{\zeta+2})f^c(x)dx &<&
\int_0^{\infty} x^{-\zeta-2}f^c(x)dx+\int_0^{\infty} x^{\zeta+2}f^c(x)dx\\
&=&\int_0^{\infty} y^{\zeta+2}g(y)dy+\int_0^{\infty} x^{\zeta+2}f^c(x)dx,
\end{eqnarray*}
where $g(y)$ denotes the density function of $\text{Gamma}(\alpha, 1/\beta)$. Notice that $$\int_0^{\infty} x^{\zeta+2}f^c(x)dx=c \int _0^{\infty}x^{\zeta+1-\alpha} \exp(-\beta/x)dx<\infty,$$ 
where $c$ is a constant and $\zeta\in (0,\alpha-2)$. Also, for any $\zeta\in (0,1)$, we have that $\int_0^{\infty} y^{\zeta+2}g(y)dy<\infty$, due to the fact that  Gamma distribution has finite third moment. 
\end{sloppypar}


\item{\bf Gamma Distribution with shape parameter $\alpha>2$:}
The density function for Gamma distribution  $f^c(x)= c x^{\alpha-1} e^{-\beta x}$, where $c$ is a constant. 

To show that Condition (b) holds, it is sufficient to show that $\int_0^{\infty} x f^c(x)dx$ and $\int_0^{\infty} |\log x|f^c(x)dx$ both are finite. The first one is obviously true, since Gamma distribution has finite mean. We have that 
$\int_0^{\infty} |\log x|f^c(x)dx=\int_0^1 |\log(x)|f^c(x)dx+\int_1^{\infty} \log(x)f^c(x)dx\leq 
\int_0^1 (1/x) f^c(x)dx+\int_1^{\infty} x f^c(x)dx=\int_1^{\infty} y g(y)dy+\int_1^{\infty} x f^c(x)dx,$
 where $g(\cdot)$ here denotes the density function of $\text{Inv-Gamma}(\alpha,\beta)$. We have that  $\int_1^{\infty} y g(y)dy<\int_0^{\infty} y g(y)dy<\infty$, since the inverse-Gamma distribution has finite mean, and $\int_1^{\infty} x f^c(x)dx<\int_0^{\infty} x f^c(x)dx<\infty$, since the Gamma distribution has finite mean. 

By the transformation of $Y=1/X$, it will be the same to show that Condition (d) is satisfied for Gamma distribution with shape parameter $\alpha>2$ as for Pearson Type V (inverse Gamma) distribution with shape parameter $\alpha>2$.

\begin{sloppypar}
\item{\bf Weibull Distribution with shape parameter $k>3$: } The  density function $f^c(x)=c x^{k-1} e^{-(x/\lambda)^k}$ where $c$ is a constant. 

We have that if $\int_0^{\infty} x^k f^c(x)dx$ and $\int_0^{\infty} |\log(x)|f^c(x)dx$ are both finite, then Conditon (b) is satisfied.  Notice that
$
\int_0^{\infty} x^k f^c(x)dx=c_1 \int_0^{\infty}x^{2k-1} e^{-(x/\lambda)^k}dx=c_2\int_0^{\infty} y^{2k-1}e^{-y^k}dy= c_3\int_1^{\infty} ze^{-z}dz=2c_3/e<\infty.
$
and
\begin{eqnarray*}
\int_0^{\infty} |\log(x)|f^c(x)dx &=& \int_0^1 |\log(x)|f^c(x)dx+\int_1^{\infty} \log(x)f^c(x)dx\\
&\leq & c \int_0^1 x e^{-(x/\lambda)^k}dx + \int_1^{\infty} x f^c(x)dx\\
&\leq  & 1+ \lambda \Gamma(1+1/k)<\infty
\end{eqnarray*}

For Condition (d), we have that Weibull distribution has finite third moment.  Therefore we only need to show that $\int_0^1 x^{-2-\zeta}f^c(x)dx<\infty$ for some $\zeta\in (0,1)$, which is equivalent to show that $\int_0^1 x^{k-3-\zeta}e^{-(x/\lambda)^k}dx<\infty$. This inequality holds when $k>3$ and  $\zeta \in (0,k-3)$. 
\end{sloppypar}

\item {\bf Inverse Weibull with shape parameter $\beta>2$:} The probability density function for inverse Weibull distribution is
$$
f^c(x)= \beta \alpha^{\beta} x^{-(\beta+1)}\exp \left (-(\alpha/x)^{\beta}\right ).
$$
We have that When $X\sim \text{Inv-Weibull}(\alpha,\beta)$, where $\text{Inv-Weibull}$ stands for inverse Weibull distribution, $1/X\sim \text{Weibull}(\alpha,\beta)$.

Notice that if $\int_0^{\infty} x^{-\beta} f^c(x)dx$ and $\int_0^{\infty} |\log(x)|f^c(x)dx$ are both finite, then Condition (b) is satisfied. By transformation $y=1/x$, we have that $\int_0^{\infty} x^{-\beta} f^c(x)dx=\int_0^{\infty} y^{\beta} g(y)dy<\infty$, where $g(y)$ denotes the Weibull density function, and the inequality holds by the same argument as for Weibull distribution.
We also have that 
\begin{eqnarray*}
	\int_0^{\infty} |\log(x)|f^c(x)dx &=& \int_0^1 |\log(x)|f^c(x)dx+\int_1^{\infty} \log(x)f^c(x)dx\\
	&\leq &  \int_0^1 (1/x) f^c(x)dx + \int_1^{\infty} x f^c(x)dx\\
	&\leq  & c+ \alpha^{-1} \Gamma(1-1/\beta)<\infty,
\end{eqnarray*}
where $c$ is a constant that followed the same argument as above, and $\alpha^{-1} \Gamma(1-1/\beta)$ is the mean of the inverse Weibull distribution, see \cite{Khan:2008:TAI:1466934.1466935} for more details.

For Condition (d), we have that inverse Weibull distribution has finite third moment. Therefore we only need to show that $\int_1^{\infty} x^{2+\zeta}f^c(x)dx<\infty$ for some $\zeta\in (0,1)$, which is equivalent to show that $\int_1^{\infty} x^{1-\beta+\zeta}e^{-(\alpha/x)^{\beta}}dx<\infty$. This inequality holds when $\beta>2$ and  $\zeta \in (0,\beta-2)$. 

\item {\bf Johnson's $S_L$ distribution with parameter $\mu$ known:}  Let $Y\sim \mathcal{N}(\mu, \sigma^3)$, then $X= \sigma\exp ((Y-\gamma)/\delta)+\mu$ follows the Johnson's $S_L$ distribution, whose probability density function is
$$
f^c(x)=\frac{e^{-\frac{1}{2}(\gamma+\delta \log(\frac{x-\mu}{\sigma}))^2}\delta}
{\sqrt{2\pi}{(x-\mu)}}.
$$

We can see that random variable $X$ can be constructed by a location-scale transformation from a log-normal distributed random variable $Z=\exp ((Y-\gamma)/\delta)$. We will discuss the situation for $\mu$ is unknown in Remark 3. With $\mu$ known, we apply the DPM prior with Gamma kernel on $X-\mu$, which is supported on $[0,\infty)$. When $X-\mu$ is only scale transformed from log-normal distributed random variable, the satisfaction of Conditions (b) and (d) follows similarly to the calculation for the log-normal distribution.


    \item{\bf Pearson Type VI distribution with parameter $\alpha_1>2$ and $\alpha_2>2$:} The probability density function for this distribution is
    $$
    f^c(x)= \frac{1}{\beta \text{B}(\alpha_1, \alpha_2)}\frac{(x/\beta)^{\alpha_1-1}}{(1+x/\beta)^{\alpha_1+\alpha_2}}.
    $$

    For the satisfaction of Condition (b), it is sufficient to show that $\int_0^{\infty} |\log x|f^c(x)dx $ and $\int_0^{\infty} \log (1+x/\beta)f^c(x)dx$ are both finite. 
    We have that $\int_0^{\infty} |\log x|f^c(x)dx \leq \int_0^1 (1/x)f^c(x)dx+\int_1^{\infty }xf^c(x)dx<\infty$ for $\alpha_2>1$, since Pearson Type VI distribution has finite mean when $\alpha_2>1$. We also have that $\int_0^{\infty} \log (1+x/\beta)f^c(x)dx<\int_0^{\infty}(x/\beta) f^c(x)dx<\infty$, when $\alpha_2>1$ for the same reason. 
	
	To verify that Condition (d)  holds, we need: (i)
	$\int_0^1\frac{x^{\alpha_1-3-\zeta}}{(1+x/\beta)^{\alpha_1+\alpha_2}}dx<\infty$, which holds true for all $\zeta\in (0, \alpha_1-2)$ since $\alpha_1>2$; and (ii)
    $\int_1^{\infty}\frac{x^{\alpha_1+1+\zeta}}{(1+x/\beta)^{\alpha_1+\alpha_2}}dx<\infty$, which holds true for all $\zeta \in(0, \alpha_2-2)$ since $\alpha_2>2$. $\Box$
	\end{enumerate}

	\end{enumerate}







\end{remark}

\begin{remark}\label{unknown}

	To focus on the main ideas of this paper, we have discussed the situations where the support of the distribution for input variables are known. As pointed out by \cite{Law_2015}, the assumption that the support of underlying input model is known could be a limitation for some cases where the support is unknown. Nevertheless, with mild modification on the DPM prior we introduced in the main part of this paper,  the modified priors can be showed to satisfy the posterior consistency property even if the support is unknown. 
	
	Although the exact support of the input distribution is unknown, it is typically reasonable to assume that we know whether the support is bounded, half-bounded, or supported on the whole real line. Under such assumption, we can extend the DPM model introduced in (\ref{DPM}) with one more layer of prior on the boundary value(s) for bounded or half-bounded supports, such that the extended new model will maintain posterior consistency for the true input distributions. Those shifted version of the commonly used distributions will be consistently estimated under such an extended version of DPM. 
	
	Specifically,  the prior is defined as expression (\ref{index}). Note that such setting is different from setting the boundary as hyper-parameters of the kernel density function.  From the modeling point of view, we can set $\xi$ as hyper-parameters of the kernel densities and set $\mu$ as its prior; see \cite{Wu_Ghosal_2008} for more details on such setting. However, though we believe such setting could still lead to consistent estimation, the proof will be dramatically complicated. By \cite{Wu_Ghosal_2008}, we will need to verify Condition A2 in their Theorem 1, and cannot directly apply their Lemma 2, due to the Condition A5, mostly. To work out this situation is totally out of the scope of this paper. Hence, we show that the approach by introducing an index parameter is consistent here, which is sufficient for demonstrating that the posterior consistency can be hold for estimating input distributions without knowing exactly its support. 
	
	To achieve this, we use Lemma 1 in \cite{Wu_Ghosal_2008}. For the completeness of our argument, we cite their Lemma 1 here:
	\begin{lemma}\label{wu08lemma1}
	(Lemma 1 in \cite{Wu_Ghosal_2008}) Let $f\mid \xi \sim \Pi^*_{\xi}$, where $\xi$ is an indexing parameter following a prior $\pi$ and let $f^c$ be the true density. Suppose that there exists a set $B$ with properties $\Pi(B)>0$ and $B\subset \{\xi: f^c\in KL(\Pi^*_{\xi})\}$. Then $f^c\in \Pi^*$, where the prior $\Pi^*=\int \Pi_{\xi}^*d\pi(\xi)$. 
	\end{lemma}
	
When the support is bounded but the boundary is unknown, we use a DPM with location-scale transformed Beta kernel as the prior. More specifically, the prior is set as \eqref{index}, where the kernel function is the transformed Beta density with parameters for boundaries 
\begin{equation}\label{transbeta}
h(x;\omega, \beta, a_1, a_2)=\frac{\Gamma(\omega,\beta)}{\Gamma(\omega)\Gamma(\beta)}\left( \frac{x-a_1}{a_2-a_1}\right)^{\omega-1}\left(1-\frac{x-a_1}{a_2-a_1}\right)^{\beta-1},
\end{equation}
 and the base distribution $G_0$ for Dirichlet Process is specified as \eqref{eq.Betaprior}.

	{\sc Proof of Corollary 1: } 
	Let $\mathcal{B}=[a_1-\delta,a_1]\times[a_2, a_2+\delta]$. Then by Lemma \ref{wu08lemma1}, it is sufficient to prove this corollary by showing that for any $\xi=(a_1-\lambda, a_2+\lambda)\in \mathcal{B}$ (with $0<\lambda\leq \delta$), $f^c\in KL(\Pi^*_{\xi})$.  Denote the true density function on  $[a_1-\lambda, a_2+\lambda]$ by $f^c(x)$, which is $f^c(x)= f^c(x)$ for $x\in [a_1,a_2]$ and $f^c(x)=0$ otherwise. If $f^c(a_1)=f^c(a_2)=0$, then we can directly apply Theorem \ref{thm:consistency} Part (iii), and prove the corollary. 
\begin{sloppypar}

When  $f(a_1)>0$ or $f(a_2)>0$, similar to Lemma \ref{wg08lemma4}, we need to construct a new density function $f_1(x)$ to approximate $f^c(x)$. 	For a given $m>0$, define $f_3(x)=[(x-a_1+m)/m]f^c(a_1)$ when $x\in [a_1-m,a_1]$, and $f_4(x)=[(a_2+m-x)/m]f^c(a_2)$ when $x\in [a_2,a_2+m]$. Let $f_2(x)=f^c(x)+f_3(x)+f_4(x)$. Then we define $f_1(x)=\frac{\max (f_2(x), m)}{\int_{a_1-\lambda}^{a_2+\lambda}\max (f_2(x), m)dx}.$
Notice that
\begin{align*}
& \scalebox{0.98}{$
\int_{a_1-\lambda}^{a_2+\lambda}\max (f_2(x), m)dx =  \int_{a_1}^{a_2}\max (f_2(x), m)dx + \int_{a_1-\lambda}^{a_1} \max(f_3(x),m) dx + \int_{a_2}^{a_2+\lambda} \max(f_4(x),m) dx $} \\
&\leq \int_{a_1}^{a_2}\max (f_2(x), m)dx + \int_{a_1-\lambda}^{a_1} (f_3(x)+m) dx + \int_{a_2}^{a_2+\lambda} (f_4(x)+m) dx \\
&= \scalebox{1.1}{$\int_{a_1}^{a_2}\max (f_2(x), m)dx + \frac{m^2-[\max(m-\lambda,0)]^2}{2m}f^c(a_1) + m\lambda + \frac{m^2-[\max(m-\lambda,0)]^2}{2m}f^c(a_2) + m\lambda $} \\
&\leq \int_{a_1}^{a_2}\max (f_2(x), m)dx+(f^c(a_1)+f^c(a_2))m/2+2m\lambda.
\end{align*}
If we let $c=\int_{a_1}^{a_2}\max (f_2(x), m)dx+(f^c(a_1)+f^c(a_2))m/2+2 m\lambda$, then by this construction, the denominator in the definition of $f_1(x)$ is less than or equal to $c$, which means that $f^c(x) \leq f_2(x) \leq \max(f_2(x),m)\leq cf_1(x)$. Moreover, $c\to 1$ when $m\to 0$. This implies that: (i) $f_1>m>0$ is continuous on $[a_1-\lambda, a_2+\lambda]$ for some $m>0$, and (ii) for any $\epsilon>0$, there exists a $c>0$, such that $(1+c)\log c<\epsilon$ and $cf_1(x)>f^c(x)$. Then such $f_1(x)$ will satisfy the requirements in the proof for Theorem \ref{thm:consistency} Part (iii). The conclusion of Corollary \ref{c-b} follows from the proof of Theorem \ref{thm:consistency} Part (iii). 
$\Box$
\end{sloppypar}

Before we show that weak consistency for the DPM prior will hold under some mild conditions when the true density functions have half bounded support and unknown end points, we need to extend the Lemma \ref{wg08lemma4} to  the following form:

\begin{lemma}\label{lemma:kl3}
	Let $f^c(x)$  be a continuous  and bounded density with support on $[a_0, \infty)$. Let $p$ be a prior on $\mathscr D_{a_0-\delta}$, the space of all densities supported on $[a_0-\delta, \infty)$, where $\delta>0$. For any $\epsilon>0$, there exist $m>0$ and $f_1(x)\geq m>0$ for $x\in [a_0-\delta,a_0]$ such that $p(\mathcal K_{\epsilon}(f_1))>0$ implies that $p(\mathcal K_{3(\epsilon+\sqrt{\epsilon})}(f^c))>0$.
\end{lemma} 
	{\sc Proof. } 	Let $f_3(x)=[(x-a_0+m)/m]f^c(a_0)$ when $x\in [a_1-m,a_1]$, and let $f_2(x)=f^c(x)+f_3(x)$. Then we define
	$$
	f_1(x)=\left \{
	\begin{array} {lr}
	\frac{\max (f_2(x), m)}{\int_{a_0-\delta}^{a_0}\max (f_2(x), m)dx+\int_{a_0}^{\infty}f^c(x)dx}, &  \mbox{for } x\in [a_0-\delta,a_0];\\
	\frac{f^c(x)}{\int_{a_0-\delta}^{a_0}\max (f_2(x), m)dx+\int_{a_0}^{\infty}f^c(x)dx},& \mbox{for } x \in (a_0, \infty).
	\end{array}
	\right.
	$$
	By Lemma 5.1 in \cite{ghosal1999consistent}, we have $\mathcal K(f^c; f)\leq (c+1)\log c + c[\mathcal K(f_1; f)+\sqrt {\mathcal K(f_1; f)}]$, where $\mathcal K(f_1;f_2)=\int f_1\log (f_1/f_2)$ is the Kullback-Leibler divergence and $c=\int_{a_0-\delta}^{a_0}\max (f_2(x), m)dx+\int_{a_0}^{\infty}f^c(x)dx\leq 1+ f^c(a_0)m/2+\delta m$. We have that $c\to 1$ when $m\to 0$. Therefore, for any given $\epsilon>0$, there exists sufficiently small $m>0$ such that $(c+1)\log c<\epsilon$ and $c<2$. As a result, for any $\epsilon>0$, $\mathcal K(f_1; f)<\epsilon$ implies that $\mathcal K(f^c; f) < \epsilon+2(\epsilon+\sqrt{\epsilon}) < 3(\epsilon+\sqrt{\epsilon})$, which implies that $\mathcal K_{3(\epsilon+\sqrt{\epsilon})}(f^c) \supseteq \mathcal K_{\epsilon}(f_1)$ and hence $p(\mathcal K_{3(\epsilon+\sqrt{\epsilon})}(f^c))\geq p(\mathcal K_{\epsilon}(f_1)) >0$.
	$\Box$
	
When the support is half bounded but the boundary is unknown, we use a DPM with location shifted Gamma kernel as the prior. More specifically, the prior is set as  \eqref{index}, where the kernel function is the shifted Gamma density with parameters for boundaries 
\begin{equation}\label{shiftGamma}
h(x;V,u, a_0)=\frac{1}{\Gamma(V)(u/V)^V}(x-a_0)^{V-1}e^{-xV/u},
\end{equation}
and the base distribution $G_0$ for Dirichlet Process is specified as (\ref{eq.priorGamma}).

 The proof of Corollary \ref{C-G} follows the same approach as the one for Corollary \ref{c-b}, by using Lemma \ref{lemma:kl3} (which is adapted from Lemma \ref{wg08lemma4}), and hence we omit the detailed proof here.  Also notice that this corollary will apply to all commonly used distributions on $[0,\infty)$ discussed above in their shifted version. 	
\end{remark}

\begin{remark} \label{remark:4}	Pareto distribution is also commonly used in simulation for modeling input data. It has a location parameter, which makes it not be supported on $[0,\infty)$. Moreover, to obtain the consistency result, we need some limits on the parameters of this distribution and some extra conditions on the parameters chosen for DPM prior with Gamma density kernel.  We use the Pareto distribution as an example to demonstrate that distributions like this can still be consistently estimated through a DPM prior without knowing the exact boundary value. If we have reasonably strong suspect that the true distribution is Pareto distribution or of similar type, then the exponential kernel and the location-scale uniform kernel are better choices than the Gamma kernel in the DPM prior.

{\bf Pareto Distribution: } The probability density function for this distribution is  
	$$
	f^c(x)=\frac{\alpha x_m^{\alpha}}{x^{\alpha+1}}, \quad \text{ for } x>x_m.
	$$
Assume that $x_m$ is known. To show the consistency holds for this distribution when the DPM kernel function is either exponential or scaled uniform density, we need to verify the conditions in Theorems 16 and 17 in \cite{Wu_Ghosal_2008}. First, we need $\int |\log f^c(x)|f^c(x)dx<\infty$, which is obviously true for the Pareto distribution. While using exponential density kernel, we need $\int xf^c(x)dx<\infty$, which is true when $\alpha>1$, since Pareto distribution has finite mean when $\alpha>1$. Otherwise, we can use the scaled uniform density kernel, which does not require finite mean for consistency, but requires that the true density function must be continuous and decreasing, which is satisfied by the Pareto distribution. For using the DPM with exponential kernel, one last requirement is that $\overline F^c(x)=1-F^c(x)$ need to be completely monotone, which is also true for the Pareto distribution.

Furthermore, all the following distributions have completely monotone $\overline F(x)$: the exponential distribution, the Weibull distributions with shape parameter less than 1, and the Gamma distributions with shape parameter less than or equal to 1. If the true density $f^c$ is one of these distributions, then using the exponential kernel in our DPM model can lead to posterior consistency at $f^c$.  See	 \cite{chiu2014complete} for more about the completely monotone property. 

To extend the result to the situation when $x_m$ is unknown, we can use the same approach as in Remark \ref{unknown}. More specifically, we can treat $x_m$ as an index parameter, assign a prior $\pi$ on it, and set the complete prior as \eqref{index} with either the exponential or the scaled uniform density as the kernel. Then as long as $\pi[a_0-\delta,a_0]>0$, where $a_0$ is the true value for the parameter $x_m$, the posterior is weakly consistent at $F^c(x)$.

\end{remark}

\section{{Appendix: Asymptotic Properties of Bayesian Nonparametric Framework}}
\label{subsec:AppendixAsympticPropertiesProof}
To prove Theorem \ref{thm:cribound}, we first introduce some useful definitions and lemmas. In the following, with a little bit abuse of notation, for a generic random variable $U$, we use $F_U$ to denote both its distribution and its cdf. We first introduce the Wasserstein-$p$ distance (or Mallow's metric).
\begin{definition}\label{wasserstein}
(\cite{Vil08} Definition 6.1) Let $U$ and $V$ be two random variables whose marginal distributions are $F_U$ and $F_V$. Then the Wasserstein-$p$ distance is
\begin{align*}
d_p(U,V) & = \left(\inf_{G \in \mathcal{C}(F_U,F_V)} \int_{\mathbb{R}\times \mathbb{R}} |u-v|^p d G(u,v) \right)^{1/p},
\end{align*}
where $\mathcal{C}(F_U,F_V)$ represents the set of all joint distributions on $\mathbb{R}\times \mathbb{R}$ with marginal distributions $F_U$ and $F_V$.
\end{definition}
Such a joint distribution $G(u,v)$ is usually called a \textit{coupling} of $(F_U,F_V)$. Based on this definition, if $F$ represents a generic distribution on $\mathbb{R}$, one can define the Wasserstein-$p$ space $\mathcal{M}_p=\{F: \int_{\mathbb{R}}|x-x_0|^p dF(x)<\infty\}$ for some $x_0\in \mathbb{R}$ according to Definition 6.4 of \cite{Vil08}. Furthermore, \cite{Vil08} has shown that this definition does not depend on the choice of $x_0$. In fact, $\mathcal{M}_p$ is the set of all distributions on $\mathbb{R}$ with finite $p$th moment.

It is well known that the Wasserstein-$p$ distance metricizes the space of $\mathcal{M}_p$. In particular, we have the following lemma.
\begin{lemma}\label{villemma}
(\cite{Vil08} Theorem 6.8) Let $\{F_{\ell}:\ell=1,2,\ldots\}$ be a sequence of distributions on $\mathbb{R}$. Let $F_0$ be a distribution on $\mathbb{R}$. If $F_0,F_1,F_2,\ldots, \in \mathcal{M}_p$ with $1\leq p<\infty$, then the weak convergence of $F_{\ell}$ to $F_0$ is equivalent to $d_p(F_{\ell},F_0)\to 0$ as $\ell\to\infty$.
\end{lemma}

Now we cite a key result that relate the Wasserstein distance on $\mathbb{R}$ to the quantile function. For any $u\in [0,1]$, let $F^{-1}(u) = \inf\{x\in \mathbb{R}: F(x)\geq u\}$ be the quantile function of $F$ (the cdf $F(x)$ is assumed to be a right-continuous function with left limit, by convention). Note that this definition works for both discrete and continuous distributions. Then we have the following property.
\begin{lemma}\label{biclemma1}
(\cite{BicFre81} Lemma 8.2) For any pairs of distributions $F_U,F_V\in \mathcal{M}_p$ with $1\leq p<\infty$,
$$d_{p} (F_U,F_V) = \left(\int_0^1 |F_U^{-1}(u)-F_V^{-1}(u)|^p du\right)^{1/p}.$$
\end{lemma}

We also need the following result on the convergence of quantile functions.
\begin{lemma}\label{quantlem}
(\cite{vvt98} Lemma 21.2) Let $\{F_{\ell}:\ell=1,2,\ldots\}$ be a sequence of distributions on $\mathbb{R}$. Let $F_0$ be a distribution on $\mathbb{R}$. Then $F_{\ell}$ converges to $F_0$ weakly as $\ell\to\infty$ if and only if $\lim_{\ell \to\infty}F_{\ell}^{-1}(u)= F_0^{-1}(u)$ for any continuity point $u\in [0,1]$ of $F_0^{-1}$.
\end{lemma}

An immediate consequence of Lemma \ref{villemma} and Lemma \ref{quantlem} is the following.
\begin{lemma}\label{translem}
Let $\{F_{\ell}:\ell=1,2,\ldots\}$ be a sequence of distributions on $\mathbb{R}$. Let $F_0$ be a distribution on $\mathbb{R}$. Suppose that $F_0,F_1,F_2,\ldots\in \mathcal{M}_p$ with $1\leq p<\infty$. Then $d_p(F_{\ell},F_0)\to 0$ as $\ell\to\infty$ if and only if $\lim_{\ell \to\infty}F_{\ell}^{-1}(u)= F_0^{-1}(u)$ for any continuity point $u\in [0,1]$ of $F_0^{-1}$.
\end{lemma}

Lemma \ref{translem} allows us to directly connect the convergence in Wasserstein-$p$ distance to the convergence of quantiles.

We cite another important theorem from \cite{Jing2018}.
\begin{lemma}\label{jinglem}
(A special case of \cite{Jing2018} Theorem 3.1) Let $U_1,\ldots,U_N$ be i.i.d. random variables from a distribution $F_0$ on $\mathbb{R}$. Let $\hat F_N$ be the empirical distribution of $U_1,\ldots,U_N$. Suppose that $\int_{\mathbb{R}} |u|^2 dF_0(u)<\infty$. Then $\mbox{E}[d_1(\hat F_N,F_0)] \leq C_1\int_{\mathbb{R}} |u|^2 dF_0(u) \cdot N^{-1/2} $, where $d_1$ is the Wasserstein-1 distance, and $C_1$ is an absolute positive constant that does not depend on $F_0$.
\end{lemma}
Lemma \ref{jinglem} is a special case of Theorem 3.1 of \cite{Jing2018}. In particular, we take $d=1$, $p=1$, and $q=2$ and simplify the upper bounds in their theorem.

%
%
%

\begin{lemma}\label{lemma:levy}
If the posterior $p(F|\mathbf{X}_m)$ is weakly consistent at $F^c$, then for any given $\epsilon_1>0, \epsilon_2>0,\epsilon_3>0$, there exists a sufficiently large integer $M_0$, such that for all $m>M_0$,
\begin{equation}
	\text{P}_{F^c}\left[\text{P}\Big(\left\{d_{LP}(\widetilde{F},F^c) > \epsilon_1\right\}~\big|~\mathbf{X}_m\Big)>\epsilon_2 \right]<\epsilon_3, \nonumber
\end{equation}	
where $\widetilde{F}\sim p\left(F\big|\mathbf{X}_m\right)$ and $\text{P}_{F^c}$ denotes the probability measure of $\mathbf{X}_m$, i.e., the measure of the true input distribution $F^c$.
\end{lemma}

\noindent {\bf Proof of Lemma \ref{lemma:levy}:}

We show that the consistency in weak neighborhood is sufficient to imply the consistency in L\'{e}vy-Prokhorov metric, which is what we need for showing asymptotic properties of the proposed simulation method based on nonparametric Bayesian framework.
For the true probability distribution $F^c$ (with density $f^c$) on $\Re$, let $W_{\epsilon}=\{F : d_{LP}(F,F^c)<\epsilon\}$ denote an open neighborhood of $F^c$ in L\'{e}vy-Prokhorov metric. By the definition of  L\'{e}vy-Prokhorov metric, we have that
\begin{eqnarray}
W_{\epsilon} &=& \cap_{A \in \mathcal{B}(\Re)}\{P: P_{f^c}(A)+\epsilon \geq P(A^{\epsilon}) \text{ and }  P(A^{\epsilon}) \geq P_{f^c}(A)-\epsilon  \} \nn\\
&=&\cap_{A \in \mathcal{B}(\Re)}\{P: P_{f^c}(A)+\epsilon+P(A^{\epsilon}\backslash A) \geq P(A) \geq P_{f^c}(A)-\epsilon - P(A^{\epsilon}\backslash A) \}\nn\\
&\supseteq& \cap _{A \in \mathcal{B}(\Re)} \{P: P_{f^c}(A)+\epsilon \geq P(A) \geq P_{f^c}(A)-\epsilon \} \equiv W_{\epsilon}^*,
\end{eqnarray}
where $\mathcal{B}(\Re)$ is the collection of all Borel sets on $\Re$, $P_{f^c}$ and $P$ are the probability associated with distributions $F^c$ and $F$ respectively, and $W_{\epsilon}^*$ defined as above is a weak neighborhood of $F^c$. Since weak consistency implies that $p(W_{\epsilon}^*\mid X_1,\ldots, X_n)$ converges to $1$ in $P_{f^c}$-probability as $n\to \infty$ for any $\epsilon>0$, we have that for any $\epsilon>0$, $1\geq p(W_{\epsilon}\mid X_1,\ldots, X_n) \geq p(W_{\epsilon}^*\mid X_1,\ldots, X_n)$,  which implies that the posterior probability on any  L\'{e}vy-Prokhorov neighborhood of the true distribution $F^c$ converges to $1$ in $P_{f^c}$-probability as $n\to \infty$. We write this relation in $\epsilon-
 \delta$ language and the lemma follows. \hfill $\Box$

\vspace{6mm}

\noindent {\bf {Proof of Theorem \ref{thm:cribound} (i):}}

For abbreviation, we write $\mu_b=\mu(\widetilde{F}^{(b)})$ and $\sigma_b^2=\sigma^2_e(\widetilde{F}^{(b)})$ for $b=1,\ldots,B$. We first rank the means of simulation outputs $\{\bar Y_{b}\}_{b=1}^B$ as $\bar Y_{(1)}< \bar Y_{(2)}< \ldots < \bar Y_{(B)}$.  Suppose $(k_1,k_2,\ldots,k_B)$ is the permutation of integers $(1,2,\ldots,B)$ such that $\bar Y_{k_b}=\bar Y_{(b)}$ for $b=1,2,\ldots,B$. In other words, $k_b$ is the original subscript of $\bar Y_{(b)}$ before they are ranked. We define a sequence with subscript ``$(b)$" as the same sequence with the original subscript $k_b$, i.e., $\mu_{(b)}=\mu_{k_b}, ~\sigma^2_{(b)} =\sigma^2_{k_b}, ~ n_{(b)}=n_{k_b}$ for $b=1,2,\ldots,B$. In this way, we have that for $b=1,\ldots,B$,
\begin{align}\label{tmunorm2}
{
\text{E}\left[\bar Y_{(b)} | \mathbf{X}_m,\mathcal{F}_B\right] = \mu_{(b)},\quad  \text{Var}\left[\bar Y_{(b)} | \mathbf{X}_m,\mathcal{F}_B\right] =\frac{\sigma^2_{(b)}}{n_{(b)}}.
}
\end{align}
Let $F_{\bar Y,B}$ be the empirical distribution of the ``sample" $\{\bar Y_{b}\}_{b=1}^B$. Let $F_{\mu,B}$ be the empirical distribution of the ``sample" $\{\mu_b\}_{b=1}^B$. By Condition (1), since the posterior distribution of $W$ is continuous, we have that with probability 1, all values of $\{\mu_b\}_{b=1}^B$ are distinct. 

Then both $F_{\bar Y,B}$ and $F_{\mu,B}$ are discrete distributions supported on at most $B$ points. Hence both $F_{\bar Y,B}$ and $F_{\mu,B}$ have finite 2nd moments almost surely, which means that they lie in $\mathcal{M}_2$. We can see that the set of all couplings of $(F_{\bar Y,B},F_{\mu,B})$ is given by the set of all bivariate probability distributions $G(u,v)$ in the set
\begin{align}
\mathcal{C}(F_{\bar Y,B},F_{\mu,B}) = &\Big\{G(u,v)=\sum_{b_1=1}^B\sum_{b_2=1}^Bw_{b_1b_2}\delta_{(\bar Y_{b_1}, \mu_{b_2})}(u,v): \nonumber \\
&~w_{b_1b_2}\geq 0, \text{ for } b_1=1,\ldots,B,\text{ and }b_2=1,\ldots,B, \nonumber \\
&\text{and } \sum_{b_1=1}^B w_{b_1b_2} = \frac{1}{B}, \text{ for } b_2=1,\ldots,B, \nonumber \\
&\text{and } \sum_{b_2=1}^B w_{b_1b_2} = \frac{1}{B}, \text{ for } b_1=1,\ldots,B\Big\}.
\end{align}
In other words, any $G\in \mathcal{C}(F_{\bar Y,B},F_{\mu,B})$ is supported on at most $B^2$ points. Now we look at a particular coupling $G_{\circ}(u,v)=\frac{1}{B}\sum_{b=1}^B \delta_{(\bar Y_b, \mu_b)}$, i.e., $G_{\circ}$ is the empirical measure supported on the $B$ original pairs $\{(\bar Y_b, \mu_b)\}_{b=1}^B$ with no misalignment (here $\delta_x$ stands for the Dirac measure at the point $x$). Then $G_{\circ}\in \mathcal{C}(F_{\bar Y,B},F_{\mu,B})$. By Definition \ref{wasserstein}, we have that for $p=1$,
\begin{align}\label{w2diff1}
& d_1 \left(F_{\bar Y,B},F_{\mu,B}\right)  =\inf_{G \in \mathcal{C}(F_{\bar Y,B},F_{\mu,B})} \int_{\mathbb{R}\times \mathbb{R}} |u-v| d G(u,v) \nonumber \\
&\leq \int_{\mathbb{R}\times \mathbb{R}} |u-v| d G_{\circ}(u,v)  =\frac{1}{B} \sum_{b=1}^B \left|\bar Y_b - \mu_b\right|
\leq \left(\frac{1}{B} \sum_{b=1}^B \left|\bar Y_b - \mu_b\right|^2 \right)^{1/2}.
\end{align}
The first inequality follows because the infimum over $\mathcal{C}(F_{\bar Y,B},F_{\mu,B})$ is always no larger than one particular element in $\mathcal{C}(F_{\bar Y,B},F_{\mu,B})$ (in this case, $G_{\circ}$). The second inequality is a simple application of Cauchy-Schwarz inequality.

According to Condition (2), $\sigma^2_b\leq C^2_{\sigma}$ for all $b=1,\ldots,B$, \eqref{tmunorm2} and \eqref{w2diff1} together imply that
\begin{align}\label{w2diff2}
&\text{E}\left[d_1^2 \left(F_{\bar Y,B},F_{\mu,B}\right) ~\Big|~ \mathbf{X}_m,\mathcal{F}_B \right]
\leq \text{E}\left[ \frac{1}{B} \sum_{b=1}^B \left|\bar Y_b - \mu_b\right|^2~\Big|~ \mathbf{X}_m,\mathcal{F}_B \right] \nonumber \\
&=  \frac{1}{B} \sum_{b=1}^B \text{E}\left[ \left|\bar Y_b - \mu_b\right|^2~\Big|~ \mathbf{X}_m,\mathcal{F}_B\right]
= \frac{1}{B} \sum_{b=1}^B \text{E}\left[ \frac{\sigma^2_b}{n_b} ~\Big|~ \mathbf{X}_m,\mathcal{F}_B\right] \leq \frac{C^2_{\sigma}}{n_{\min}}.
\end{align}
Since this upper bound does not depend on $\mathcal{F}_B$, we can remove the condition on $\mathcal{F}_B$ by taking iterated expectation:
\begin{align}\label{w2diff3}
&\text{E}\left[d_1^2 \left(F_{\bar Y,B},F_{\mu,B}\right) ~\Big|~ \mathbf{X}_m \right] = \text{E} \left\{\text{E}\left[d_1^2 \left(F_{\bar Y,B},F_{\mu,B}\right) ~\Big|~ \mathbf{X}_m,\mathcal{F}_B \right] \right\}
 \leq \frac{C^2_{\sigma}}{n_{\min}}.
\end{align}
Since $\mu_b=\mu(\tilde F^{(b)})$ and $\tilde F^{(b)}$ ($b=1,\ldots,B$) are random draws from the posterior $p(F|\mathbf{X}_m)$, we have that $\{\mu_b\}_{b=1}^B$ is a random sample of the random variable $W=\mu(\tilde F)$ with $\tilde F\sim p(F|\mathbf{X}_m)$. Now we invoke Lemma \ref{jinglem} and obtain that conditional on $\mathbf{X}_m$,
\begin{align}\label{w2diff4}
\text{E}\left[ d_1(F_{\mu,B}, F_W(\cdot|\mathbf{X}_m)) ~\big|~ \mathbf{X}_m\right] \leq \frac{C_1\int_{\mathbb{R}} w^2 dF_W(w|\mathbf{X}_m)}{\sqrt{B}}.
\end{align}
According to Condition (1), $\int_{\mathbb{R}} w^2 dF_W(w|\mathbf{X}_m)$ is almost surely finite, so is the upper bound in \eqref{w2diff4}.

Now we combine \eqref{w2diff3} and \eqref{w2diff4}, and use the triangle inequality and Cauchy-Schwarz inequality to obtain that
\begin{align}\label{w2diff5}
& \text{E}\left[ d_1(F_{\bar Y,B}, F_W(\cdot|\mathbf{X}_m)) ~\big|~ \mathbf{X}_m\right] \leq \text{E}\left[ d_1(F_{\bar Y,B}, F_{\mu,B}) ~\big|~ \mathbf{X}_m\right] + \text{E}\left[ d_1(F_{\mu,B}, F_W(\cdot|\mathbf{X}_m)) ~\big|~ \mathbf{X}_m\right] \nonumber \\
&\leq \sqrt{\text{E}\left[ d_1^2(F_{\bar Y,B}, F_{\mu,B}) ~\big|~ \mathbf{X}_m\right]} + \text{E}\left[ d_1(F_{\mu,B}, F_W(\cdot|\mathbf{X}_m)) ~\big|~ \mathbf{X}_m\right] \nonumber \\
&\leq\frac{C_{\sigma}}{\sqrt{n_{\min}}} + \frac{C_1\int_{\mathbb{R}} w^2 dF_W(w|\mathbf{X}_m)}{\sqrt{B}}.
\end{align}
We note that for any $u\in (0,1)$, $F_{\bar Y,B}^{-1} (u) = \bar Y_{\lceil uB\rceil}$ since $F_{\bar Y,B}$ is a discrete distribution supported on $\{\bar Y_b\}_{b=1}^B$, and $F_W^{-1}(u|\mathbf{X}_m) = q_W\left(u|\mathbf{X}_m\right)$. By Lemma \ref{biclemma1}, we have that
\begin{align}\label{w2diff6}
& \text{E}\left[ d_1(F_{\bar Y,B}, F_W(\cdot|\mathbf{X}_m)) ~\big|~ \mathbf{X}_m\right] = \text{E}\left[\int_0^1 \left|F_{\bar Y,B}^{-1} (u)- F_W^{-1}(u|\mathbf{X}_m)\right| du ~\Big|~ \mathbf{X}_m\right]  \nonumber \\
&= \text{E}\left[\int_0^1 \left|\bar Y_{\lceil uB\rceil} - q_W(u|\mathbf{X}_m)\right| du ~\Big|~ \mathbf{X}_m\right] \leq
\frac{C_{\sigma}}{\sqrt{n_{\min}}} + \frac{C_1\int_{\mathbb{R}} w^2 dF_W(w|\mathbf{X}_m)}{\sqrt{B}},
\end{align}
where the last inequality follows from \eqref{w2diff5}. 

Now for the Hausdorff distance we have considered, we have the following relation
\begin{align}\label{haus1}
& \int_0^{1} d_{H}\left(\left[\bar Y_{\left(\lceil(\alpha^*/2)B\rceil\right)},
\bar Y_{\left(\lceil(1-\alpha^*/2)B\rceil\right)}\right], \left[q_W\left(\alpha^*/2|\mathbf{X}_m\right) ,q_W\left(1-\alpha^*/2|\mathbf{X}_m\right)\right] \right) d\alpha^* \nonumber \\
& = \int_0^{1} \left[\left|\bar Y_{\left(\lceil(\alpha^*/2)B\rceil\right)} - q_W\left(\alpha^*/2|\mathbf{X}_m\right)\right|\right] d\alpha^* + \int_0^{1} \left[\left|\bar Y_{\left(\lceil(1-\alpha^*/2)B\rceil\right)} - q_W\left(1-\alpha^*/2|\mathbf{X}_m\right)\right|\right] d\alpha^*  \nonumber \\
& = 2\int_0^{1/2} \left[\left|\bar Y_{\left(\lceil uB\rceil\right)} - q_W\left(u|\mathbf{X}_m\right)\right|\right] du + 2\int_{1/2}^1 \left[\left|\bar Y_{\left(\lceil vB\rceil\right)} - q_W\left(v|\mathbf{X}_m\right)\right|\right] dv  \nonumber \\
& = 2\int_0^{1} \left[\left|\bar Y_{\left(\lceil uB\rceil\right)} - q_W\left(u|\mathbf{X}_m\right)\right|\right] du.
\end{align}
Therefore, \eqref{w2diff6} and \eqref{haus1} together imply that
\begin{align}\label{w2diff7}
&\text{E}\left[ \int_0^{1} d_{H}\left(\left[\bar Y_{\left(\lceil(\alpha^*/2)B\rceil\right)},
\bar Y_{\left(\lceil(1-\alpha^*/2)B\rceil\right)}\right], \left[q_W\left(\alpha^*/2|\mathbf{X}_m\right) ,q_W\left(1-\alpha^*/2|\mathbf{X}_m\right)\right] \right) d\alpha^* ~\Big|~ \mathbf{X}_m\right] \nonumber\\
&\leq \frac{2C_{\sigma}}{\sqrt{n_{\min}}} + \frac{2C_1\int_{\mathbb{R}} w^2 dF_W(w|\mathbf{X}_m)}{\sqrt{B}}.
\end{align}
This has proved \eqref{eq.dhorder} in Part (i) of Theorem \ref{thm:cribound}.
\vspace{3mm}


\begin{sloppypar}
Next we prove \eqref{eq.CrIconverge}. We first notice that according to Condition (1), given $\mathbf{X}_m$, $F_W(w|\mathbf{X}_m)$ is a strictly increasing continuous cdf on its support. Therefore, its inverse $F_W^{-1}(u|\mathbf{X}_m)$ is also a strictly increasing function for all $u\in(0,1)$. By Lemma \ref{translem}, for any given number $\delta>0$ and any given $\gamma \in (0,1)$, there exists $\zeta=\zeta(\delta,\gamma,\mathbf{X}_m)>0$, such that
\begin{align}\label{w2diff8}
&d_1(F_{\bar Y,B}, F_W(\cdot|\mathbf{X}_m))< \zeta \implies |\bar Y_{(\lceil\gamma B\rceil)} - q_W(\gamma|\mathbf{X}_m)|< \frac{\delta}{2}.
\end{align}
Conditional on the input data $\mathbf{X}_m$, for any $\zeta$ as above and any given $\epsilon>0$, we can set $B_0=B_0(\zeta(\delta,\gamma,\mathbf{X}_m),\epsilon,\mathbf{X}_m)=\lceil 16C_1^2(\int_{\mathbb{R}} w^2dF_W(w|\mathbf{X}_m))^2/(\zeta^2\epsilon^2)\rceil$ and $n_{\min,0}=n_{\min,0}(\zeta(\delta,\gamma,\mathbf{X}_m),\epsilon,\mathbf{X}_m)=\lceil 16C_{\sigma}^2/(\zeta^2\epsilon^2)\rceil$, such that by Markov's inequality and Equation \eqref{w2diff6}, for all $B>B_0$ and $n_{\min}>n_{\min,0}$,
\begin{align}\label{w2diff9}
&\text{P}\left(d_1(F_{\bar Y,B}, F_W(\cdot|\mathbf{X}_m)) \geq  \zeta ~\Big|~ \mathbf{X}_m \right) \leq \frac{\text{E}\left[d_1(F_{\bar Y,B}, F_W(\cdot|\mathbf{X}_m))~\Big|~ \mathbf{X}_m\right]}{\zeta} \nonumber \\
&\leq \frac{1}{\zeta} \left(\frac{2C_{\sigma}}{\sqrt{n_{\min}}} + \frac{2C_1\int_{\mathbb{R}} w^2 dF_W(w|\mathbf{X}_m)}{\sqrt{B}}\right) \nonumber \\
&< \frac{1}{\zeta} \left(\frac{C_{\sigma}}{\sqrt{16 C_{\sigma}^2/(\zeta^2\epsilon^2)}} + \frac{C_1\int_{\mathbb{R}} w^2 dF_W(w|\mathbf{X}_m)}{\sqrt{16 C_1^2(\int_{\mathbb{R}} w^2dF_W(w|\mathbf{X}_m))^2/(\zeta^2\epsilon^2)}}\right) \nonumber\\
&= \frac{\epsilon}{4} + \frac{\epsilon}{4} = \frac{\epsilon}{2}.
\end{align}
\eqref{w2diff8} and \eqref{w2diff9} together imply that for all $B>B_0$ and $n_{\min}>n_{\min,0}$,
\begin{align}\label{w2diff10}
&\text{P}\left(|\bar Y_{(\lceil\gamma B\rceil)} - q_W(\gamma|\mathbf{X}_m)| \geq  \frac{\delta}{2} ~\Big|~ \mathbf{X}_m \right) \leq \text{P}\left(d_1(F_{\bar Y,B}, F_W(\cdot|\mathbf{X}_m)) \geq \zeta ~\Big|~ \mathbf{X}_m \right) <\frac{\epsilon}{2}.
\end{align}
Now in \eqref{w2diff10}, we replace $\gamma$ by both $\alpha^*/2$ and $1-\alpha^*/2$, and let $B_1=\max\{B_0(\zeta(\delta,\alpha^*/2,\mathbf{X}_m),\epsilon,\mathbf{X}_m),B_0(\zeta(\delta,1-\alpha^*/2,\mathbf{X}_m),\epsilon,\mathbf{X}_m)\}$, and\\ $n_{\min,1}=\max\{n_{\min,0}(\zeta(\delta,\alpha^*/2,\mathbf{X}_m),\epsilon,\mathbf{X}_m),n_{\min,0}(\zeta(\delta,1-\alpha^*/2,\mathbf{X}_m),\epsilon,\mathbf{X}_m)\}$. Then from \eqref{w2diff10}, we have that for all $B>B_1$ and $n_{\min}>n_{\min,1}$,
\begin{align}\label{CrIdiff1}
&~~~ \text{P}\Big( d_{H}\Big(\big[\bar Y_{\left(\lceil(\alpha^*/2)B\rceil\right)},
\bar Y_{\left(\lceil(1-\alpha^*/2)B\rceil\right)}\big], \big[q_W\left(\alpha^*/2|\mathbf{X}_m\right) ,q_W\left(1-\alpha^*/2|\mathbf{X}_m\right)\big] \Big) \geq  \delta ~\Big|~ \mathbf{X}_m \Big) \nonumber \\
& = \text{P}\Big( \big|\bar Y_{\left(\lceil(\alpha^*/2)B\rceil\right)}-q_W\left(\alpha^*/2|\mathbf{X}_m\right)\big|
+ \big| \bar Y_{\left(\lceil(1-\alpha^*/2)B\rceil\right)} -q_W\left(1-\alpha^*/2|\mathbf{X}_m\right)\big| \geq  \delta ~\Big|~ \mathbf{X}_m \Big) \nonumber \\
& \leq \text{P}\Big( \big|\bar Y_{\left(\lceil(\alpha^*/2)B\rceil\right)}-q_W\left(\alpha^*/2|\mathbf{X}_m\right)\big| \geq  \frac{\delta}{2} ~\Big|~ \mathbf{X}_m \Big)  \nonumber \\
&\quad +  \text{P}\Big( \big| \bar Y_{\left(\lceil(1-\alpha^*/2)B\rceil\right)} -q_W\left(1-\alpha^*/2|\mathbf{X}_m\right)\big| \geq  \frac{\delta}{2} ~\Big|~ \mathbf{X}_m \Big) \nonumber \\
&< \frac{\epsilon}{2} + \frac{\epsilon}{2} = \epsilon.
\end{align}
Hence \eqref{eq.CrIconverge} has been proved. \hfill $\Box$
\end{sloppypar}
\vspace{8mm}




\noindent {\bf {Proof of Theorem \ref{thm:cribound} (ii):}}

According to Condition (3), for any $\delta>0$, there exists $\epsilon_1>0$, such that $|\mu(F)-\mu(F^c)|<\delta/2$ if $d_{LP}(F,F^c)<2\epsilon_1$. In other words,
\begin{align}\label{mulp}
& |\mu(F)-\mu(F^c)|> \delta \implies |\mu(F)-\mu(F^c)|\geq \delta/2  \nonumber \\
& \implies d_{LP}(F,F^c)\geq 2\epsilon_1 \implies d_{LP}(F,F^c) > \epsilon_1.
\end{align}
Based on Lemma \ref{lemma:levy}, for any given $\epsilon_1>0, \epsilon_2>0,\epsilon_3>0$, there exists a sufficiently large integer $M_0$, such that for all $m>M_0$,
\begin{equation}\label{weakcon}
\text{P}_{F^c}\left[\text{P}\left(\left\{d_{LP}(\widetilde{F},F^c) > \epsilon_1\right\}~\big|~\mathbf{X}_m\right)>\epsilon_2 \right]<\epsilon_3,
\end{equation}
where $\widetilde{F}\sim {p}\left(F\big|\mathbf{X}_m\right)$ and $\text{P}_{F^c}$ denotes the probability measure of $\mathbf{X}_m$. Now the weak consistency of $p(F|\mathbf{X}_m)$ at $F^c$ as defined in \eqref{weakcon} is assumed in Condition (4). Hence, from \eqref{mulp} and \eqref{weakcon}, we have that for any $\delta>0,\epsilon_2>0,\epsilon_3>0$, there exists a large integer $M_0$ {that depends on $\delta,\epsilon_2,\epsilon_3$}, such that for all $m>M_0$,
\begin{align} 
& \text{P}_{F^c}\left[\text{P}\left(\left\{ \left|\mu(\widetilde{F})-\mu(F^c)\right| > \delta\right\}~\big|~\mathbf{X}_m\right)>\epsilon_2 \right] \nonumber \\
& \leq \text{P}_{F^c}\left[\text{P}\left(\left\{ d_{LP}(\widetilde{F},F^c) > \epsilon_1\right\}~\big|~\mathbf{X}_m\right)>\epsilon_2 \right]<\epsilon_3, \nonumber
\end{align}
or equivalently
\begin{align}\label{weakconmu}
&~~ \text{P}_{F^c}\left[\text{P}\left(\left\{ \mu(F^c)-\delta \leq \mu(\widetilde{F})\leq \mu(F^c)+ \delta\right\}~\big|~\mathbf{X}_m\right) > 1-\epsilon_2 \right] \nonumber \\
& = \text{P}_{F^c}\left[\text{P}\left(\left\{ |\mu(\widetilde{F})-\mu(F^c)| \leq \delta\right\}~\big|~\mathbf{X}_m\right) > 1-\epsilon_2 \right] \geq 1-\epsilon_3.
\end{align}
According to Condition (1), the conditional posterior distribution $F_W(\cdot|\mathbf{X}_m)$ for $W=\mu(\widetilde F)$ with $\widetilde F\sim p(F|\mathbf{X}_m)$ has a positive density on its support. Therefore, in its support, $F_W(\cdot|\mathbf{X}_m)$ is a strictly monotone continuous cumulative distribution function, and its quantile function (as its inverse) $q_W(\gamma|\mathbf{X}_m)$ is also continuous. Based on this relation, we have that for $W=\mu(\widetilde{F})$,
\begin{align}\label{cdfbound1}
& \text{P}\left(\left\{ \mu(F^c)-\delta \leq \mu(\widetilde{F})\leq \mu(F^c)+ \delta\right\}~\big|~\mathbf{X}_m\right) > 1-\epsilon_2 \nonumber \\
&\implies \text{P}\left(\left\{\mu(F^c)-\delta \leq \mu(\widetilde{F}) \right\}~\big|~\mathbf{X}_m\right) > 1-\epsilon_2 \nonumber \\
&\implies \text{P}\left(\left\{ \mu(F^c)-\delta \leq W \right\}~\big|~\mathbf{X}_m\right) > 1-\epsilon_2 \nonumber \\
&\implies 1 - F_W(\mu(F^c)-\delta ~\big|~\mathbf{X}_m) > 1- \epsilon_2 \nonumber \\
&\implies F_W(\mu(F^c)-\delta ~\big|~\mathbf{X}_m) \leq \epsilon_2.
\end{align}
And similarly
\begin{align}\label{cdfbound2}	
& \text{P}\left(\left\{ \mu(F^c)-\delta \leq \mu(\widetilde{F})\leq \mu(F^c)+ \delta\right\}~\big|~\mathbf{X}_m\right) > 1-\epsilon_2 \nonumber \\
&\implies \text{P}\left(\left\{ \mu(\widetilde{F})\leq \mu(F^c)+ \delta \right\}~\big|~\mathbf{X}_m\right) > 1-\epsilon_2 \nonumber \\
&\implies \text{P}\left(\left\{W \leq \mu(F^c)+ \delta\right\}~\big|~\mathbf{X}_m\right) > 1-\epsilon_2 \nonumber \\
&\implies F_W(\mu(F^c)+\delta ~\big|~\mathbf{X}_m) > 1- \epsilon_2.
\end{align}
Now for any given quantile $\gamma\in (0,1)$, if $0<\epsilon_2<\min\{\gamma,1-\gamma\}$, then $\epsilon_2 < \gamma < 1-\epsilon_2$. For such small $\epsilon_2$, the continuity of the quantile function $q_W(\gamma|\mathbf{X}_m)$, \eqref{cdfbound1} and \eqref{cdfbound2} imply that
\begin{align}\label{quant2side}
& F_W(\mu(F^c)-\delta ~\big|~\mathbf{X}_m) \leq \epsilon_2 < \gamma < 1-\epsilon_2 < F_W(\mu(F^c)+\delta ~\big|~\mathbf{X}_m) \nonumber \\
& \implies \mu(F^c) -\delta < q_W(\gamma|\mathbf{X}_m) < \mu(F^c) + \delta \implies \left|q_W(\gamma|\mathbf{X}_m)- \mu(F^c) \right|<\delta.
\end{align}
If we combine the relations from \eqref{weakconmu}, \eqref{cdfbound1}, \eqref{cdfbound2}, and \eqref{quant2side}, then we have shown that for any given $\delta>0,\epsilon_3>0,\gamma\in (0,1)$, $\epsilon_2\in(0,\min\{\gamma,1-\gamma\})$, there exists a sufficiently large integer $M_0$ {that depends on $\delta,\epsilon_2,\epsilon_3$}, such that for all $m>M_0$,
\begin{align}\label{quantconv1}
& ~~  \text{P}_{F^c}\left[\left|q_W(\gamma|\mathbf{X}_m)- \mu(F^c) \right|<\delta\right] \nonumber \\
&\geq \text{P}_{F^c}\left[F_W(\mu(F^c)-\delta ~\big|~\mathbf{X}_m) \leq \epsilon_2 < \gamma < 1-\epsilon_2 < F_W(\mu(F^c)+\delta\left|~\mathbf{X}_m\right.)\right] \nonumber \\
&= \text{P}_{F^c}\left[F_W(\mu(F^c)-\delta ~\big|~\mathbf{X}_m) \leq \epsilon_2 \text{ and } F_W(\mu(F^c)+\delta\left|~\mathbf{X}_m\right) > 1- \epsilon_2 \right] \nonumber \\
&\geq \text{P}_{F^c}\left[\text{P}\left(\left\{ \mu(F^c)-\delta \leq \mu(\widetilde{F})\leq \mu(F^c)+ \delta\right\}~\big|~\mathbf{X}_m\right) > 1-\epsilon_2 \right] \nonumber \\
&\geq 1-\epsilon_3.
\end{align}

From Equation \eqref{w2diff10} in the proof of Part (i), we have that for the $\delta>0$, $\epsilon_2>0$, and $\gamma\in(0,1)$ given as above, there exist $B_2$ and $n_{\min,2}$ that only depend on $\epsilon_2,\delta,\gamma,C_{\sigma},\mathbf{X}_m$ and the function $\mu(\cdot)$, such that for all $B>B_2$ and $n_{\min}>n_{\min,2}$,
\begin{align}
& ~~ \text{P}\left( \left|\bar Y_{(\lceil\gamma B\rceil)}-q_W(\gamma|\mathbf{X}_m)\right| > \delta ~\Big|~ \mathbf{X}_m \right) < \epsilon_2. \nonumber
\end{align}
Since this relation always holds true conditional on $\mathbf{X}_m$, it implies that
\begin{align}\label{quantconv2}
& ~~ \text{P}_{F^c}\left[ \text{P}\left( \left|\bar Y_{(\lceil\gamma B\rceil)}-q_W(\gamma|\mathbf{X}_m)\right| > \delta ~\Big|~ \mathbf{X}_m \right) > \epsilon_2 \right] = 0.
\end{align}
Finally, based on \eqref{quantconv1} and \eqref{quantconv2}, we have that
\begin{align}\label{quantconv3}
&~~  \text{P}_{F^c}\left[\text{P}\left( \left|\bar Y_{(\lceil\gamma B\rceil)}-\mu(F^c)\right| \geq 2\delta ~\Big|~ \mathbf{X}_m \right) > 2\epsilon_2 \right] \nonumber \\
&\leq  \text{P}_{F^c}\left[\text{P}\left( \left|\bar Y_{(\lceil\gamma B\rceil)}-q_W(\gamma|\mathbf{X}_m)\right| + \left|q_W(\gamma|\mathbf{X}_m)-\mu(F^c)\right| \geq 2\delta ~\Big|~ \mathbf{X}_m \right) > 2\epsilon_2 \right] \nonumber \\
&\leq  \text{P}_{F^c}\left[\text{P}\left( \left|\bar Y_{(\lceil\gamma B\rceil)}-q_W(\gamma|\mathbf{X}_m)\right| > \delta ~\Big|~ \mathbf{X}_m \right) + \text{P}\left(\left|q_W(\gamma|\mathbf{X}_m)-\mu(F^c)\right| \geq \delta ~\Big|~ \mathbf{X}_m \right) > 2\epsilon_2 \right] \nonumber \\
&\scalebox{0.95}{$
\leq \text{P}_{F^c}\left[\text{P}\left( \left|\bar Y_{(\lceil\gamma B\rceil)}-q_W(\gamma|\mathbf{X}_m)\right| > \delta ~\Big|~ \mathbf{X}_m \right)  > \epsilon_2 \right] + \text{P}_{F^c}\left[\text{P}\left(\left|q_W(\gamma|\mathbf{X}_m)-\mu(F^c)\right| \geq \delta ~\Big|~ \mathbf{X}_m \right) > \epsilon_2 \right]$} \nonumber \\
&\stackrel{(*)}{\leq} \text{P}_{F^c}\left[\text{P}\left( \left|\bar Y_{(\lceil\gamma B\rceil)}-q_W(\gamma|\mathbf{X}_m)\right| > \delta ~\Big|~ \mathbf{X}_m \right)  > \epsilon_2 \right]+ \text{P}_{F^c}\left[ \left|q_W(\gamma|\mathbf{X}_m)-\mu(F^c)\right| \geq \delta \right] \nonumber \\
&\stackrel{(**)}{<} 0 + \epsilon_3 = \epsilon_3,
\end{align}
where (*) follows because given $\mathbf{X}_m$, $\{\left|q_W(\gamma|\mathbf{X}_m)-\mu(F^c)\right| \geq \delta\}$ is a deterministic event (with conditional probability either 0 or 1); (**) follows because of \eqref{quantconv1}. Note that since $B_2(\zeta,\epsilon_2,\mathbf{X}_m)$ and $n_{\min,2}(\zeta,\epsilon_2,\mathbf{X}_m)$ depend on $\zeta,\epsilon_2,\mathbf{X}_m$, and $\zeta$ depends on $\delta,\gamma,\mathbf{X}_m$, so $B_2$ and $n_{\min,2}$ depend on $\delta,\gamma,\epsilon_2,\mathbf{X}_m$. Therefore, the conclusion of Theorem \ref{thm:cribound} (ii) follows by renaming $2\delta$ by $\delta$, $2\epsilon_2$ by $\eta$ (such that $\eta<\in (0, 2\min\{\gamma,1-\gamma\}) $), and $2\epsilon_3$ by $\epsilon$. \hfill $\Box$

	
	\section{{Appendix: Variance Decomposition of System Performance Estimation}}
	\label{subsec:AppendixVarianceDecomposition}

	\noindent {\bf {Proof of Theorem \ref{thm:vardecomp}(i):}}
		
		Given the real-world data $\mathbf{X}_m$, the variance of $\bar{Y}(\widetilde{F})$ quantifies the overall estimation uncertainty of the system mean response $\mu^c=\mu(F^c)$. Here, we decompose this variance to measure the relative contributions from the input and simulation uncertainties,
		\begin{eqnarray}
		\mbox{Var}(\bar{Y}(\widetilde{F})|\mathbf{X}_m) 
		&= & \mbox{E}_{\widetilde{F}^{(b)}}[\mbox{Var}(\bar{Y}_b|\mathbf{X}_m,\widetilde{F}^{(b)})|\mathbf{X}_m]+\mbox{Var}_{\widetilde{F}^{(b)}}[\mbox{E}(\bar{Y}_b|\mathbf{X}_m,\widetilde{F}^{(b)})|\mathbf{X}_m]  
		\nonumber\\
		&=& \mbox{E}_{\widetilde{F}^{(b)}}\left[\left.\frac{\sigma_b^2}{n_b}\right|\mathbf{X}_m\right] + \mbox{Var}_{\widetilde{F}^{(b)}}[\mu_b|\mathbf{X}_m] 
		\label{eq.VarDecomp1}	\\
		&\approx & \frac{1}{B} \sum_{b=1}^B \frac{S_b^2}{n_b}
		+ \frac{1}{B} \sum_{b=1}^B (\bar{Y}_b-\bar{\bar{Y}})^2.
		\nonumber
		\end{eqnarray}
		On the right side of Equation~(\ref{eq.VarDecomp1}), the first term $\sigma_S^2\equiv\mbox{E}_{\widetilde{F}^{(b)}}\left[\left.\frac{\sigma_b^2}{n_b}\right|\mathbf{X}_m\right]$ measures the impact from simulation uncertainty and the second term $\sigma_I^2\equiv \mbox{Var}_{\widetilde{F}^{(b)}}[\mu_b|\mathbf{X}_m] $ measures the impact from input uncertainty. 
		Since the sample mean and variance $\bar{Y}_b$ and $S^2_b$ are the consistent estimators for $\mu_b$ and $\sigma^2_b$, we estimate $\sigma_S^2$ with $\widehat{\sigma}_S^2=\frac{1}{B} \sum_{b=1}^B \frac{S_b^2}{n_b}$ and estimate $\sigma_I^2$ with $\widehat{\sigma}_I^2= \frac{1}{B} \sum_{b=1}^B (\bar{Y}_b-\bar{\bar{Y}})^2$, where $\bar{\bar{Y}}=\frac{1}{B}\sum_{b=1}^B \bar{Y}_b$. $\Box$

		\vspace{.3cm}
	
	\noindent {\bf {Proof of Theorem \ref{thm:vardecomp}(ii):}}
	
We first prove 
	$\sigma_S^2 \stackrel{p}\rightarrow 0$ as $n_{\min}\rightarrow \infty$. For any $\delta>0,\epsilon>0,C_{\sigma}>0$, let $n_{\min} >C_{\sigma}/(\delta\epsilon)$ such that
	\begin{align*}
	&\mbox{P}[\sigma_S^2(\mathbf{X}_m)\geq \delta] \stackrel{(*)}\leq \frac{1}{\delta}\mbox{E}\left[\sigma_S^2(\mathbf{X}_m)\right]  
	= \frac{1}{\delta}\mbox{E}\left[
	\mbox{E}\left[\left. \frac{\sigma_b^2}{n_b}\right|\mathbf{X}_m\right]\right]  \stackrel{(**)}\leq\frac{1}{\delta}\frac{C_{\sigma}}{n_{\min}}  < \epsilon,   
	\end{align*}
	where (*) follows by the Markov's inequality and (**) follows according to Condition (2) of Theorem~\ref{thm:cribound}. Thus, $\sigma_S^2 \stackrel{p}\rightarrow 0$ as $n_{\min}\rightarrow \infty$.

Then, we prove $\sigma_I^2 \stackrel{p}\rightarrow 0$ as $m \rightarrow \infty$. By the Markov's inequality, we have for any $\delta>0$,
	\begin{align} \label{eq.midstep1}
	\mbox{P} \left[\sigma_I^2(\mathbf{X}_m)\geq \delta \right] \leq \frac{1}{\delta}\mbox{E}\left[\sigma_I^2(\mathbf{X}_m)\right] 
	= \frac{1}{\delta}
	\mbox{E}\left[
	\mbox{E}\left[(\mu_b-\bar{\mu})^2 | \mathbf{X}_m\right] \right] 
	\stackrel{(*)}\leq \frac{1}{\delta}\mbox{E}\left[
	\mbox{E}[(\mu_b-\mu^c)^2 | \mathbf{X}_m ]\right], 	
	\end{align}
	where (*) follows because the sample mean $\bar{\mu} = \arg\min_{\mu} \mbox{E} (\mu_b-\mu)^2$.

	 Since $|\mu(\widetilde F)|\leq C_{\mu}$ for almost surely all $\widetilde F\sim p(F|\mathbf{X}_m)$, we have that $(\mu_b-\mu^c)^2\leq 2[\mu_b^2+(\mu^c)^2]\leq 2[C_{\mu}^2+ (\mu^c)^2]$ for all $b=1,2,\ldots,B$. Then for any $\delta_1>0$, 
	\begin{align} \label{probeqn1}
	\text{E} \left[(\mu_b-\mu^c)^2 ~\big|~\mathbf{X}_m\right] =& \text{E} \left[(\mu_b-\mu^c)^2\cdot \mbox{I}(|\mu_b-\mu^c|\leq \delta_1) ~\big|~\mathbf{X}_m\right] \nonumber  \\
	&+ \text{E} \left[(\mu_b-\mu^c)^2\cdot \mbox{I}(|\mu_b-\mu^c|> \delta_1) ~\big|~\mathbf{X}_m\right] \nonumber \\
	\leq & \delta_1^2 + 2\left[C_{\mu}^2+ (\mu^c)^2\right]\text{P} \left[|\mu_b-\mu^c|> \delta_1 ~\big|~\mathbf{X}_m\right],
	\end{align} 
 	where $\mbox{I}(\cdot)$ is the indicator function. 	
		From \eqref{mulp} obtained by applying Condition~(3) of Theorem~\ref{thm:cribound}, for any $\delta_1>0$, there exists $\epsilon_1>0$ such that 
$|\mu(F)-\mu(F^c)|> \delta_1 \implies d_{LP}(F,F^c) > \epsilon_1$. 
 By Condition (4) of Theorem \ref{thm:cribound}, for this $\epsilon_1$ and any $\epsilon_2>0$, $\epsilon_3>0$, there exists a large integer $m_0$ that depends on $\epsilon_1,\epsilon_2,\epsilon_3$, such that for all $m>m_0$,  
	\begin{align*}
	\text{P}_{F^c}\left[\text{P}\left(\left\{d_{LP}(\widetilde{F},F^c) > \epsilon_1\right\}~\big|~\mathbf{X}_m\right) > \epsilon_2 \right] < \epsilon_3. 
	\end{align*}
Thus, for all $m>m_0$,
	\begin{align}\label{probeqn2} 
	& \text{P}_{F^c}\left[\text{P}\left(\left\{ \left|\mu_b -\mu^c\right| > \delta_1 \right\}~\big|~\mathbf{X}_m\right)>\epsilon_2 \right] \nonumber \\
	& \leq \text{P}_{F^c}\left[\text{P}\left(\left\{ d_{LP}(\widetilde{F},F^c) > \epsilon_1\right\}~\big|~\mathbf{X}_m\right)>\epsilon_2 \right]<\epsilon_3. 
	\end{align}
    Then, \eqref{probeqn1} and \eqref{probeqn2} together imply that for all $m>m_0$,
	\begin{align} \label{probeqn3}
	& \text{E}\left[(\mu_b-\mu^c)^2\right] = \text{E}_{\mathbf{X}_m} \text{E}\left[(\mu_b-\mu^c)^2 ~\big|~\mathbf{X}_m\right] \nonumber \\
	& \leq \delta_1^2 + 2\left[C_{\mu}^2+ (\mu^c)^2\right]\text{E}_{\mathbf{X}_m} \left\{\text{P} \left[|\mu_b-\mu^c|> \delta_1 ~\big|~\mathbf{X}_m\right]\right\} \nonumber \\
	& \leq \delta_1^2 + 2\left[C_{\mu}^2+ (\mu^c)^2\right] \text{E}_{\mathbf{X}_m} \left\{\text{P} \left[|\mu_b-\mu^c|> \delta_1 ~\big|~\mathbf{X}_m\right]\cdot \mbox{I}(\text{P} \left[|\mu_b-\mu^c|> \delta_1 ~\big|~\mathbf{X}_m\right] \leq \epsilon_2)\right\} \nonumber \\
	& ~~ + 2\left[C_{\mu}^2+ (\mu^c)^2\right] \text{E}_{\mathbf{X}_m} \left\{\text{P} \left[|\mu_b-\mu^c|> \delta_1 ~\big|~\mathbf{X}_m\right]\cdot \mbox{I}(\text{P} \left[|\mu_b-\mu^c|> \delta_1 ~\big|~\mathbf{X}_m\right] > \epsilon_2)\right\} \nonumber \\
	& \leq \delta_1^2 + 2\left[C_{\mu}^2+ 2(\mu^c)^2\right] \text{E}_{\mathbf{X}_m} \left\{\epsilon_2\cdot \mbox{I}(\text{P} \left[|\mu_b-\mu^c|> \delta_1 ~\big|~\mathbf{X}_m\right] \leq \epsilon_2)\right\} \nonumber \\
	& ~~ + 2\left[C_{\mu}^2+ (\mu^c)^2\right] \text{E}_{\mathbf{X}_m} \left\{1\cdot \mbox{I}(\text{P} \left[|\mu_b-\mu^c|> \delta_1 ~\big|~\mathbf{X}_m\right] > \epsilon_2)\right\} \nonumber \\
	&\leq \delta_1^2 + 2\left[C_{\mu}^2+ (\mu^c)^2\right] \epsilon_2 + 2\left[C_{\mu}^2+ (\mu^c)^2\right] \text{P}_{F^c}\left[\text{P}\left(\left\{\left|\mu_b -\mu^c\right| > \delta_1 \right\}~\big|~\mathbf{X}_m\right)>\epsilon_2 \right] \nonumber \\
	&\stackrel{(*)}{\leq} \delta_1^2 + 2\left[C_{\mu}^2+ (\mu^c)^2\right] (\epsilon_2 + \epsilon_3),
	\end{align} 
	where (*) follows from \eqref{probeqn2}.
	For any $\epsilon>0$ and the $\delta>0$ given in \eqref{eq.midstep1}, we can choose $\delta_1=\sqrt{\epsilon\delta/2}$, $\epsilon_2=\epsilon_3 = \epsilon\delta/\left[8C_{\mu}^2+ 8(\mu^c)^2\right]$, such that from \eqref{eq.midstep1} and \eqref{probeqn3},
	\begin{align*}
	& \mbox{P} \left[\sigma_I^2(\mathbf{X}_m,\mathcal{F}_B)\geq \delta \right] \leq  \frac{1}{\delta}  \mbox{E}\left[(\mu_b-\mu^c)^2\right] \leq \frac{1}{\delta} \left\{ \delta_1^2 + 2\left[C_{\mu}^2+ (\mu^c)^2\right] (\epsilon_2 + \epsilon_3)\right\} \\ 
	&\leq \frac{1}{\delta} \left\{ \frac{\epsilon\delta}{2} + 2\left[C_{\mu}^2+ (\mu^c)^2\right] \cdot \frac{2\epsilon\delta}{8\left[C_{\mu}^2+ (\mu^c)^2\right]}\right\} = \epsilon,
	\end{align*}
	as long as $m>m_0$, where $m_0$ depends on $\epsilon_1,\epsilon_2,\epsilon_3$, or equivalently, $m_0$ depends on $\epsilon$ and $\delta$. 
	This has shown that $\sigma_I^2 \stackrel{p}\rightarrow 0$ as $m \rightarrow \infty$. $\Box$

	
	\section{Appendix: Sensitivity Analysis of Hyper-parameters for $\pmb{\theta}_{\alpha}$}
	\label{subsec:AppendixSensitivityAnalysis}
	We use examples listed in Table~\ref{table:example} with sample size $m=50$ to study the sensitivity to the values of hyper-parameters $\pmb{\theta}_{\alpha}$. DPM with appropriate kernel densities are used for different examples.
	That means DPM with Gamma kernel used for Example~1 and 2, DPM with Gaussian kernel used for Examples~3, and DPM with Beta kernel used for Example~4.
	Table~\ref{table:alpha} records 95\% symmetric CIs of KS and AD distances obtained from 1000 macro-replications. The results indicate that the values of hyper-parameters $\pmb{\theta}_{\alpha}$ have an insignificant impact on the input model estimation, where $\mbox{Gamma}(2,4)$ prior was used in \cite{Escobar_West_1995} and the discrete $\mbox{Gamma}(1,1)$ prior was used in \cite{Wang2011}. The choice of hyper-parameters does not have significant impact on the density estimation accuracy.

	\begin{table}[]
		\centering
		\caption{KS and AD distances for Examples~1--4 with different hyper-parameters $\pmb{\theta}_{\alpha}$}
		\label{table:alpha}
		\scalebox{0.75}{
			\begin{tabular}{|c|c|c|c|c|c|}
				\hline
				\multicolumn{2}{|c|}{$m=50$}              & Example 1        & Example 2       & Example 3       & Example 4       \\ \hline
				\multirow{2}{*}{$\mbox{Gamma}(0.5,0.5)$} & $D_m$ & 0.106$\pm$0.002  & 0.073$\pm$0.001 & 0.075$\pm$0.001 & 0.070$\pm$0.001 \\ \cline{2-6}
				& $A_m$ & 11.870$\pm$0.175 & 7.594$\pm$0.097 & 6.365$\pm$0.096 & 8.808$\pm$0.097 \\ \hline
				\multirow{2}{*}{$\mbox{Gamma}(1,1)$}     & $D_m$ & 0.102$\pm$0.002  & 0.071$\pm$0.001 & 0.072$\pm$0.001 & 0.068$\pm$0.001 \\ \cline{2-6}
				& $A_m$ & 11.278$\pm$0.158 & 7.203$\pm$0.088 & 6.083$\pm$0.093 & 8.253$\pm$0.092 \\ \hline
				\multirow{2}{*}{$\mbox{Gamma}(4,4)$}     & $D_m$ & 0.104$\pm$0.002  & 0.074$\pm$0.001 & 0.075$\pm$0.001 & 0.069$\pm$0.001 \\ \cline{2-6}
				& $A_m$ & 11.495$\pm$0.166 & 7.787$\pm$0.104 & 6.484$\pm$0.098 & 8.490$\pm$0.095 \\ \hline
				\multirow{2}{*}{$\mbox{Gamma}(2,4)$}     & $D_m$ & 0.105$\pm$0.002  & 0.072$\pm$0.001 & 0.073$\pm$0.001 & 0.068$\pm$0.001 \\ \cline{2-6}
				& $A_m$ & 11.762$\pm$0.174 & 7.419$\pm$0.092 & 6.207$\pm$0.094 & 8.337$\pm$0.094 \\ \hline
			\end{tabular}
		}
	\end{table}

\end{document}